\pdfoutput=1

\documentclass[iop, apj, twocolappendix, numberedappendix]{emulateapj}
\usepackage{xspace}
\usepackage{amsmath}
\usepackage{framed} 
\usepackage{txfonts}
\usepackage{epstopdf}
\usepackage{color}
\usepackage{rotating}
\usepackage{natbib}
\usepackage{ulem}
\usepackage{xspace}
\usepackage[colorlinks=true,urlcolor=black,linkcolor=blue,citecolor=blue]{hyperref}
\usepackage{sistyle}
\SIthousandsep{,}

\def\figdir{.}

\newcommand{\msun}{\>{M_{\odot}}}

\newcommand{\msunyr}{\>{M_{\odot}/{\rm yr}}}

\def\gcm3{\mathrm{g} / \mathrm{cm}^3}
\def\m200m{M_{\rm 200m}}

\def\gtsima{$\; \buildrel > \over \sim \;$}
\def\ltsima{$\; \buildrel < \over \sim \;$}
\def\prosima{$\; \buildrel \propto \over \sim \;$}
\def\gsim{\lower.7ex\hbox{\gtsima}}
\def\lsim{\lower.7ex\hbox{\ltsima}}
\def\simgt{\lower.7ex\hbox{\gtsima}}
\def\simlt{\lower.7ex\hbox{\ltsima}}
\def\simpr{\lower.7ex\hbox{\prosima}}

\def\illustris{Illustris\xspace}
\def\fret{f_{\rm ret}}
\def\tpeak{t_{\rm peak}}
\def\mlimit{M_{\rm final}}
\def\width{\sigma_{\rm SFR}}

\usepackage{etoolbox}
\makeatletter
\patchcmd{\NAT@citex}
  {\@citea\NAT@hyper@{\NAT@nmfmt{\NAT@nm}\NAT@date}}
  {\@citea\NAT@nmfmt{\NAT@nm}\NAT@hyper@{\NAT@date}}
  {}
  {}
\patchcmd{\NAT@citex}
  {\@citea\NAT@hyper@{%
     \NAT@nmfmt{\NAT@nm}%
     \hyper@natlinkbreak{\NAT@aysep\NAT@spacechar}{\@citeb\@extra@b@citeb}%
     \NAT@date}}
  {\@citea\NAT@nmfmt{\NAT@nm}%
   \NAT@aysep\NAT@spacechar%
   \NAT@hyper@{\NAT@date}}
  {}
  {}
\patchcmd{\NAT@citex}
  {\@citea\NAT@hyper@{%
     \NAT@nmfmt{\NAT@nm}%
     \hyper@natlinkbreak{\NAT@spacechar\NAT@@open\if*#1*\else#1\NAT@spacechar\fi}%
       {\@citeb\@extra@b@citeb}%
     \NAT@date}}
  {\@citea\NAT@nmfmt{\NAT@nm}%
   \NAT@spacechar\NAT@@open\if*#1*\else#1\NAT@spacechar\fi%
   \NAT@hyper@{\NAT@date}}
  {}
  {}
\makeatother

\shorttitle{Star Formation Histories}
\shortauthors{Diemer et al.}

\journalinfo{The Astrophysical Journal, {\rm 839:26 (20pp), 2017 April 10}}
\submitted{Received 2017 January 9; revised 2017 March 2; accepted 2017 March 21; published 2017 April 11}

\begin{document}


\defcitealias{gladders_13_icbs4}{G13}
\defcitealias{pacifici_16}{P16}


\title{Log-normal star formation histories in simulated and observed galaxies}
\author{Benedikt Diemer\altaffilmark{1}, Martin Sparre\altaffilmark{2,3}, Louis E. Abramson\altaffilmark{4}, and Paul Torrey\altaffilmark{5,6}}

\affil{
$^1$ Institute for Theory and Computation, Harvard-Smithsonian Center for Astrophysics, 60 Garden St., Cambridge, MA 02138, USA; \href{mailto:benedikt.diemer@cfa.harvard.edu}{benedikt.diemer@cfa.harvard.edu} \\
$^2$ Heidelberger Institut f\"ur Theoretische Studien, Schloss-Wolfsbrunnenweg 35, 69118 Heidelberg, Germany \\
$^3$ Dark Cosmology Centre, Niels Bohr Institute, University of Copenhagen, Juliane Maries Vej 30, 2100 Copenhagen, Denmark \\
$^4$ Department of Physics \& Astronomy, UCLA, 430 Portola Plaza, Los Angeles, CA 90095-1547, USA \\
$^5$ MIT Kavli Institute for Astrophysics and Space Research, 77 Massachusetts Ave. 37-241, Cambridge MA 02139,USA \\
$^6$ California Institute of Technology, Pasadena, CA 91125, USA \\
}


\begin{abstract}
Gladders et al. have recently suggested that the star formation histories (SFHs) of individual galaxies are characterized by a log-normal function in time, implying a slow decline rather than rapid quenching. We test their conjecture on theoretical SFHs from the cosmological simulation \illustris and on observationally inferred SFHs. While the log-normal form necessarily ignores short-lived features such as starbursts, it fits the overall shape of the majority of SFHs very well. In particular, 85\% of the cumulative SFHs are fitted to within a maximum error of 5\% of the total stellar mass formed, and 99\% to within 10\%. The log-normal performs systematically better than the commonly used delayed-$\tau$ model, and is superseded only by functions with more than three free parameters. Poor fits are mostly found in galaxies that were rapidly quenched after becoming satellites. We explore the log-normal parameter space of normalization, peak time, and full width at half maximum, and find that the simulated and observed samples occupy similar regions, though \illustris predicts wider, later-forming SFHs on average. The ensemble of log-normal fits correctly reproduces complex metrics such as the evolution of \illustris galaxies across the star formation main sequence, but overpredicts their quenching timescales. SFHs in \illustris are a diverse population not determined by any one physical property of galaxies, but follow a tight relation, where $\mathrm{width}\propto\mathrm{(peak\ time)}^{3/2}$. We show that such a relation can be explained qualitatively (though not quantitatively) by a close connection between the growth of dark matter halos and their galaxies.
\end{abstract}



\section{Introduction}
\label{sec:intro}

One of the most fundamental aspects of galaxy formation is the star formation history (SFH), both in individual galaxies and in the universe overall. Unfortunately, the time-resolved SFHs of individual galaxies are difficult to measure observationally in all but the most local galaxies where we have access to resolved stellar populations \citep{weisz_11, weisz_14, skillman_14, williams_11, lewis_15}. In more remote systems, we need to rely on stellar archaeology, i.e. measurements of stellar ages based on photometric or spectroscopic observations and stellar population synthesis models \citep{tinsley_68, gallagher_84, sandage_86, kauffmann_03, thomas_05, kennicutt_12, mcdermid_15, leja_16}. The number of independent time bins in such SFH measurements is usually small, though better time resolution can be achieved with more flexible parameterizations of the SFH \citep{tojeiro_07, tojeiro_09, pacifici_13, pacifici_16_timing, pacifici_16}. Instead of focusing on individual galaxies, one can try to measure the star formation rate (SFR) of galaxy populations at different redshifts and connect them to their progenitors statistically, but such inferences are complicated by scatter and merging in a $\Lambda$CDM universe \citep{behroozi_13_numberdensity, torrey_15_numberdensity, torrey_16_numberdensity, wellons_16_numberdensity}.

In a  global sense, however, many fundamental aspects of star formation in the universe are now well established. The star formation rate density (SFRD) peaks around $z \approx 2$ and declines thereafter \citep{lilly_96, madau_98, hopkins_06, madau_14}. At any redshift along this global trajectory, star-forming galaxies exhibit a correlation between their stellar mass and SFR, leading to the concepts of a ``star formation main sequence'' and a quiescent population \citep{brinchmann_04, noeske_07, elbaz_07, karim_11, whitaker_12, speagle_14}. Massive galaxies tend to form their stars earlier, a trend known as downsizing \citep{cowie_96, heavens_04, treu_05, bundy_06, neistein_06, kriek_07, conroy_09}. Finally, the SFR has been connected to a number of physical properties of galaxies such as  morphology \citep{postman_84, wuyts_11} and environment, which turns out to be closely related to whether a galaxy is a satellite or central \citep{oemler_74, dressler_80, peng_10, peng_12}.

One might hope that these global observations would strongly constrain the SFHs of individual galaxies, but this connection is not easily established. For example, average SFHs can be inferred by integrating the main-sequence SFR over time \citep[e.g.,][]{leitner_12}, but this approach leads to inconsistencies \citep{leja_15}. Instead, the most successful theoretical models link the growth of stellar mass to the growth of the dark matter halos that galaxies inhabit, for example, via subhalo abundance matching \citep[e.g.,][]{kravtsov_04, conroy_06, moster_13, behroozi_13_shmr}, halo occupation distributions \citep[e.g.,][]{peacock_00, seljak_00_wl, hearin_16_decoratedhod}, semi-analytic models \citep{kauffmann_93, somerville_01, guo_11}, or other assumptions \citep{bouche_10, dave_12, lilly_13, tacchella_13, mitra_17}. One important conclusion from these models is that there has to be significant scatter between halo and galaxy masses (and thus growth histories) in order to explain observations \citep{more_09_ii, behroozi_13_shmr, reddick_13, gu_16}. As a result, even models that agree on global constraints can lead to orthogonal interpretations of the evolution of individual galaxies. A good example for such disagreement is the ``rapid quenching\footnote{The term ``quenching'' is somewhat ambiguous. In this paper, we use it to mean the cessation of star formation, without any presumption as to whether the decrease happens quickly or slowly, and whether it happens due to a diminishing gas supply or other physical processes.}'' framework where galaxies follow the main sequence until they sharply fall below the main sequence \citep{peng_12, wetzel_13, tinker_16, tacchella_16_ms} and the ``stochastic'' framework where correlated scatter and the central limit theorem lead to the main sequence \citep{kelson_14, kelson_16}.

Recently, another rather different picture has been proposed. Inspired by the fact that the global SFRD is well fit by a log-normal in time, \citet[][hereafter \citetalias{gladders_13_icbs4}]{gladders_13_icbs4} suggested that this form might also describe the SFHs of individual galaxies \citep[\citetalias{gladders_13_icbs4};][]{dressler_13_icbs2, dressler_16, oemler_13_icbs3, abramson_15, abramson_16}. The log-normal SFR is given by the expression
\begin{equation}
\label{eq:lognormal}
\rm SFR(t) = \frac{A}{\sqrt{2 \pi \tau^2} \times t} \exp \left(-\frac{(\ln t - T_0)^2}{2 \tau^2} \right)
\end{equation}
where $A$, $T_0$, and $\tau$ are free parameters (throughout the paper, $t$ refers to the time since the big bang, not lookback time). \citetalias{gladders_13_icbs4} emphasize that the key assumption need not be the exact functional form of the log-normal, but rather its steep rise and slow decline in linear time, suggesting a physical picture different from main sequence star formation interrupted by sudden quenching. While there is some evidence that the majority of galaxies cease their star formation gradually \citep{noeske_07, schawinski_14, peng_15, eales_17, gutcke_17}, the log-normal SFR is no more than an assumption. The appeal of this assumption is that it allows for wide-ranging predictions if the log-normal parameters for a sample of galaxies can be inferred. This procedure was implemented by \citetalias{gladders_13_icbs4} who found surprisingly good agreement of the predicted stellar mass functions and their evolution, the star formation main sequence, downsizing, and many more complicated metrics \citep[\citetalias{gladders_13_icbs4};][]{abramson_16}. These successes cannot trivially be reproduced with symmetric forms of the SFH such as a Gaussian in linear time \citepalias{gladders_13_icbs4}.

Given the scarcity of reliable, time-resolved SFH observations, cosmological simulations of galaxy formation can help to differentiate between the different physical pictures. These simulations have recently reached a reasonable agreement with a number of observables, providing some level of confidence in their predictions for individual galaxies \citep{vogelsberger_14_nature, torrey_14, schaye_15, dave_16}. More specifically, \citet{sparre_15} showed that the galaxy population of the \illustris simulation broadly matches the observed main sequence, and investigated individual SFHs using principal component analysis (see also the related analyses of \citealt{simha_14} and \citealt{cohn_15}).

In this paper, we systematically investigate the fundamental assumption of \citetalias{gladders_13_icbs4}, namely whether the log-normal functional form is a good fit for SFHs in the \illustris simulation and for the inferred SFHs of \citet[][hereafter \citetalias{pacifici_16}]{pacifici_16}. We find that log-normals fit the majority of \illustris galaxies very well, particularly in the mass range studied by \citetalias{gladders_13_icbs4}. We investigate the log-normal parameter space of normalization, peak time, and width as a common language for simulated and observed SFHs. Our goal is not to test the fitting procedure of \citetalias{gladders_13_icbs4} in detail or to compare the galaxy population in \illustris to that of \citetalias{gladders_13_icbs4}. Instead, we study which physical properties of \illustris galaxies translate into particular values of the log-normal parameters. We also compare the log-normal to other fitting functions and discuss the implications of the log-normal framework in terms of quenching and the global star formation properties of the universe. 

\begin{figure}
\centering
\includegraphics[trim = 2mm 1mm 3mm 0mm, clip, scale=0.72]{\figdir/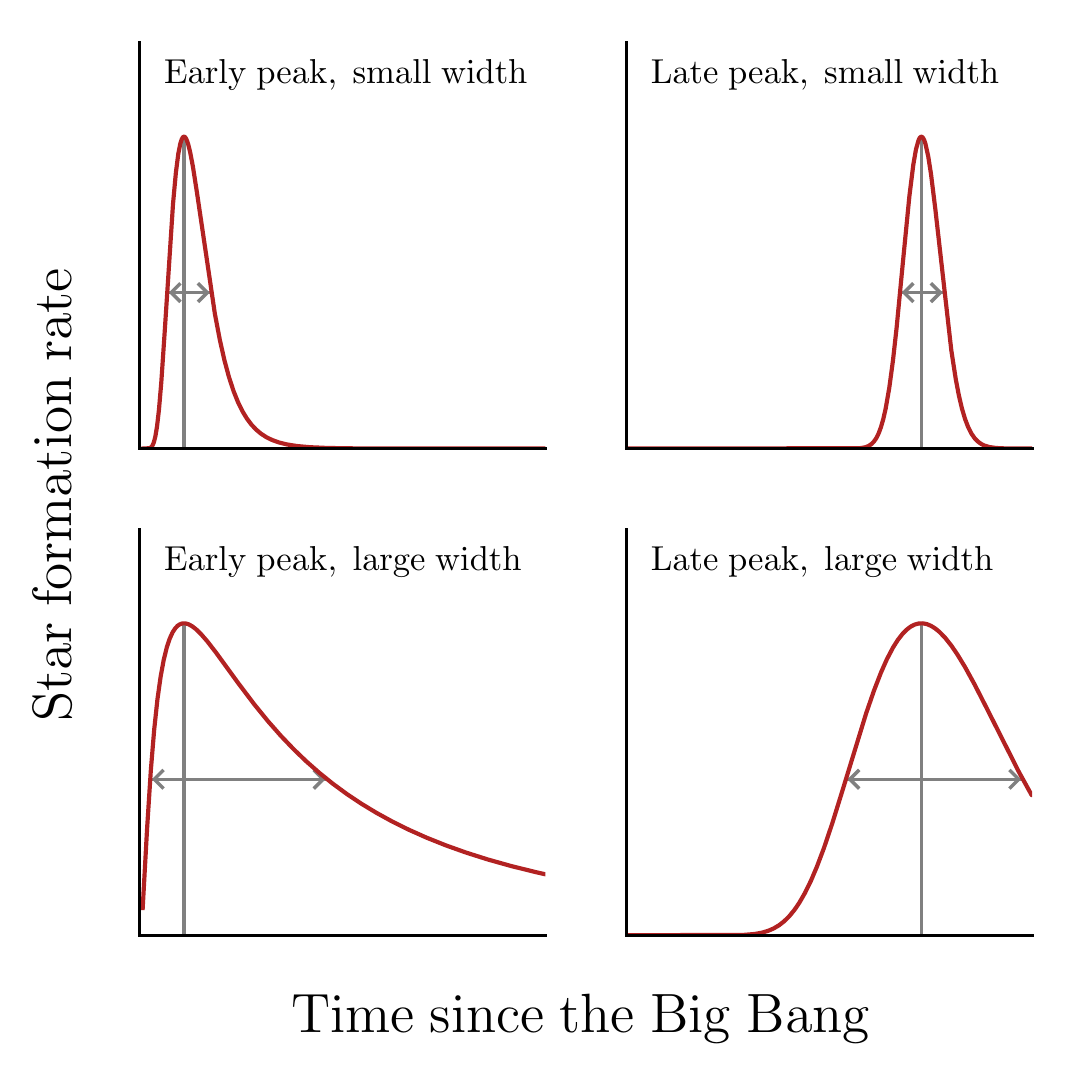}
\caption{Illustration of the log-normal SFH. The shape is determined by the peak time $\tpeak$ and width $\width$ (defined as full width at half maximum). The key feature of the log-normal is that, in linear time, it rises quickly and declines slowly, with the two timescales inextricably linked.}
\label{fig:toy}
\end{figure}

The paper is organized as follows. In Section \ref{sec:methods}, we describe the simulated and observed data our analysis is based on. In Section \ref{sec:results}, we investigate the quality of the log-normal fits and the resulting parameter space. In Section \ref{sec:discussion}, we discuss the implications of our findings for the global star formation properties of the universe and the quenching of star formation. We summarize our results in Section \ref{sec:conclusions}. In Appendices \ref{sec:app:funcs} and \ref{sec:app:fits} we give mathematical expressions of various fitting functions and discuss additional details regarding fits and fit results.

Throughout the paper, we extensively use the language of log-normal functions (Equation \ref{eq:lognormal}). This functional form imposes the constraint that the rise and decline of star formation are symmetric in logarithmic time. Hence, the parameters $T_0$ and $\tau$ are in units of logarithmic time, and can be interpreted as the time when half the stars have formed and the duration of the galaxy's star formation. However, the log-normal does not peak at $t = \exp(T_0)$, and we thus substitute the more intuitive parameter space of the SFR's peak time, $\tpeak$, and its full width at half maximum, $\width$. Figure~\ref{fig:toy} illustrates how these parameters determine the shape of the log-normal. Finally, the normalization $A$ corresponds to the total integrated star formation, a quantity we re-cast as the total stellar mass formed,
\begin{equation}
\label{eq:mlimit}
\mlimit = A \times 10^9 \times f_{\rm ret}
\end{equation}
where $\fret = 0.6$ is the retention factor due to stellar evolution (see Section \ref{sec:methods:illustris}). The numerical factor accounts for the fact that SFRs are measured in $\msunyr$ whereas times are in gigayears. Detailed expressions for a number of useful properties of log-normal SFHs are given in Appendix~\ref{sec:app:funcs:lognormal}.


\section{Simulation, Data, and Methods}
\label{sec:methods}

In this section, we introduce the observed and simulated datasets our analysis is based on, as well as our method for extracting and fitting SFHs from simulations.

\subsection{The Gladders et al. Galaxy Sample}
\label{sec:methods:g13}

Our primary observed dataset consists of the stellar masses, SFRs, and best-fit log-normal parameters of \citetalias{gladders_13_icbs4}. The underlying galaxy observations were taken from a number of galaxy surveys. In particular, the stellar masses and SFRs for galaxies with $M_* > 4\times10^{10} \msun$ were drawn from the Sloan Digital Sky Survey \citep[see][and references therein]{oemler_13_icbs1}, whereas the PG2MC survey \citep{calvi_11} was used for galaxies with $M_* < 4\times10^{10} \msun$. Due to the smaller volume of the latter survey, the samples were re-normalized to create equal weights. The resulting sample contains $2094$ galaxies with a mean redshift of $0.0678$, and is complete above $M_* = 10^{10} \msun$. We refer the reader to \citetalias{gladders_13_icbs4} and the references therein for details on how the stellar masses and SFRs were computed.

The fundamental assertion of \citetalias{gladders_13_icbs4} is that the log-normal, a functional form that describes the global SFRD of the universe, might also be a good description of the SFHs of individual galaxies. However, for each observed galaxy, only two variables are known ($M_*$ and SFR) whereas the log-normal has three free parameters (Equation \ref{eq:lognormal}). Thus, \citetalias{gladders_13_icbs4} used additional global constraints, namely the SFRD (accounting for the contribution from galaxies with $M_* < 10^{10} \msun$) and SFR distributions back to $z \approx 1$. Some of these observations were drawn from the IMACS Cluster Building Survey \citep[ICBS,][]{oemler_13_icbs1, dressler_13_icbs2, oemler_13_icbs3}. In a combined fit over the global constraints and the stellar masses and SFRs of each galaxy, they found the best-fit $T_0$ and $\tau$ for the $2094$ galaxies (see their Figure 9). The normalization of the log-normals was set such that the integrated SFR matches the observed $M_*$, meaning that one has to assume a retention factor, $\fret$, i.e. the ratio between the stellar mass initially formed to the stellar mass that survives until the galaxy is observed. \citetalias{gladders_13_icbs4} assumed a retention factor of $0.6$ which we adopt throughout this paper.

\subsection{The Pacifici et al. Inferred SFHs}
\label{sec:methods:p16}

\citetalias{pacifici_16} derived SFHs by fitting the multi-band photometry of 845 quiescent galaxies with spectral energy distributions (SEDs) computed from a large library of theoretical SFHs. The SED library consists of 500,000 simulated galaxy SEDs, created by applying the semi-analytic model of \citet{delucia_07} to the merger trees from the Millennium Simulation \citep[][see also \citealt{pacifici_12}]{springel_05_millennium}, and predicting the emission using the stellar population synthesis model of \citet{bruzual_03}. This library is flexible enough not to introduce a particular shape of the SFH a priori. \citetalias{pacifici_16} restricted themselves to quiescent galaxies, defined to have specific star formation rates
\begin{equation}
\label{eq:p16_sfr}
{\rm sSFR(z_{\rm obs})} \leq \frac{0.2}{10^9 \times t_{\rm obs}}
\end{equation}
where the sSFR is in units of $\rm yr^{-1}$ and $t_{\rm obs}$ is the cosmic time at observation in gigayears. For the comparisons with the \illustris and \citetalias{gladders_13_icbs4} samples, we set $z_{\rm obs} = 0$. We consider the median SFHs in six bins in redshift and six bins in mass given by \citetalias{pacifici_16} (see their Figure 5).

\subsection{The Illustris Simulation}
\label{sec:methods:illustris}

The \illustris simulation \citep{vogelsberger_14_nature} follows a comoving cosmological volume of 106.5 cubic Mpc. The cosmological parameters were set according to the WMAP9 cosmology of \citet{hinshaw_13}, and the same cosmology was adapted for all calculations in this paper. The simulation was run using the moving-mesh code \textsc{Arepo} \citep{springel_10} which includes physical models for gas cooling, star formation, metal enrichment, black hole growth, as well as feedback from stellar winds, supernovae, and AGNs \citep{vogelsberger_13, vogelsberger_14_illustris, vogelsberger_14_nature, genel_14, torrey_14, sijacki_15}. For the purposes of this paper, the star formation prescription is of particular interest. In \illustris, stars are formed according to a sub-grid model, which stochastically places ``star'' particles in gas cells that exceed a threshold number density of $0.13\ \rm cm^{-3}$ \citep{springel_03, vogelsberger_13}. Each star particle represents a population of stars born with a Chabrier initial mass function \citep{chabrier_03}. The timescale of star formation depends on the inverse root of the density, leading to a relation between surface density and star formation that roughly follows the observed Kennicutt-Schmidt relation \citep[e.g.,][]{kennicutt_98_review}. 

As they age, simulation star particles return mass to the interstellar medium via stellar winds and supernovae \citep{vogelsberger_13}. Star particles therefore have masses that evolve with time to values lower than their initial birth mass. For any given galaxy, this decrease causes the integrated SFR to be higher than the sum of stellar particle masses. Throughout the paper, we distinguish these two stellar mass definitions carefully. The mass decrease in a stellar population can be expressed as a retention factor, which we find to vary between $0.54$ for the oldest and $0.62$ for the youngest stellar populations in \illustris, with a mean of $0.57$. We note that this is very close to the value of $0.6$ assumed by \citetalias{gladders_13_icbs4}, meaning that we can directly compare the cumulative SFRs of the observed and simulated galaxy samples without incurring a large error.

Dark matter halos, and the galaxies that inhabit them, are identified using the \textsc{Subfind} algorithm \citep{davis_85, springel_01_subfind, dolag_09}. The various properties of galaxies are computed based on the bound matter within twice the the stellar half-mass radius of a subhalo as identified by \textsc{Subfind} \citep{vogelsberger_14_nature}. We use a number of galaxy properties listed in the \textsc{Subfind} catalogs, including various mass and size definitions, metallicity, black hole mass, and properties of the parent group if applicable. Furthermore, we use the stellar assembly data provided by \citet{rodriguezgomez_16}, for example, the fraction of a galaxy's stellar mass that was formed in situ, ex situ, and in various types of mergers. 

We consider all galaxies with $M_* \geq 10^{9} \msun$, a sample of \num{29203} galaxies (\num{19375} of which are centrals at $z = 0$, and $9828$ satellites). For the comparisons with the observational galaxy sample described in Section \ref{sec:methods:g13}, we use the sub-set of high-mass galaxies with $M_* \geq 10^{10} \msun$, a sample of $6947$ galaxies ($4643$ centrals and $2304$ satellites). In both samples, the satellite fraction is almost exactly one-third. Given that the initial mass of star particles in \illustris is about $1.6 \times 10^6 \msun$, all galaxies are resolved by $600$ or more particles, and the high-mass sample by $6000$ or more particles. \citet{sparre_15} tested the resolution dependence of star formation in \illustris using a lower-resolution run \citep{vogelsberger_14_illustris}, and found good convergence in key properties such as the star formation main sequence, the SFHs, and their variance.

\begin{figure*}
\centering
\includegraphics[trim = 0mm 2mm 2mm 0mm, clip, scale=0.5]{\figdir/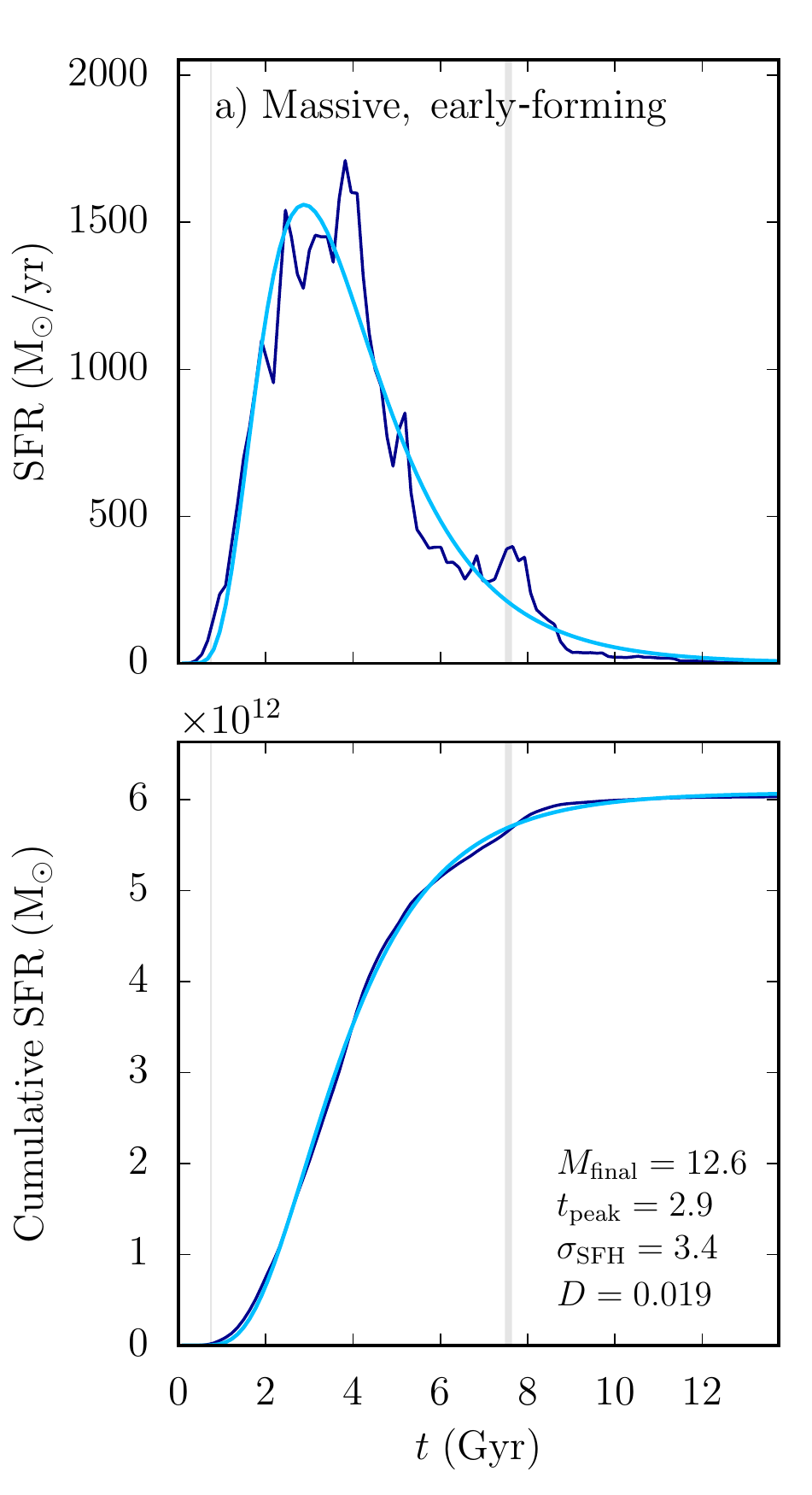}
\includegraphics[trim = 8mm 2mm 2mm 0mm, clip, scale=0.5]{\figdir/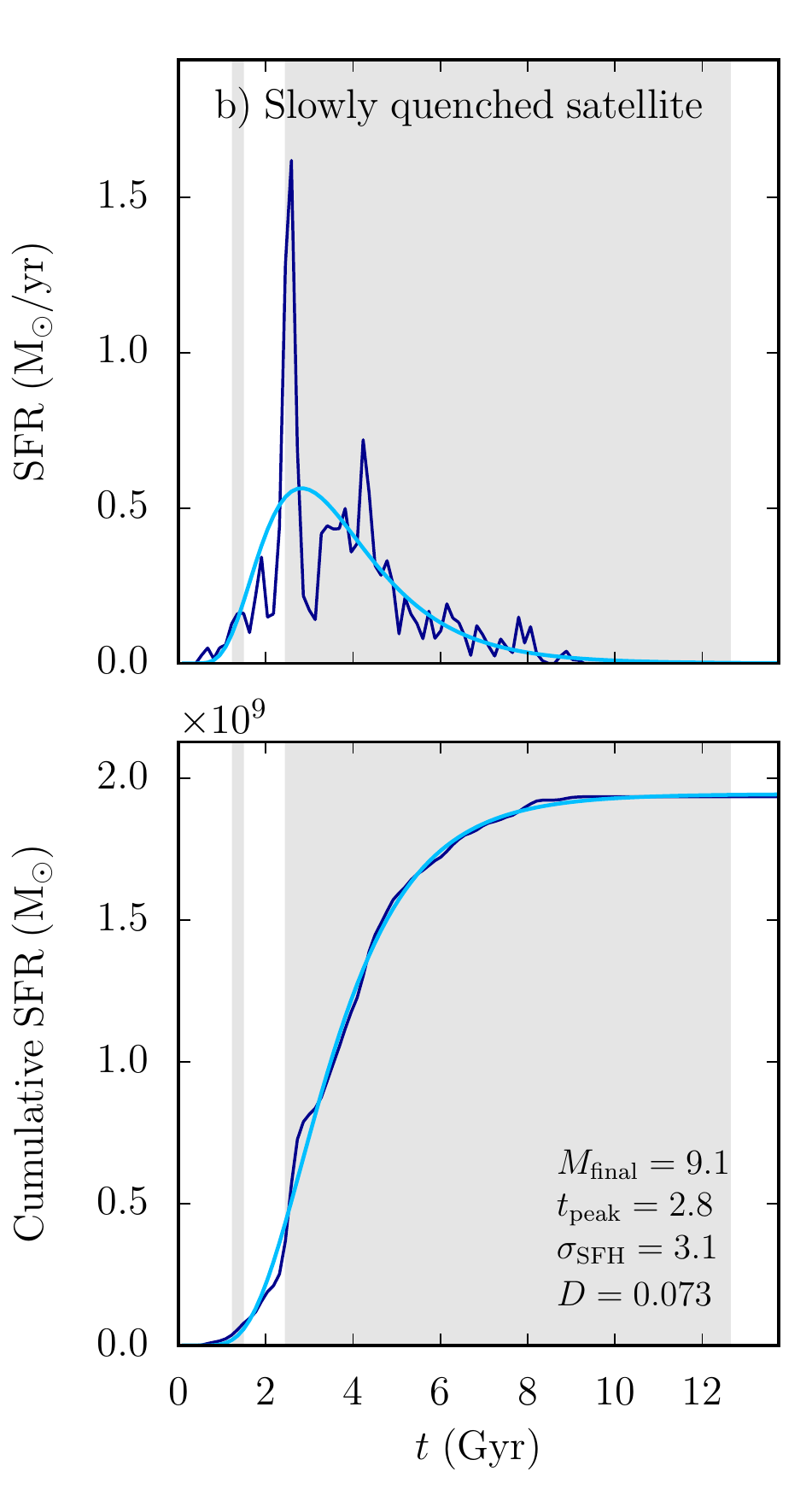}
\includegraphics[trim = 8mm 2mm 2mm 0mm, clip, scale=0.5]{\figdir/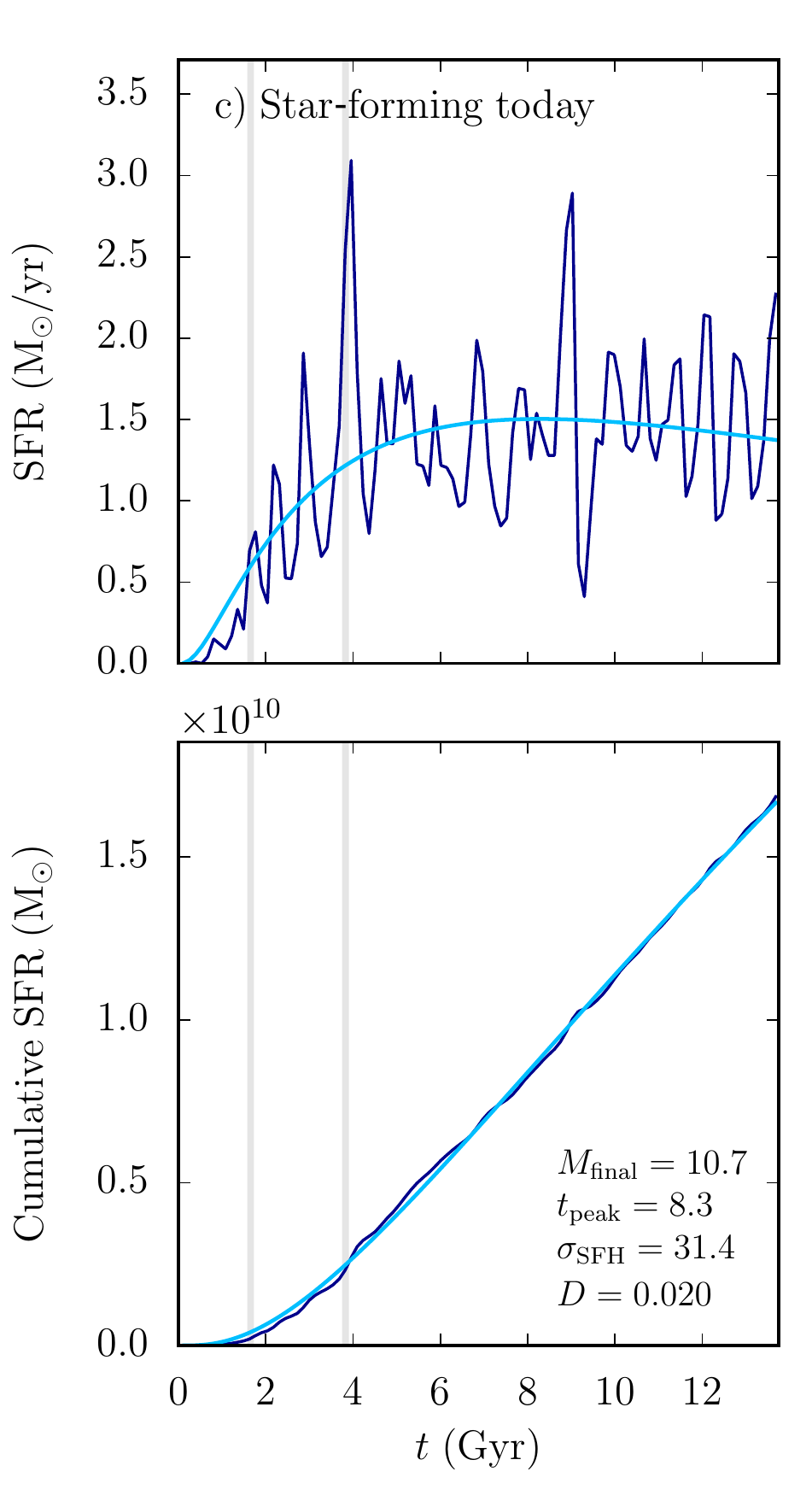}
\includegraphics[trim = 8mm 2mm 2mm 0mm, clip, scale=0.5]{\figdir/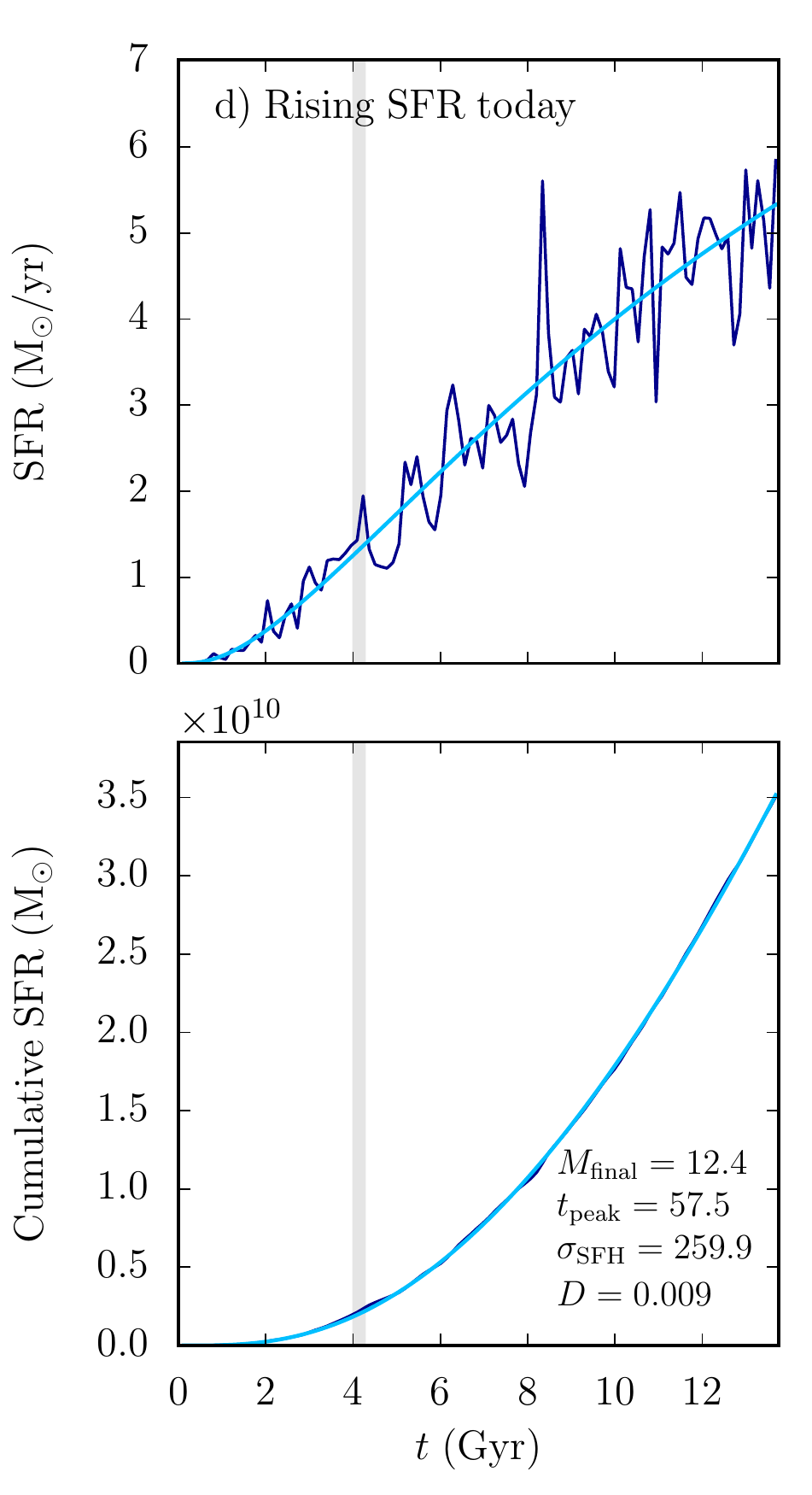}
\includegraphics[trim = 0mm 2mm 2mm 0mm, clip, scale=0.5]{\figdir/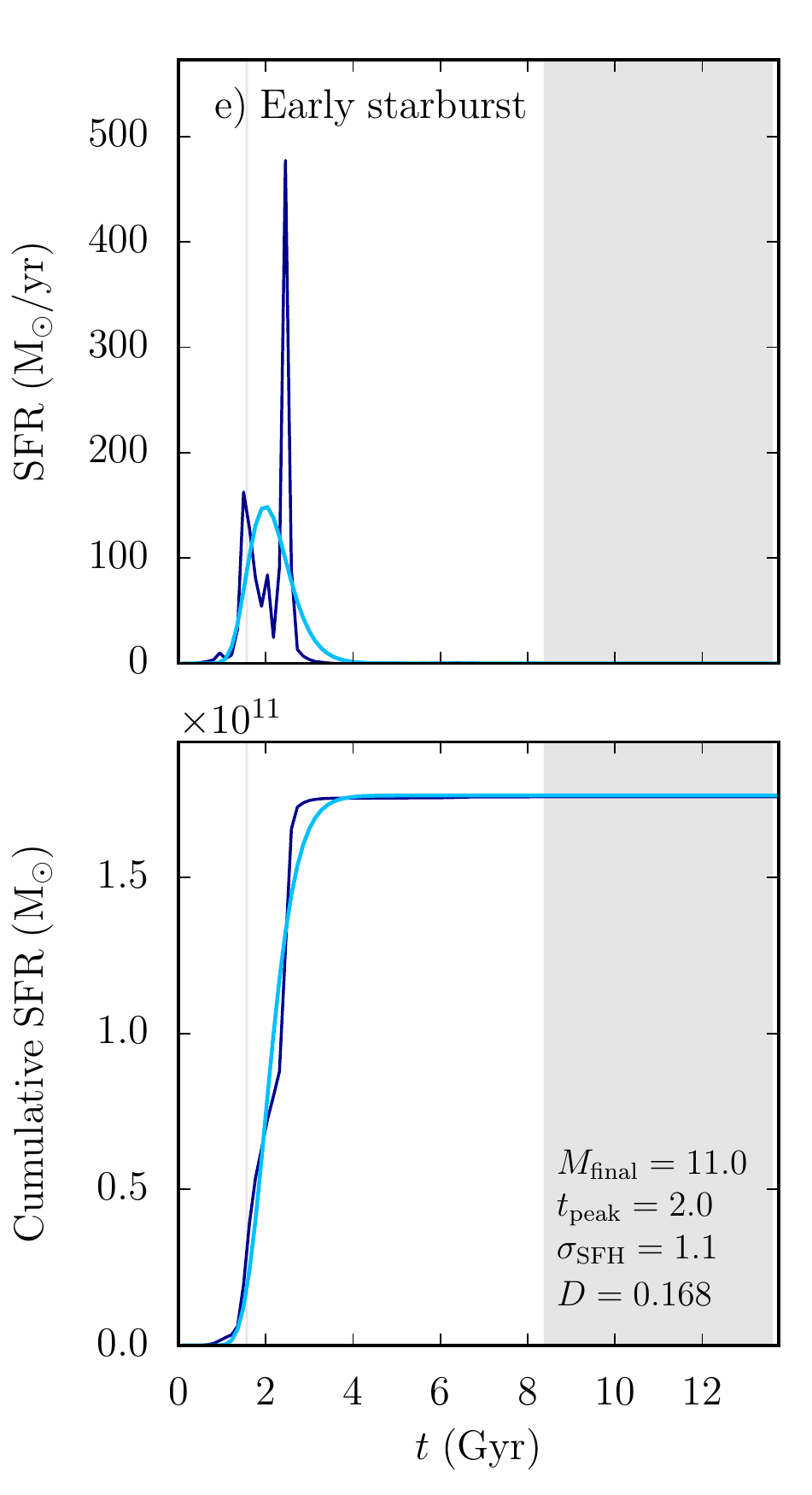}
\includegraphics[trim = 8mm 2mm 2mm 0mm, clip, scale=0.5]{\figdir/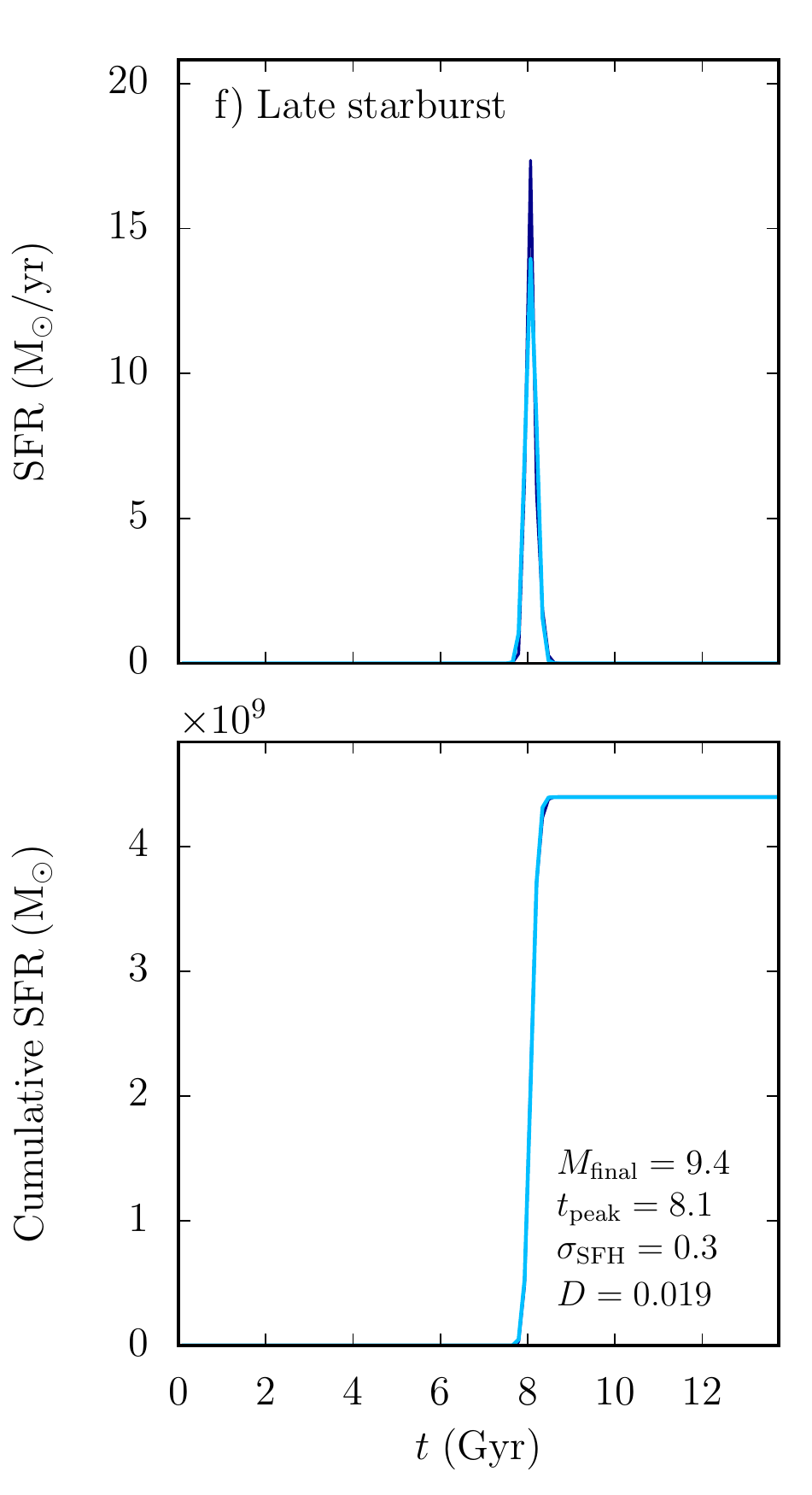}
\includegraphics[trim = 8mm 2mm 2mm 0mm, clip, scale=0.5]{\figdir/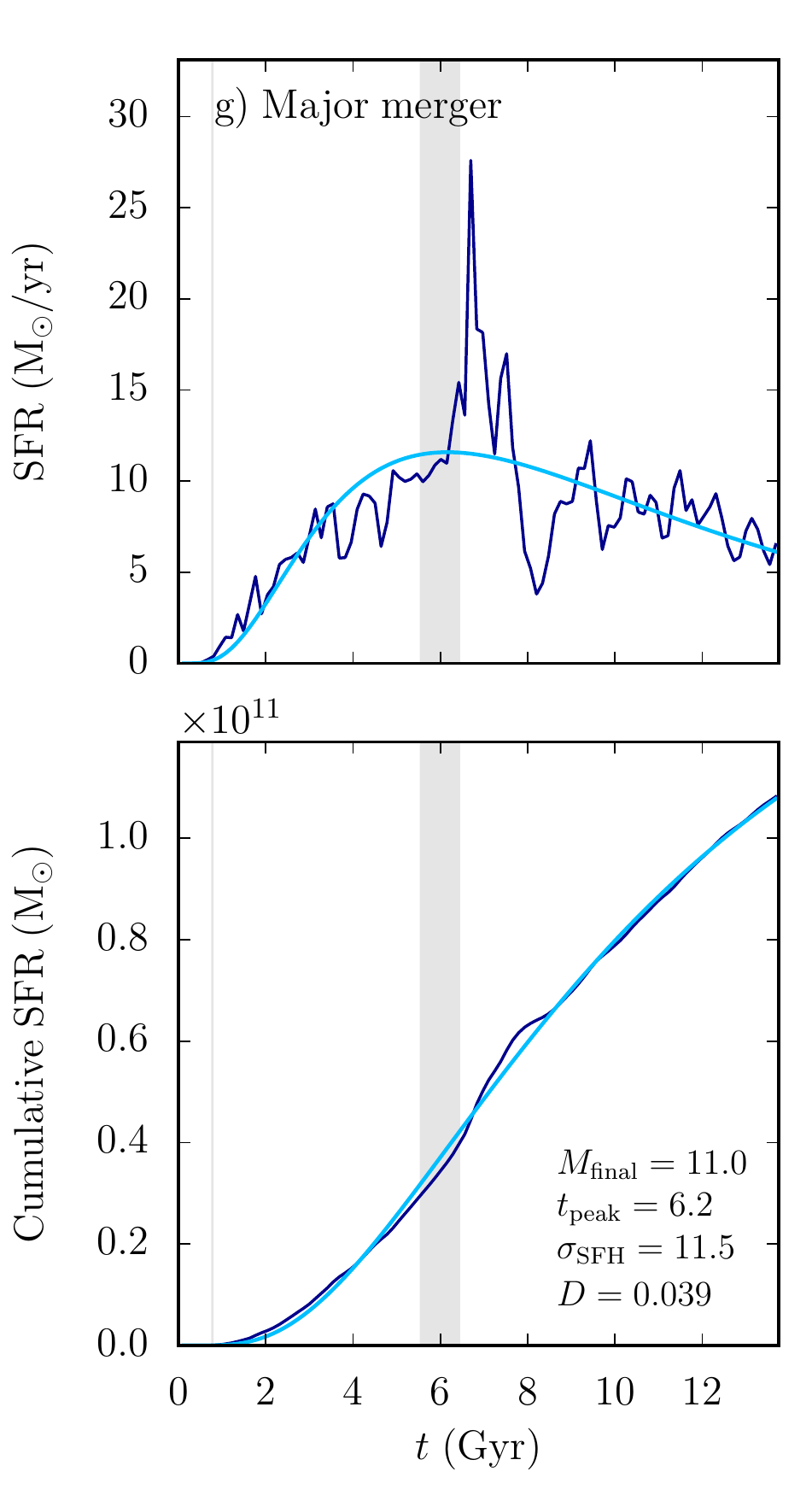}
\includegraphics[trim = 8mm 2mm 2mm 0mm, clip, scale=0.5]{\figdir/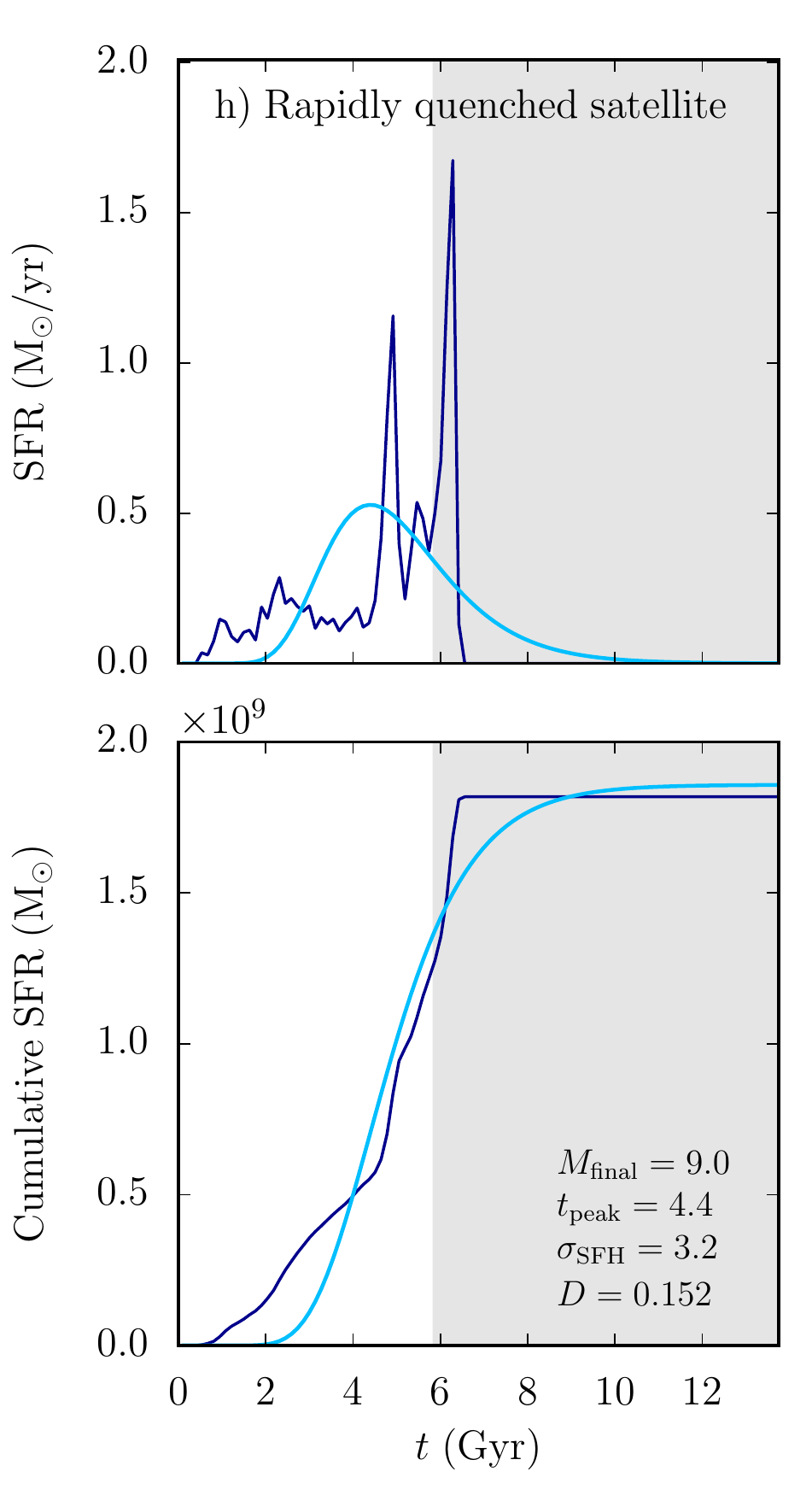}
\caption{Representative log-normal fits to SFHs from \illustris. The dark blue lines show the SFR and cumulative SFH of a simulated galaxy (top and bottom of each set of panels), and the light blue line the best-fit log-normal. The gray shaded areas indicate times when the galaxy was a satellite. The best-fit parameters are listed in the bottom right corners of the SFH panels, where $\mlimit$ is in $\log_{10}$ units and $\tpeak$ and $\width$ are in gigayears. From top left to right bottom: 
(a) a massive, early forming galaxy that stopped forming stars a few gigayears ago;
(b) a satellite that experienced a burst of star formation after infall and slowly quenched thereafter;
(c) a very broad SFH;
(d) a SFH that is still rising today, resulting in a peak time in the far future;
(e) the worst fitted SFH in the high-mass sample, a system that forms almost all of its stars in two early starbursts;
(f) a ``late bloomer'' that formed in one intense starburst;
(g) a galaxy that experienced a starburst due to a major merger; 
(h) a satellite that experienced a starburst upon being accreted, and then quenched abruptly. 
Except for the case of rapid quenching, a large variety of cumulative SFHs is well fit by the log-normal form, even if the SFR is noisy or bursty.}
\label{fig:fits}
\end{figure*}

\subsection{SFHs and Log-normal Fitting}
\label{sec:methods:sfh}

We extract the SFHs of \illustris galaxies using the same procedure as \citet{sparre_15}. In particular, we consider all stellar particles in the corresponding friends-of-friends subgroup and bin them by their birth times, weighted by their initial mass, in $100$ equal time bins between the first and last snapshots of the simulation (at $t = 54$ Myr and $t = 13.75$ Gyr, respectively). Defined in this way, the SFH includes the star formation in all progenitors that have merged with a galaxy during its lifetime. This definition matches observations of stellar mass and SFR at $z = 0$, which do not distinguish between stars formed {\it in situ} (in the galaxy in question) or {\it ex situ} (in a smaller galaxy that merged). However, it does not match the observational definitions at higher redshift if a galaxy is to accrete a significant amount of stars at later times, though the effect of such mergers is significant only for high-mass galaxies \citep{conroy_09, rodriguezgomez_16}. In some figures, we use the SFR at $z = 0$, which was computed from the SFR in all gas cells in a simulated galaxy. We have verified that this instantaneous SFR matches the SFH averaged over $20$ Myr very well.

We now investigate whether the tabulated SFHs are well fit by log-normal functions. We note that the term ``fit'' takes on a slightly unusual meaning in this context because the log-normal form is not intended to describe the details of an SFH. First, the SFHs in cosmological simulations suffer from shot noise due to the limited number of stellar particles formed in a given time bin. This noise grows as the bin size decreases, making it difficult to define a meaningful goodness-of-fit statistic such as $\chi^2$. Second, the spatial resolution (or force softening length) in \illustris is such that giant molecular clouds are not resolved, which is why the sub-grid model described in Section \ref{sec:methods:illustris} is employed. The short-term structure of an SFH is thus significantly influenced by the characteristics of the sub-grid ISM model and resolution, as shown by high-resolution zoom-in simulations \citep[e.g.,][]{hayward_11, torrey_12, sparre_17}. We note that other cosmological hydrodynamical simulations use similarly motivated sub-grid star formation models \citep{schaye_15, dave_16}, meaning their SFHs may be subject to comparable uncertainties on short timescales.

For these reasons, we disregard the short-term behavior of the simulated SFHs and instead focus on the overall, cumulative evolution of a galaxy's stellar mass. We fit the cumulative SFR,
\begin{equation}
\rm cSFR(t) \equiv \int_0^t \rm SFR(t') dt'
\end{equation}
with the integral of the log-normal SFR, which turns out to be given by a complementary error function (Appendix~\ref{sec:app:funcs:lognormal}). We use the Levenberg--Marquardt algorithm to minimize the square residuals between this functional form and the cumulative SFR in $100$ time bins. The values in these bins are highly correlated, which a standard $\chi^2$ statistic would not account for. We ignore this issue in the fit itself, but quantify the quality of the fit by computing the Kolmogorov--Smirnov statistic $D$, equivalent to the maximum difference of the cumulative functions at any time. We normalize this number by the total stellar mass created in a galaxy such that
\begin{equation}
D \equiv \frac{\rm max \left( \left| cSFR_{\rm sim}(t) - cSFR_{\rm fit}(t) \right| \right)}{{\rm cSFR}(t_0)}
\end{equation}
where $t_0$ is the age of the universe today. We use $D$ to quantify the quality of our fits in Section \ref{sec:results:fitquality}.

Finally, we note an inconvenient feature of the log-normal functional form: when only data from times well before the peak is available, the peak time and width are not well constrained, resulting in a strong degeneracy between $T_0$ and $\tau$. This issue manifests itself in a small fraction of \illustris galaxies that experience rising SFRs $z=0$ and are sometimes assigned unphysically high values of $T_0$ and $\tau$. We shall later return to the question of whether such galaxies exist in the real universe, but in a fitting sense, they cause inconvenient artifacts such as SFHs that peak hundreds of gigayears in the future. 

Thus, we impose a prior on $\tpeak$ such that the fit is unaffected at $\tpeak < t_0$ (the time at $z = 0$). Later peak times are penalized by multiplying the residuals by the term $(1 + (\log_{10} \tpeak - \log_{10} t_0)^2)$ which effectively cuts off the distribution at 50 Gyr, corresponding to a scale factor $a_{\rm peak,max} = 10$ in the \illustris cosmology. A similar, less informative prior penalizes extreme widths of $\width > 10^{2.5}$. The vast majority, over 90\% of the galaxies in our sample, are unaffected by these priors. Galaxies in the high-$\tpeak$ tail experience drastically different best-fit parameters, namely changes of 50\% or greater for 4\% of the galaxies. However, due to the strong degeneracy of $T_0$ and $\tau$, the actual fitted SFHs of these galaxies are almost identical to those fit without a prior. Thus, the fit quality $D$ is barely affected: only 3\% of the SFHs experience a change in fit quality of 20\% or greater. The prior is even less important for the high-mass sample. For the inferences and comparisons shown in this paper, the tails in the $T_0$ and $\tau$ distributions are not important.

\subsection{Halo Mass Accretion Histories}
\label{sec:methods:mah}

Besides the SFH, we also consider the mass accretion history (MAH) of a galaxy's dark matter halo, which we extract from the \textsc{Sublink} merger trees provided by \citet{rodriguezgomez_15}. For each halo at $z = 0$, we follow its most massive (or main branch) progenitor back in time and record its mass. In order to reduce the complex shape of the MAHs to characteristics such as a halo's formation time, we fit the MAHs with the exponential form of \citet{wechsler_02_halo_assembly},
\begin{equation}
\label{eq:wechsler}
M(z)=M_0 e^{-\alpha z}
\end{equation}
where $M_0$ is the halo mass at $z = 0$. This function was designed to fit the MAHs of isolated halos and is not equipped to capture the effects of subhalo accretion onto a larger host. Thus, we restrict the fits to epochs when the halos were not subhalos, which improves the correlation between halo formation redshift, and the SFH of their respective galaxies (Section \ref{sec:results:correlations}).  Following \citet{wechsler_02_halo_assembly}, we define a formation redshift where the halo has formed half its mass, $z_{\rm wechsler} = \ln(2) / \alpha$. We have also experimented with the more flexible, three-parameter function of \citet{tasitsiomi_04_clusterprof} and \citet{mcbride_09}, but find that the resulting formation redshift correlates less strongly with the SFHs.

Finally, we attempt to draw a more direct connection to the SFH fits by fitting the MAHs with a log-normal in time. We find that the log-normal is not particularly adapted to the shape of MAHs. The exponential rise at early times is generally well fit, but the additional free parameter compared to the \citet{wechsler_02_halo_assembly} function rarely improves the fit significantly. If the MAH flattens or even decreases, the power-law term in the \citet{tasitsiomi_04_clusterprof} formula is a better description, and if the MAH is well described by a pure exponential, the third free parameter is unconstrained. For this reason, about a quarter of the fits fail to converge, even with the priors described in Section \ref{sec:methods:sfh}. We return to the relation between SFH and MAH log-normal fits in Section \ref{sec:discussion:galaxyhalo}.


\begin{figure}
\centering
\includegraphics[trim = 0mm 5mm 5mm 0mm, clip, scale=0.6]{\figdir/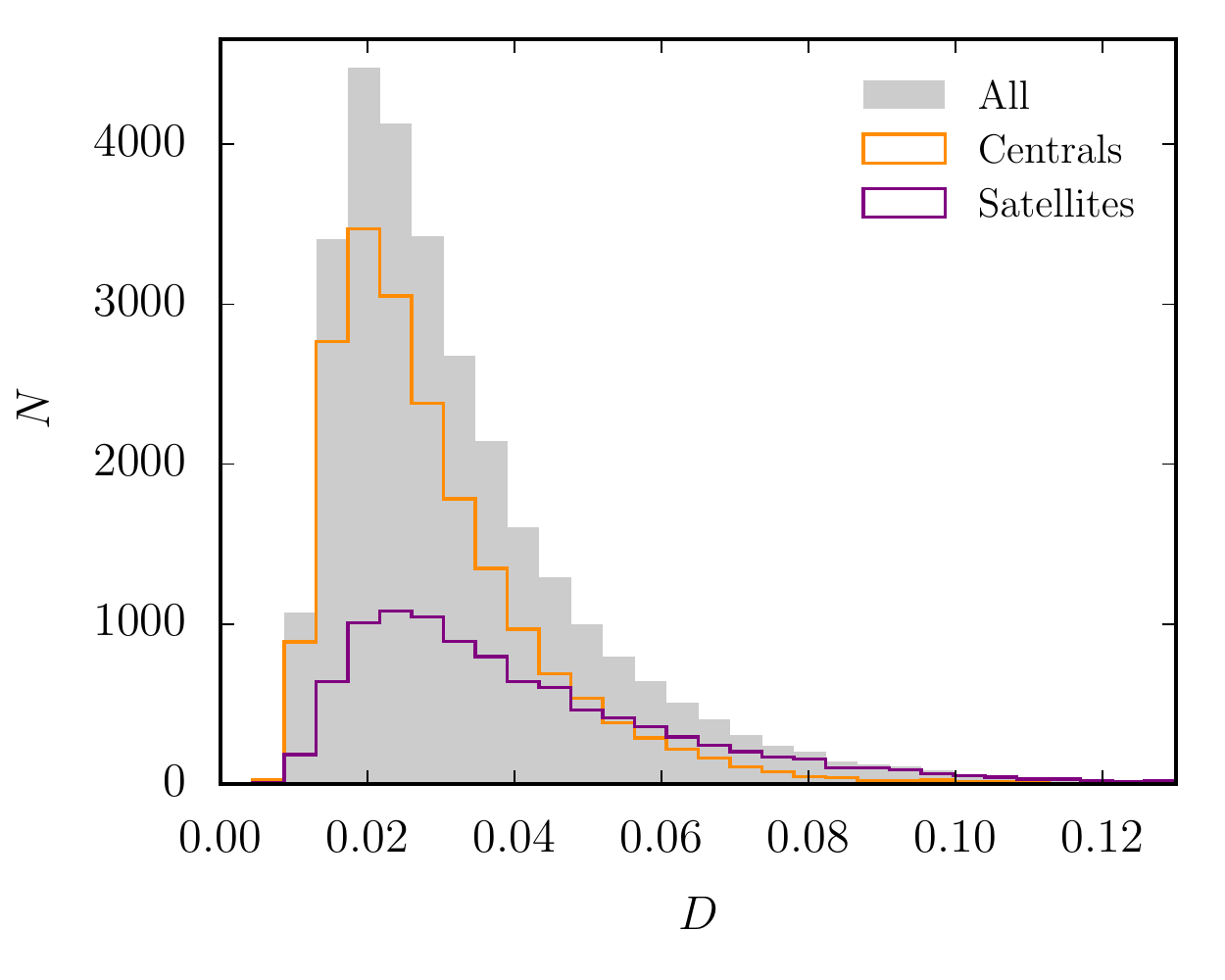}
\caption{Distribution of fit quality, defined as the maximum deviation from the cumulative SFH, $D$. The plot cuts out a small fraction of outliers with $D > 0.13$). The vast majority of SFHs is fit well by the log-normal form. The tail at $D > 0.08$ is almost entirely due to low-mass satellites that experienced rapid quenching.}
\label{fig:fit_quality}
\end{figure}

\section{Results}
\label{sec:results}

In this section, we consider the quality of log-normal fits to \illustris and observationally inferred SFHs, and compare the best-fit parameters from the observational and simulated galaxy samples.

\subsection{The Quality of Log-normal Fits}
\label{sec:results:fitquality}

Figure~\ref{fig:fits} shows an array of example log-normal fits to a range of types of SFH, namely those of early and late-forming galaxies, centrals and satellites, slow and fast decliners, and galaxies that underwent a major merger. Despite the extremely different shapes of the corresponding SFHs, the log-normal captures the main features in the cumulative distribution while glossing over the noise and spikes in the SFR. Figure~\ref{fig:fits} highlights a large range of fit qualities, including the worst fit to any SFH in the high-mass sample ($M_* > 10^{10} \msun$). In the lower-mass sample ($M_* < 10^{10} \msun$), one particularly difficult case for the log-normal function is a sharp cut-off in star formation, for example the rapid quenching of satellites after infall into their host (see Figure~\ref{fig:fits} for an example, or \citealt{mistani_16} for an analysis of this effect in \illustris).

Figure~\ref{fig:fit_quality} shows the distribution of $D$, the maximum fractional residual between the fit and the cumulative SFR. While $D$ ranges from almost zero to $34\%$, the vast majority of galaxies, namely $85\%$ ($92\%$ of centrals and $73\%$ of satellites),  are fit to $5\%$ or better at all times. Only $1\%$ of the fits are off by $10\%$ or more at any time ($0.4\%$ of centrals and $2.7\%$ of satellites). The statistics are even better in the high-mass sample, where only $0.3\%$ of galaxies ($0.1\%$ of centrals and $0.8\%$ of satellites) exhibit a $D$ of $10\%$ or greater. Figure~\ref{fig:fit_quality} also demonstrates that the tail toward poor fit qualities consists almost entirely of satellites. Visual inspection of these objects confirms that most of them are low-mass satellites that experienced rapid quenching after being accreted into a larger halo.

The generally excellent quality of the log-normal fits is by no means guaranteed, as we are fitting $100$ time bins with a function of only three free parameters.  For comparison, we also consider the delayed-$\tau$ model \citep[e.g.,][]{gavazzi_02} and the double power law \citep[e.g.,][]{behroozi_13_shmr} in Appendix~\ref{sec:app:funcs}. We show that, on average, the log-normal function performs slightly better than the delayed-$\tau$ model, which has the same number of free parameters. In contrast, the double power law has four free parameters (normalization, peak time, and rising and falling slope), leading to a significantly more accurate fit, particularly for rapidly quenched satellites.

We also apply our fitting procedure to the inferred median SFHs of \citetalias{pacifici_16} (see Appendix~\ref{sec:app:fits:p16} and Figure~\ref{fig:pacifici_fits}). We find fit qualities very similar to those of the \illustris SFHs, with $D$ between 1.4\% and 5\%, and a median of 2.8\% (compared to 3.1\% for the \illustris sample).

In summary, the log-normal functional form provides an excellent description of the majority of the simulated and observationally inferred cumulative SFHs we considered. The key prediction of the log-normal is a steep rise and slow decline in linear time, and that the two timescales are coupled. This fundamental characteristic is shared with the delayed-$\tau$ model, meaning that many of the following results could probably be obtained based on delayed-$\tau$ fits as well. We conclude that a steep rise and slow decline are key features of the SFHs considered here, except for galaxies whose star formation is suddenly truncated by external processes such as satellite quenching. Models where the rise and decline timescales are independent (such as the double power law) can achieve a better fit for some galaxies, but have less predictive power: by fitting to only the rising or declining part of the SFH, one cannot infer the timescale of the other part. We further discuss this issue in Appendix~\ref{sec:app:fits:comp}.

\begin{figure}
\centering
\includegraphics[trim = 3mm 2mm 3mm 0mm, clip, scale=0.54]{\figdir/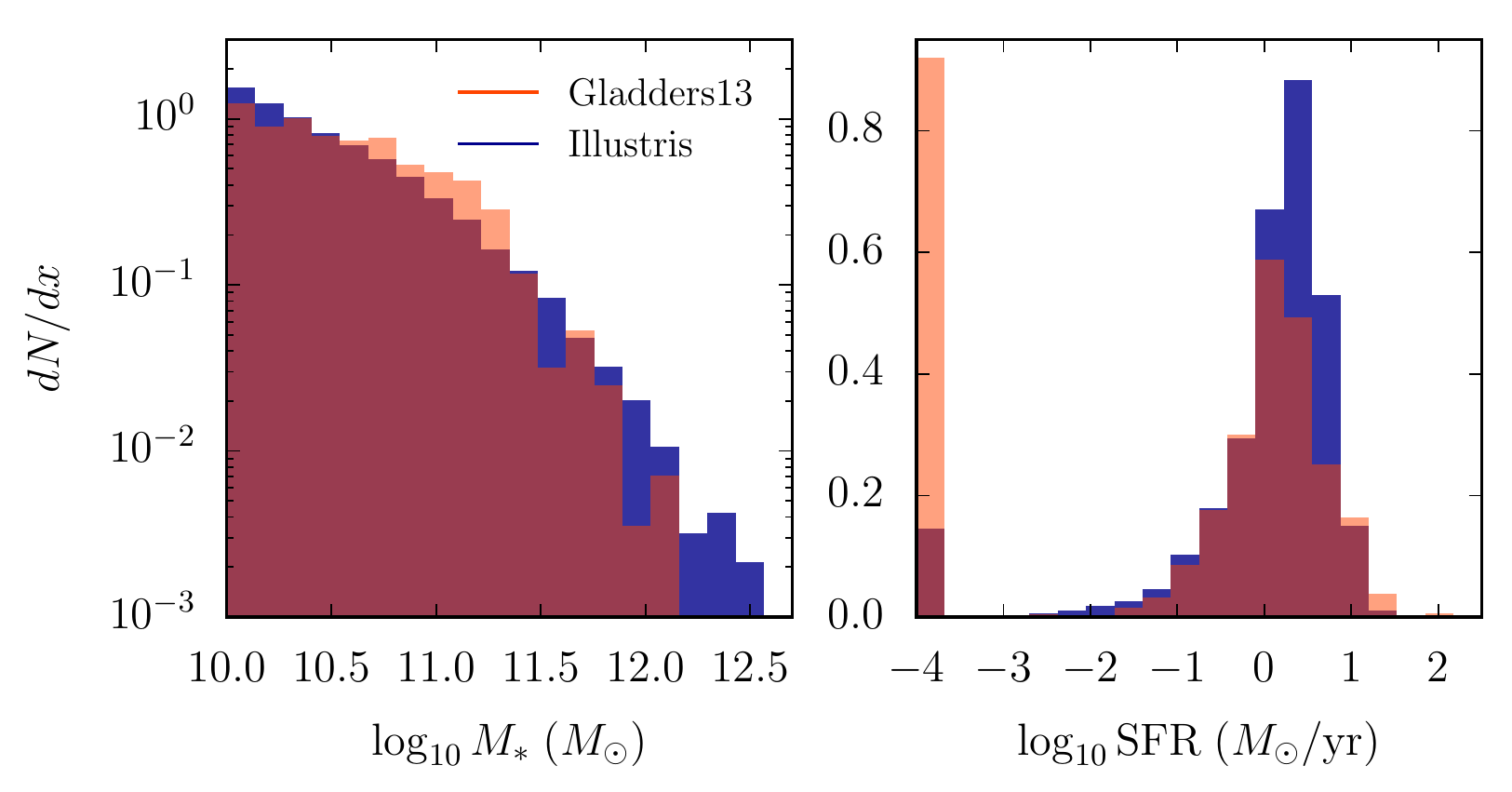}
\caption{Comparison of the galaxy populations in the \citetalias{gladders_13_icbs4} (orange) and \illustris (blue) samples (with $M_* > 10^{10} \msun$). While the stellar mass functions (left panel) are similar, \illustris contains fewer quiescent and more star-forming galaxies than the \citetalias{gladders_13_icbs4} sample (right panel).}
\label{fig:gal_pop}
\end{figure}

\begin{figure*}
\centering
\includegraphics[trim = 3mm 8mm 3mm 0mm, clip, scale=0.55]{\figdir/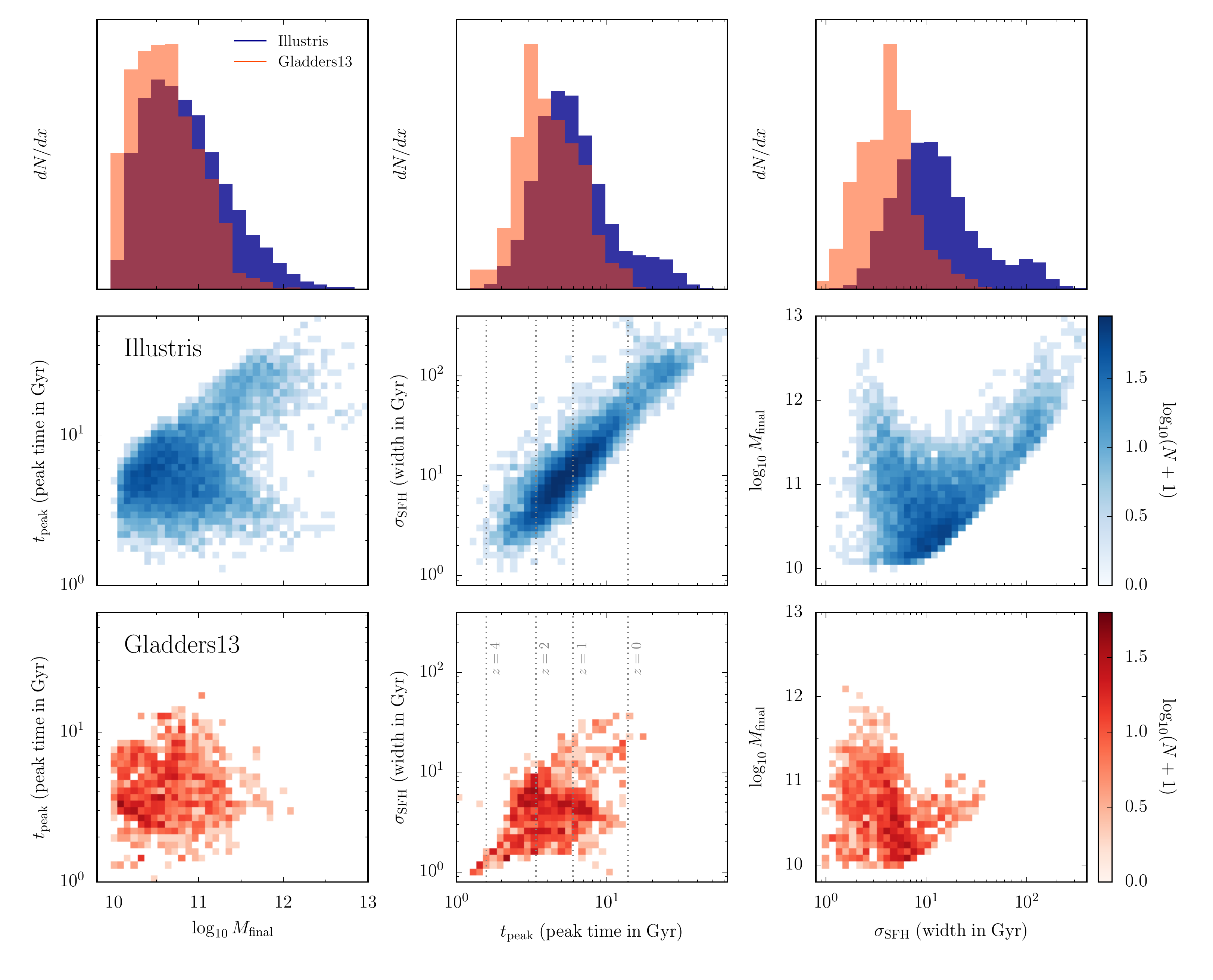}
\caption{Log-normal parameters of the \illustris (blue, $M_* > 10^{10} \msun$) and \citetalias{gladders_13_icbs4} (orange) samples. The top row shows the distributions of the three fit parameters, namely the total stellar mass formed, peak time, and full width at half maximum. The histograms are normalized to the same sample size. The bottom two rows show the density of galaxies as a function of pairs of these parameters. The \illustris and \citetalias{gladders_13_icbs4} samples occupy similar areas in this parameter space, but \illustris SFHs exhibit larger widths and a more pronounced correlation between $\tpeak$ and $\width$. The tail of the \illustris population toward high $\tpeak$ and $\width$ (i.e., SFHs, which will peak in the future) is influenced by our prior and should not be taken too seriously (Section \ref{sec:methods:sfh}). See Appendix~\ref{sec:app:fits:stdparams} for a comparison in the original $T_0$--$\tau$ parameter space of \citetalias{gladders_13_icbs4}.}
\label{fig:params_alt}
\end{figure*}

\begin{figure}
\centering
\includegraphics[trim = 5mm 8mm 0mm 3mm, clip, scale=0.78]{\figdir/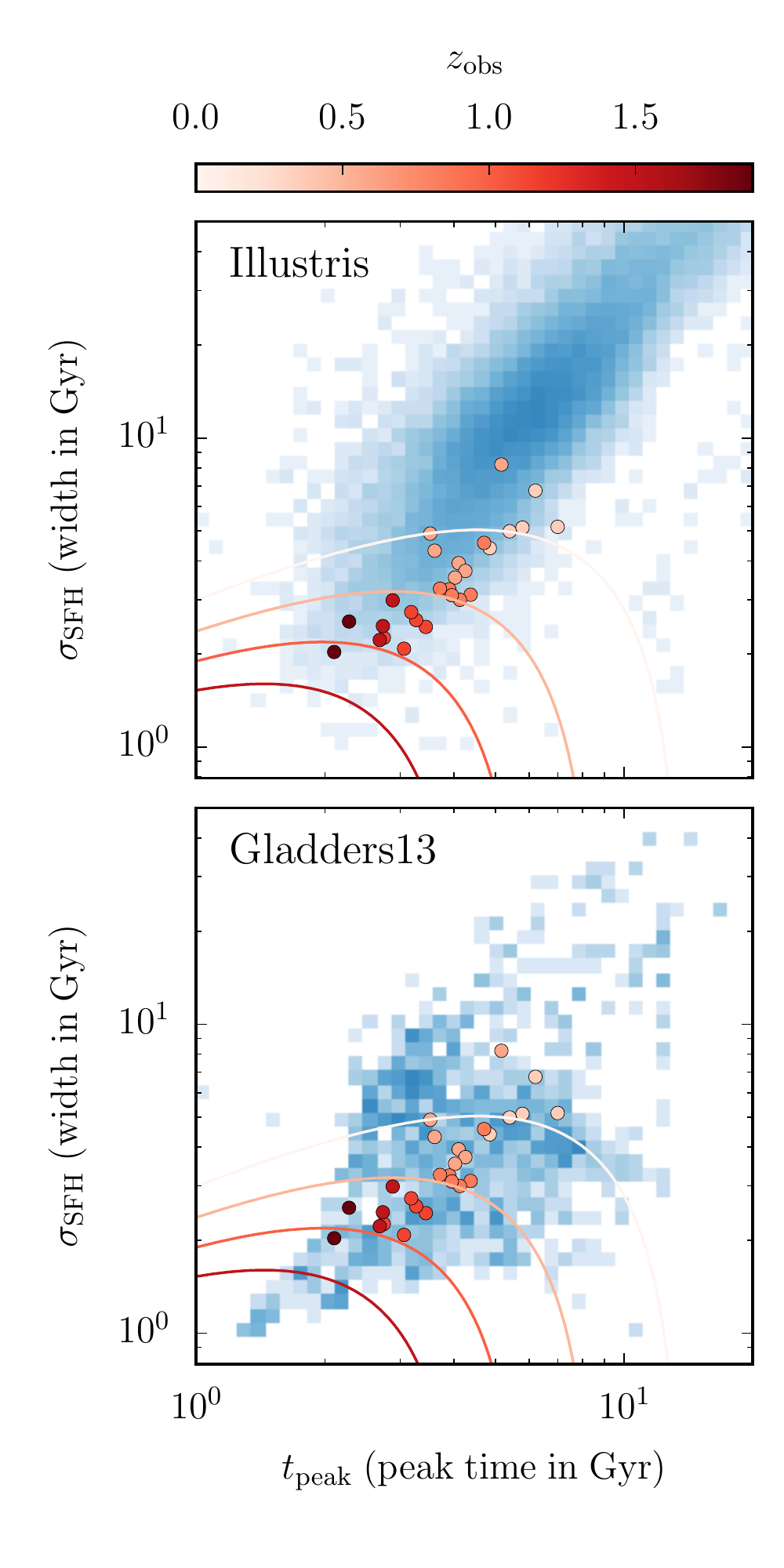}
\caption{Comparison of the log-normal parameters of the \citet[][round points]{pacifici_16} sample with those of \illustris (top) and \citetalias{gladders_13_icbs4} (bottom). The \citetalias{pacifici_16} sample is restricted to quiescent galaxies which, in log-normal space, lie below the redshift-dependent lines (same color scale as the round points). At each redshift, the SFR of the \citetalias{pacifici_16} fits slightly exceeds the quiescent cut, indicating that the log-normal fits overestimate the actually observed SFR.}
\label{fig:pacifici_params}
\end{figure}

\subsection{The Log-normal Parameter Space}
\label{sec:results:params}

We now consider the distribution of best-fit parameters in the \citetalias{gladders_13_icbs4}, \citetalias{pacifici_16}, and \illustris samples. Before attempting such a comparison, we need to check that the galaxy populations in \illustris and in the \citetalias{gladders_13_icbs4} sample are at least roughly comparable. Figure~\ref{fig:gal_pop} shows the abundance of galaxies as a function of stellar mass and SFR in \illustris (blue) and the \citetalias{gladders_13_icbs4} sample (orange). The distributions of $M_*$ are relatively similar (left panel), but the SFRs (right panel) differ significantly. The \citetalias{gladders_13_icbs4} sample contains a much larger fraction of entirely quiescent, non-star forming galaxies, whereas \illustris contains an excess of galaxies with SFRs between $1$ and $10 \msun / \rm yr$. This disagreement is not surprising because the \citetalias{gladders_13_icbs4} sample is more or less representative of the quiescent and star-forming fraction in the $z = 0$ universe \citep{abramson_15}, whereas an excess of blue, star-forming galaxies in \illustris at $z = 0$ has been noted before \citep{vogelsberger_14_illustris}. Furthermore, \citet{sparre_15} detected a paucity of starburst galaxies in \illustris compared to observations. Conversely, we remind the reader that the log-normal parameters of \citetalias{gladders_13_icbs4} are inferred from a semi-global fit rather than measured from resolved SFHs. With these caveats in mind, we emphasize that our goal is not to test the galaxy population of \illustris or the fitting procedure of \citetalias{gladders_13_icbs4}, but rather to compare the regions of log-normal parameter space occupied by the two samples, and to investigate the physical properties of galaxies that occupy those regions. For those purposes, the populations shown in Figure~\ref{fig:gal_pop} are sufficiently similar.

We compare the best-fit parameters of the \illustris and \citetalias{gladders_13_icbs4} samples in Figure~\ref{fig:params_alt}. The top row shows histograms for each parameter. While the distributions of final stellar masses roughly agree, we notice that \illustris galaxies peak slightly later (higher $\tpeak$) and have significantly larger widths (higher $\width$). We note that $\mlimit$ is a reflection of the stellar mass of galaxies, modulo their current age: for early forming galaxies, $\mlimit \approx M_*(z=0)$, whereas late-forming galaxies have yet to form some of the stars that contribute to $\mlimit$. 

The differences in the distributions are partially due to the excess of blue, star-forming galaxies in \illustris compared to the \citetalias{gladders_13_icbs4} sample (Figure~\ref{fig:gal_pop}). The fits to the most extreme of those cases tend to be poorly constrained and are influenced by our prior, which discourages $\tpeak > t_0$ (Section \ref{sec:methods:sfh}), causing tails of a few percent of the population toward very high values in all three parameters ($\mlimit \gsim 10^{12}$, $\tpeak \gsim t_0$, $\width \gsim 30$). The high values of $\mlimit$ are an artifact: although the corresponding galaxies have relatively low stellar masses at $z = 0$, their fitted log-normal SFHs predict a large amount of future star formation, increasing their $\mlimit$.

The bottom two rows of Figure~\ref{fig:params_alt} show the correlations between each pair of parameters. The $\mlimit$-$\tpeak$ (left column) and $\mlimit$-$\tau$ (right column) planes show the expected trends with stellar mass: larger galaxies form earlier (have lower $\tpeak$) and decline relatively fast (have lower $\width$), in agreement with the downsizing picture. The population of extremely late-forming galaxies in \illustris manifests itself as tails in the top right hand corners of both panels. At low stellar masses, the scatter in peak time increases. Interestingly, these trends are less significant in the \citetalias{gladders_13_icbs4} sample, especially the relation between mass and peak time.

We now consider the relation between peak time and width (center column of Figure~\ref{fig:params_alt}). Once again ignoring the \illustris population with a peak time later than $z = 0$, the \citetalias{gladders_13_icbs4} and \illustris samples occupy a similar region of parameter space. In both samples, earlier-forming galaxies form faster, but the correlation is more significant in \illustris. Furthermore, at fixed $\tpeak$ \illustris galaxies have broader widths, i.e., they form stars for a longer time than inferred by \citetalias{gladders_13_icbs4}. 

We note another potentially revealing difference between the \citetalias{gladders_13_icbs4} and \illustris samples: the absence of young galaxies or ``late bloomers'' \citep[\citetalias{gladders_13_icbs4};][]{oemler_13_icbs3, dressler_16} in \illustris. Those are galaxies that form the majority of their stars relatively quickly after $z \approx 1$. The corresponding parameter space of $\tpeak$ between $z = 1$ and $z = 0$ and $\width \lsim 5$ Gyr is visibly populated in \citetalias{gladders_13_icbs4}, but deserted in \illustris (though a few such galaxies exist in the lower-mass sample, Section~\ref{sec:results:correlations}). This difference is another manifestation of the relatively wide SFHs in \illustris: if galaxies peak as late as $z = 1$, they are most likely still active today (at least according to the log-normal fits).

As an independent, observational check, we compare the \citetalias{gladders_13_icbs4} and \illustris samples to the log-normal parameters of the \citetalias{pacifici_16} median SFHs in Figure~\ref{fig:pacifici_params}. The color scale of the \citetalias{pacifici_16} points indicates their observation redshift. The \citetalias{gladders_13_icbs4} and \citetalias{pacifici_16} samples match very well and occupy the same $\tpeak$-$\width$ relation. The lower scatter in the \citetalias{pacifici_16} sample is expected since those parameters were derived from median SFHs rather than individual galaxies. However, \citetalias{pacifici_16} considered only quiescent galaxies as defined in Equation~\ref{eq:p16_sfr}. The corresponding limits at each redshift are shown with lines in Figure~\ref{fig:pacifici_params}, and points of a given color should be compared to the \illustris and \citetalias{gladders_13_icbs4} galaxies below the corresponding line. We note that the \citetalias{pacifici_16} points themselves lie slightly above the lines at each redshift, indicating that the log-normal fits overestimate the sSFR at the redshift of observation (see Appendix~\ref{sec:app:fits:p16} and Figure~\ref{fig:pacifici_fits}). Nevertheless, the agreement between the observational samples is reassuring, and lends credibility to the inferred parameters of \citetalias{gladders_13_icbs4}.

\subsection{Which Galaxy Properties Determine the SFH?}
\label{sec:results:correlations}

\begin{figure}
\centering
\includegraphics[trim = 4mm 1mm 4mm 0mm, clip, scale=0.57]{\figdir/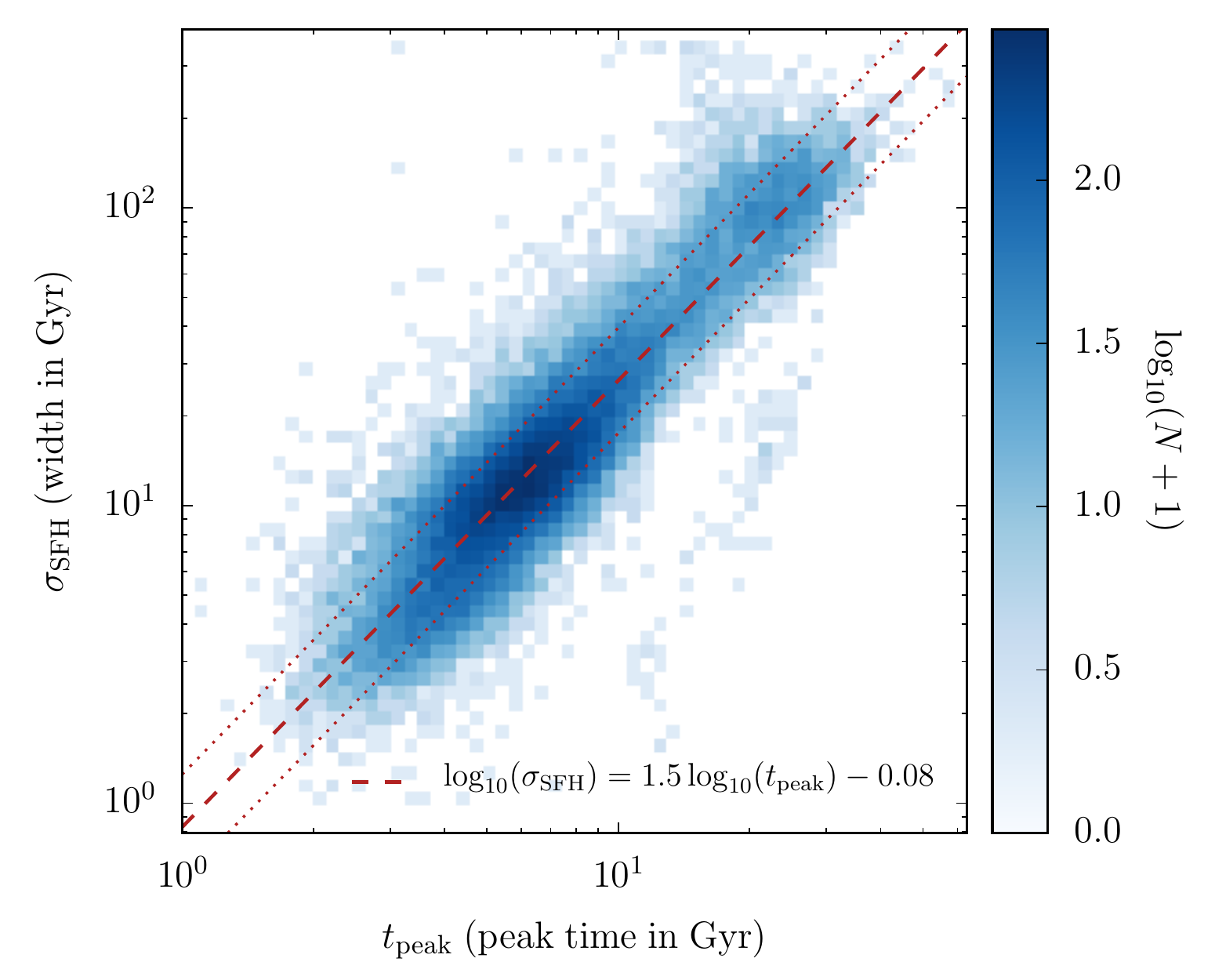}
\caption{Correlation between peak time and width for the entire \illustris sample. The dashed line shows the best-fit power law, the dotted lines the 68\% scatter of 0.18 dex (about 50\%).}
\label{fig:params_tpeakwidth}
\end{figure}

\begin{figure*}
\centering
\includegraphics[trim = 20mm 2mm 5mm 0mm, clip, scale=0.61]{\figdir/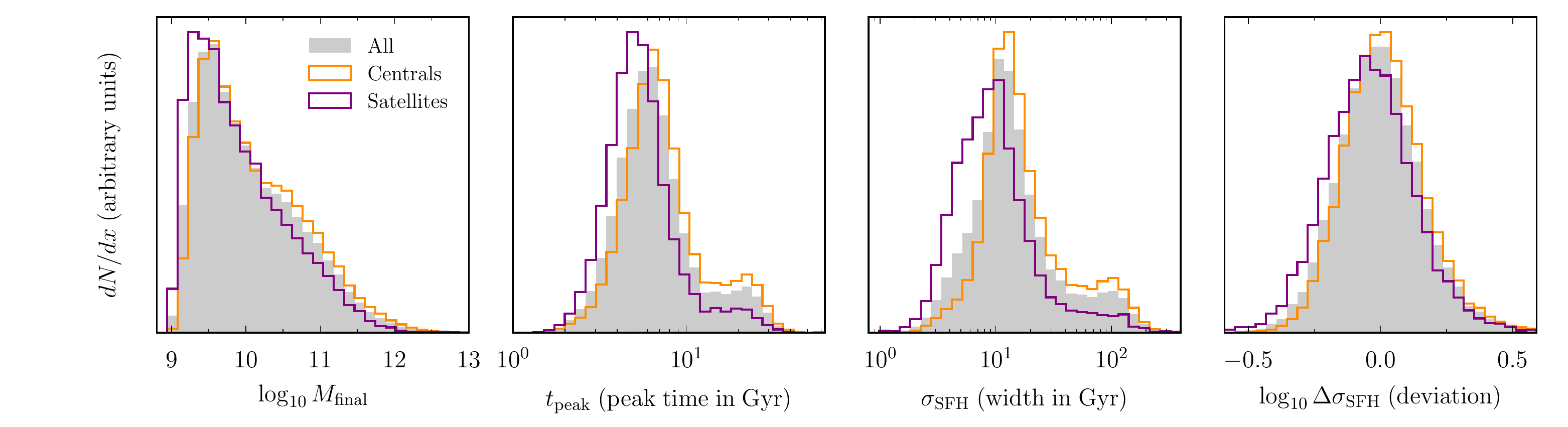}
\caption{Dependence of the log-normal parameters on whether a galaxy is a central (orange) or a satellite (purple) at $z = 0$. The entire galaxy sample ($M_* > 10^9 \msun$) is shown in gray. Though centrals are, on average, more massive than satellites, satellites form earlier and faster, contrary to the general trend that more massive galaxies form earlier. The tails toward very high $\tpeak$ and $\width$ are predominantly caused by extremely late-forming centrals. The right panel shows the deviation from the correlation between $\tpeak$ and $\width$ (Equation \ref{eq:correlation}), i.e. whether an SFH is relatively wide or narrow at fixed peak time.}
\label{fig:params_censat}
\end{figure*}

\begin{figure*}
\centering
\includegraphics[trim = 2mm 4mm 0mm 0mm, clip, scale=0.55]{\figdir/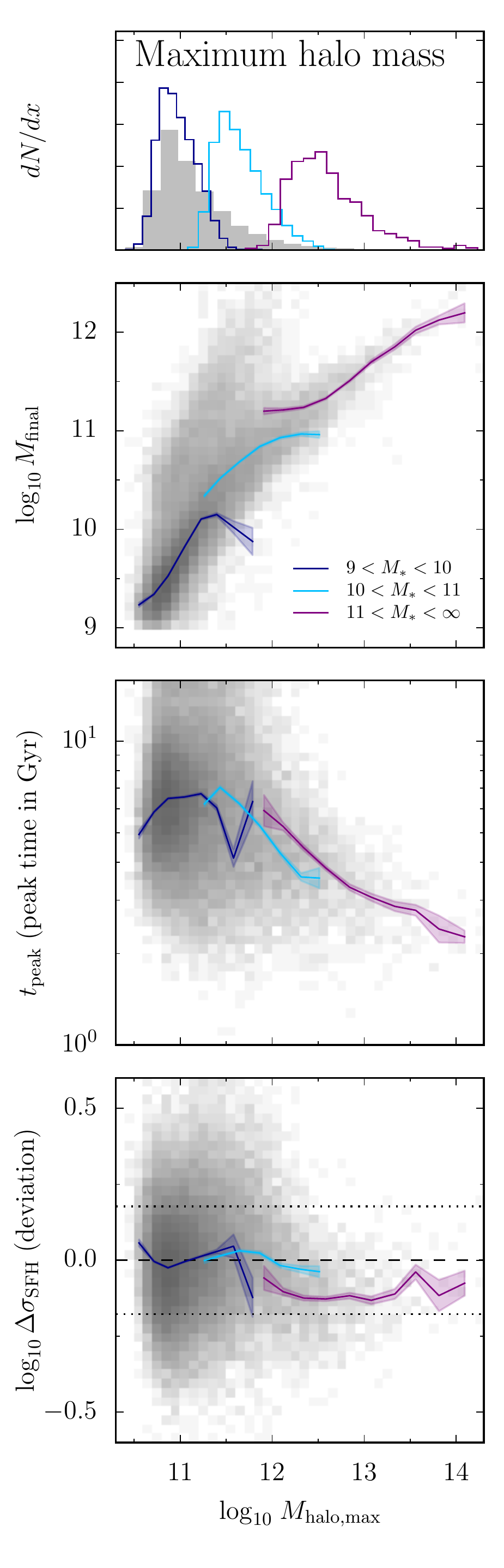}
\includegraphics[trim = 19mm 4mm 0mm 0mm, clip, scale=0.55]{\figdir/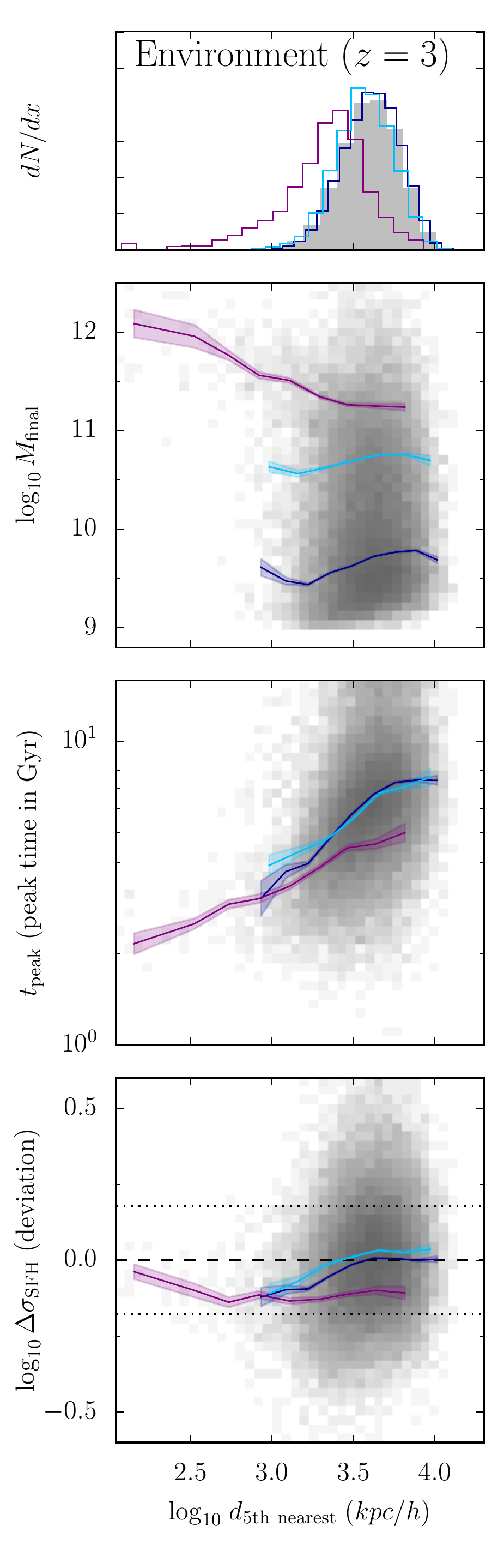}
\includegraphics[trim = 19mm 4mm 0mm 0mm, clip, scale=0.55]{\figdir/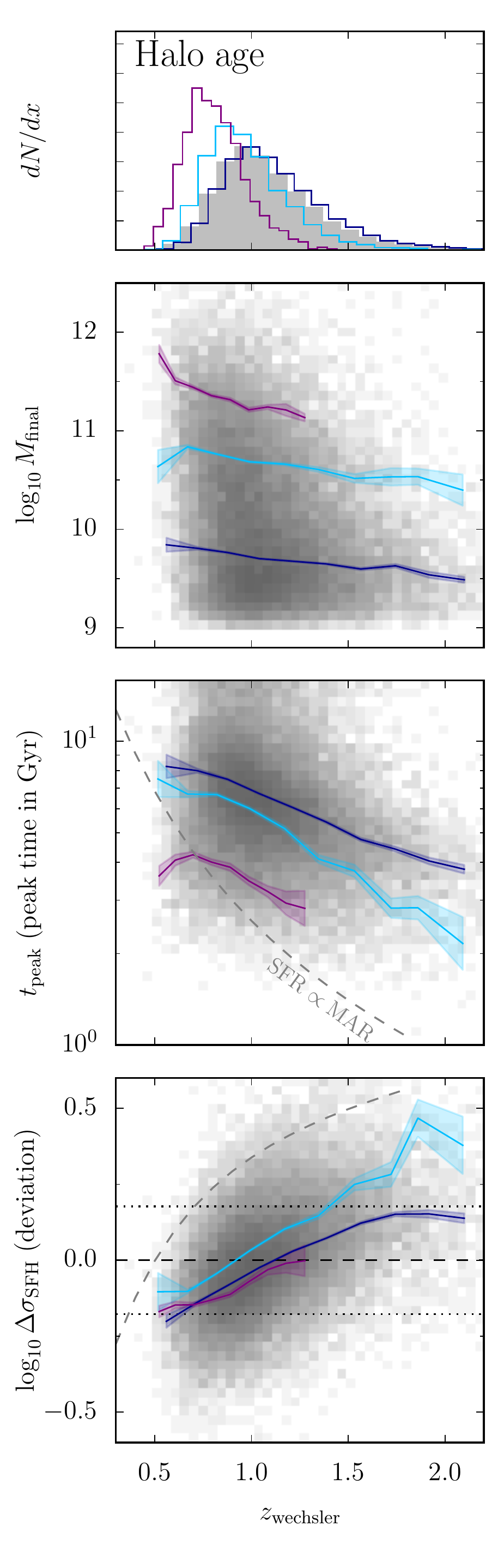}
\includegraphics[trim = 19mm 4mm 0mm 0mm, clip, scale=0.55]{\figdir/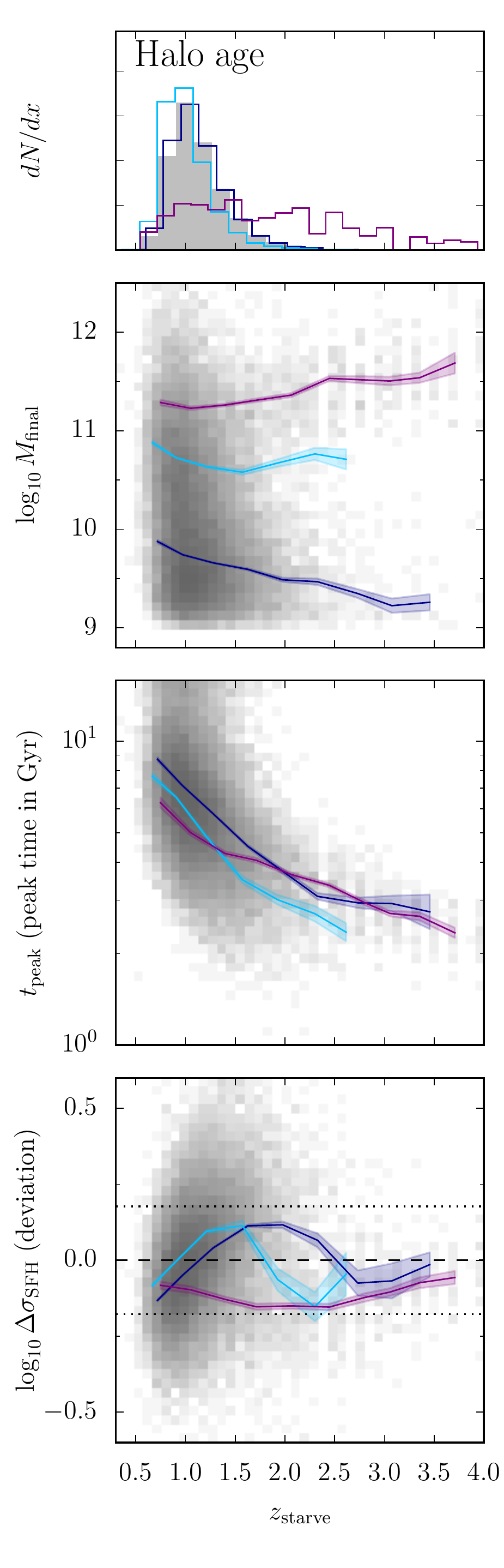}
\caption{Galaxy--halo connection in the log-normal parameter space. Each column refers to a halo or galaxy property, namely halo mass, environment (quantified by the distance to the fifth nearest neighbor at $z=3$), the halo age according to \citet{wechsler_02_halo_assembly}, and the ``starvation redshift'' where a halo is expected to quench in the age matching formalism of \citet{hearin_13}. In each column, the top panel shows the distribution of the respective property in our simulated galaxy sample (gray). The bottom three rows show the distribution of $\mlimit$, $\tpeak$, and the deviation from the $\tpeak$--$\width$ relation as a function of the halo property (gray shading). The lines and shaded areas show the median and statistical uncertainty for each of the three bins in stellar mass. The black dashed and dotted lines in the bottom panels highlight the mean and 68\% scatter of the $\tpeak$-$\width$ relation. In the third column, the gray dashed lines show the relation expected if the SFR was proportional to the halo mass accretion rate (MAR). See Sections \ref{sec:results:correlations} and \ref{sec:discussion:galaxyhalo} for a detailed interpretation of this figure.}
\label{fig:correlations}
\end{figure*}

Having established that our simulated and observed galaxy samples occupy similar areas of the log-normal parameter space, we now ask what galaxy formation physics causes particular shapes of the SFH: are they pre-determined by initial conditions such as the density environment, influenced by the history of a galaxy's halo, or do they arise from the complicated physics of galaxy formation? For this investigation, we consider the entire \illustris sample with $M_* > 10^9 \msun$. 

\subsubsection{The Peak Time--Width Relation}
\label{sec:res:correlations:peaktimewidth}

Figure~\ref{fig:params_tpeakwidth} reproduces the $\tpeak$--$\width$ relation for the entire \illustris sample. The correlations with $\mlimit$ are very similar to those in the high-mass sample shown in Figure~\ref{fig:params_alt}. Figure~\ref{fig:params_tpeakwidth} demonstrates that the $\tpeak$--$\width$ relation is well approximated by a power law, 
\begin{equation}
\label{eq:correlation}
\width = 0.83\, \tpeak^{3/2} \,,
\end{equation}
with a standard deviation of 0.18 dex (about 50\%). The line fit and scatter for the $M_* > 10^{10} \msun$ sample are virtually identical. According to this relation, the SFHs in \illustris follow a scaling relation between their peak time and width, where early forming galaxies also form quickly. We notice a few outliers, for example, relatively late, short starburst SFHs (see Figure~\ref{fig:fits} for an example), but overall the relation is surprisingly tight. \citetalias{pacifici_16} found a similar relation between SFH width and the age of the universe, but they referred to the time of observation rather than peak time (their Figure 6).

Given how well the peak time of an SFH predicts its width, the question of which galaxy properties determine the SFH can be divided into two separate questions: which properties change peak time and width along their degeneracy, and which properties (if any) influence width at fixed peak time?

\subsubsection{Centrals vs. Satellites}
\label{sec:res:correlations:satcen}

Part of the answer to these questions might be found in the differences between central and satellite galaxies. Figure~\ref{fig:params_censat} shows the distributions of their parameters compared to the overall sample. The distributions do not strongly depend on mass, except at very high masses, where most galaxies are early forming.

Figure~\ref{fig:params_censat} shows that even though satellites are, on average, slightly less massive than centrals (first panel), they form earlier (second panel), in contrast with the downsizing expectation that larger galaxies form earlier. Moreover, satellites form significantly faster than centrals (third panel), both because they are an intrinsically older population and because their star formation may be quenched after infall into their hosts \citep[e.g.,][]{vandenbosch_08, hearin_13, wetzel_13}. 

However, the differences between the central and satellite populations do not appear to be responsible for the scatter in the $\tpeak$-$\width$ relation. The right panel of Figure~\ref{fig:params_censat} shows the deviation from this relation, which is more or less symmetric for both centrals and satellites, though centrals have a slight tendency to lie above the relation (i.e., exhibit relatively wide SFHs at fixed $\tpeak$) and satellites below. The scatter around the relation is comparable for the two sub-samples, 0.16 dex for centrals and 0.2 dex for satellites.

\subsubsection{The Galaxy--Halo Connection}
\label{sec:res:correlations:halo}

As discussed in Section~\ref{sec:intro}, we expect the MAH of a halo to have significant influence on the evolutionary history of its galaxy. Thus, we compare the log-normal parameters to a number of halo properties in Figure~\ref{fig:correlations}. The first impression from the gray histograms is that all correlations are subject to large scatter, particularly at low mass. This scatter implies that galaxy SFHs are diverse and not determined by any one parameter (see \citealt{dressler_16} for an observational investigation). Nevertheless, some trends are well-defined in an average sense. Many halo and galaxy properties correlate with halo and stellar mass, introducing a trivial correlation with peak time and width. Thus, we split the overall sample into three stellar mass bins. We have also verified that the central and satellite samples are similarly distributed for the correlations shown.

The left column of Figure~\ref{fig:correlations} shows the correlations with the maximum mass a halo has obtained over its history (for subhalos, the mass today can be significantly lower than this maximum). The corresponding relations with stellar mass are very similar. For large, early forming galaxies, $\mlimit$ is roughly equal to $M_*$ today, which manifests itself in a well-defined stellar mass--halo mass relation at the high-mass end. As expected, galaxies in more massive galaxies form earlier. At low halo or stellar masses, however, the correlation disappears \citep[in agreement with][]{dressler_16}. At first sight, this correlation seems to be in conflict with the expectation that large dark matter halos form late in hierarchical structure formation. However, while the halos of massive clusters are still growing today, their central galaxies are growing by mergers rather than star formation, meaning that their stellar populations whose ages we investigate here were largely formed in other halos \citep{rodriguezgomez_16}. Thus, large halos form late whereas large central galaxies form early \citep[e.g.,][]{neistein_06}.

The next variable we consider is the environment at $z = 3$, i.e. whether a galaxy was born in an overdense or underdense region of the universe. We quantify density by the distance to the fifth nearest neighbor, but using a smoothed galaxy overdensity gives similar results. The trend that galaxies born in overdense environments form earlier holds for all stellar masses, and is a manifestation of the earlier collapse times of halos in overdense regions, known as ``assembly bias'' \citep[][see also the arguments in \citealt{dressler_80}, \citealt{abramson_16}, and \citealt{kelson_16}]{gao_05_assembly, wechsler_06, croton_07, dalal_08, zentner_14, lin_16, miyatake_16, tinker_16}. The environment at earlier and later redshifts shows similar relations with $\tpeak$, but the correlation weakens at late ($z < 1$) and very early ($z > 4$) redshifts. 

We expect that the age of a galaxy should be strongly influenced by the age of its halo. We have measured halo age in a number of ways: as the redshift where half (or a quarter) of today's mass has been accreted, by fitting with the MAH models of \citet{wechsler_02_halo_assembly} and \citet{tasitsiomi_04_clusterprof}, and as the starvation redshift $z_{\rm starve}$ suggested by \citet{hearin_13} which is defined as the maximum of $z_{\rm wechsler}$, the redshift where the halo first grew to $10^{12} \msun$, and the redshift when it was accreted if it is a satellite. The correlations with $z_{\rm wechsler}$ and $z_{\rm starve}$ are shown in the right columns of Figure~\ref{fig:correlations}. For all stellar masses, earlier halo formation does correspond to lower $\tpeak$ as expected, though the correlation exhibits large scatter. This scatter is not surprising given that the stellar mass--halo mass relation is subject to a scatter of about 0.2 dex \citep[e.g.,][]{more_09_ii, behroozi_13_shmr, reddick_13, gu_16}.

Interestingly, $z_{\rm starve}$ exhibits a more pronounced (and more or less mass-independent) relation with $\tpeak$. By construction, $z_{\rm starve}$ is supposed to correspond to the redshift when a halo's galaxy is expected to quench \citep{hearin_13}. The strong correlation with $\tpeak$ means that the \illustris galaxy population supports the age matching picture in which the SFR (or color) of a galaxy is directly connected to its halo's accretion history. However, this connection is non-trivial: massive, early forming galaxies tend to live in halos with a high $z_{\rm starve}$ (because the halo reached $10^{12} \msun$ early), whereas $z_{\rm wechsler}$ may indicate a late formation time for the same halo, leading to the large disagreement between the trends for high-mass galaxies (purple lines in Figure~\ref{fig:correlations}). All other definitions of formation redshift are less correlated with $\tpeak$ than $z_{\rm starve}$ and $z_{\rm wechsler}$.

So far, we have discussed how $\tpeak$ depends on stellar mass, environment, and halo age. However, what influences the duration of a galaxy's star formation at fixed peak time? From the bottom panels in Figure~\ref{fig:correlations}, we conclude that mass and environment have little effect on whether an SFH lies above or below the peak time--width relation. It appears that very massive galaxies ($M_* > 10^{11} \msun$) generally lie slightly below the relation, but this trend is not very strong. However, $z_{\rm wechsler}$ does correlate significantly, and for all stellar masses, with the deviation from the relation. In particular, at fixed peak time, galaxies in early forming halos tend to have relatively broad SFHs, while late-forming halos have relatively narrow SFHs. In Section~\ref{sec:discussion:galaxyhalo}, we demonstrate that this trend is generically expected if the the SFR is proportional to the halo mass accretion rate (MAR). We cannot detect a comparable correlation with $z_{\rm starve}$, indicating that $\width$ is set by the MAH at times after $z_{\rm starve}$ to which $z_{\rm wechsler}$ is sensitive.

We note that these results do not allow us to directly assess the relative importance of mass quenching and environment quenching \citep[e.g.,][]{peng_10, vogelsberger_14_illustris}. The strong correlation with $z_{\rm starve}$ (which relies on a physical halo mass scale) indicates that halo quenching plays an important role, but the formation history of halos is tied to their environment through assembly bias, making it hard to disentangle the two effects.

We conclude that the accretion history of a galaxy's halo plays a key role in determining $\tpeak$ and $\width$, in agreement with many theoretical models (Section~\ref{sec:intro}). We discuss the galaxy--halo connection further in Section~\ref{sec:discussion:galaxyhalo}.

\subsubsection{Ancillary Correlations}
\label{sec:res:correlations:other}

\begin{figure}
\centering
\includegraphics[trim = 2mm 4mm 0mm 0mm, clip, scale=0.51]{\figdir/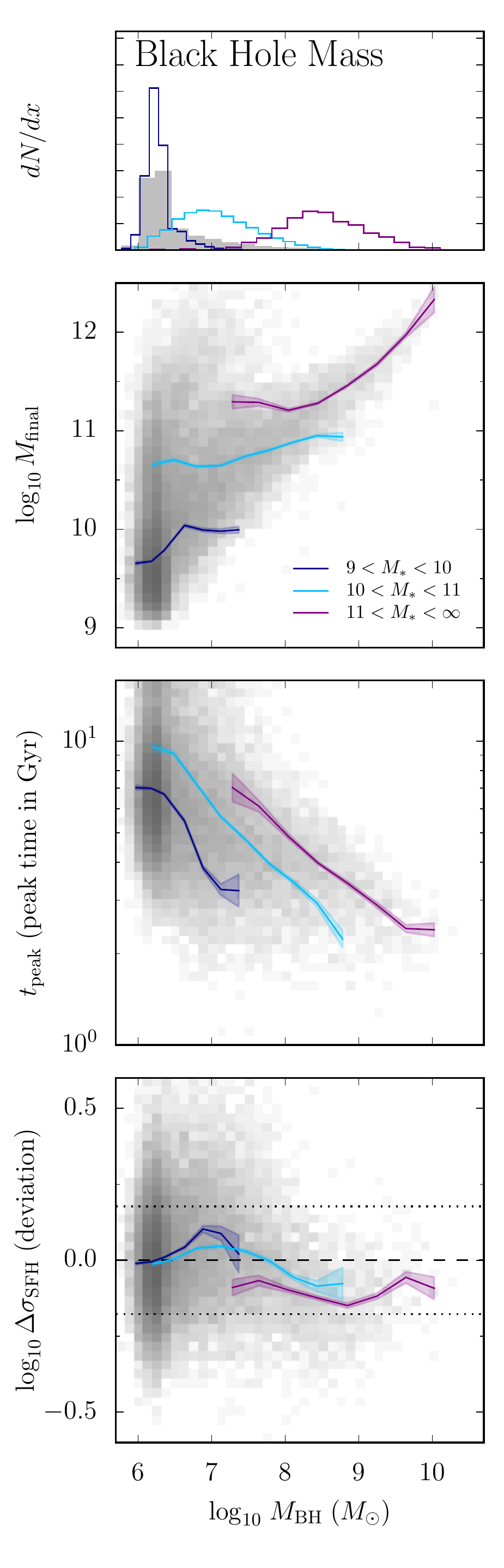}
\includegraphics[trim = 20mm 4mm 0mm 0mm, clip, scale=0.51]{\figdir/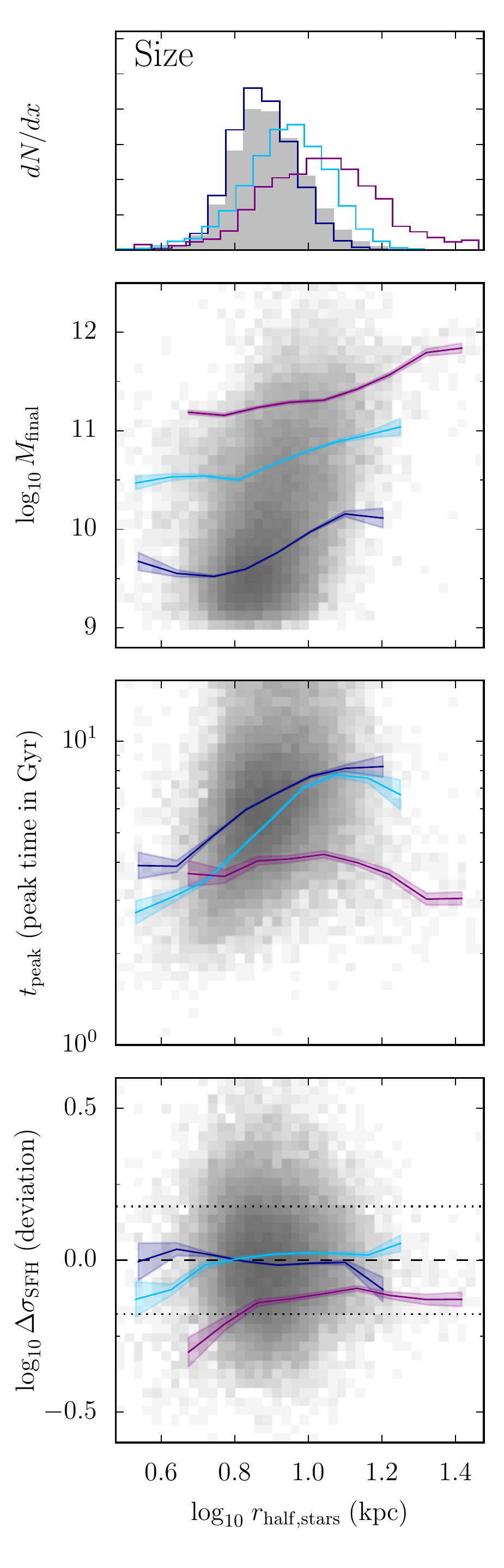}
\caption{Same as Figure~\ref{fig:correlations}, but for black hole mass (left column) and galaxy size (defined as the stellar half-mass radius, right column). \illustris predicts a tight relation between peak time and black hole mass such that older galaxies host larger black holes. Similarly, older galaxies are significantly smaller (at $M_* < 10^{11} \msun$, the trend disappears for the most massive galaxies).}
\label{fig:correlations2}
\end{figure}

We now turn to physical properties of galaxies that we do not expect to fundamentally influence the SFH, but that exhibit interesting correlations with the SFH. To that end, the left column of Figure~\ref{fig:correlations2} shows correlations of the best-fit parameters with black hole mass at $z = 0$. Even at fixed stellar mass, \illustris predicts a strong relation between black hole mass and $\tpeak$, where earlier-forming galaxies host more massive black holes. The population at high $\mlimit$ and low black hole mass is due to to extremely late-forming galaxies, which are assigned an artificially high $\mlimit$ despite their moderate stellar mass at $z = 0$ (see the discussion in Section~\ref{sec:results:params}).

The accretion onto black holes cannot be resolved in a cosmological simulation, meaning that some of the correlation may be a manifestation of the seeding of black holes and their co-evolution with galaxies in the \illustris model. However, given that \illustris reproduces the observed black hole mass function \citep{sijacki_15}, Figure~\ref{fig:correlations2} makes a clear prediction that earlier-forming galaxies host larger black holes \citep[see also][]{sijacki_15, bluck_16}.

Another property of galaxies that shows a strong correlation with $\tpeak$ is their size, defined as the stellar half-mass radius at $z = 0$ (right column of Figure~\ref{fig:correlations2}). At a given mass, earlier-forming galaxies are significantly smaller today than their late-forming counterparts \citep{oesch_10, valentinuzzi_10, poggianti_13, carollo_13, vanderwel_14, fagioli_16}. This correlation becomes insignificant for the largest \illustris galaxies with $M_* > 10^{11} \msun$, presumably because their structure is significantly influenced by their merger history \citep{rodriguezgomez_17_spin}. We have also investigated a number of other galaxy properties such as metallicity, host halo mass (for satellites), merger history, and the fraction of stars formed in situ. Many of those properties correlate strongly with stellar mass, which drives their correlations with $\tpeak$. 


\section{Discussion}
\label{sec:discussion}

\begin{figure}
\centering
\includegraphics[trim = 0mm 2mm 5mm 0mm, clip, scale=0.66]{\figdir/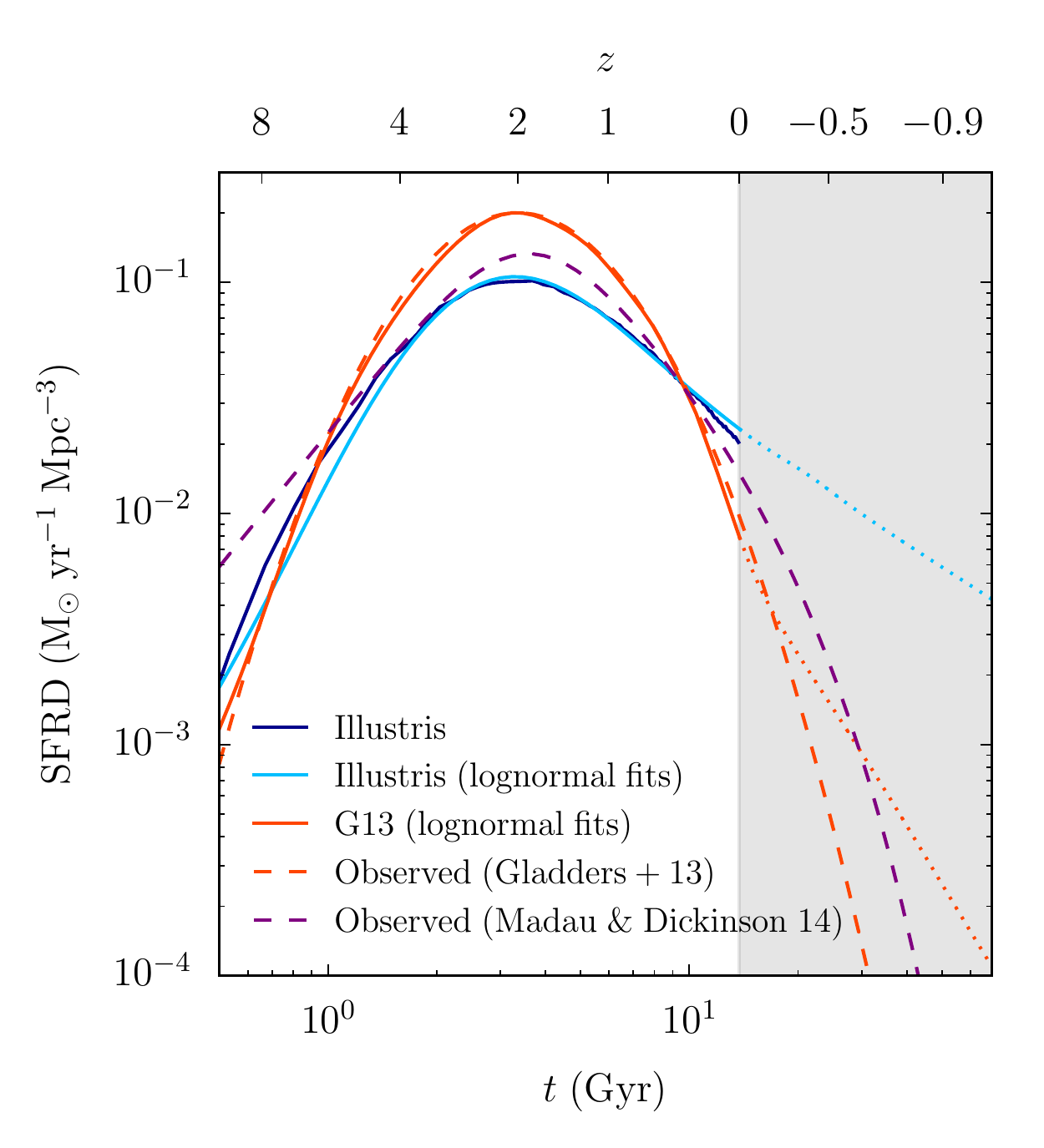}
\caption{Global star formation rate density from observations, simulation, and log-normal fits. The SFRD of all \illustris galaxies in our sample is shown in dark blue, as well as the corresponding log-normal fits (light blue), the \citetalias{gladders_13_icbs4} log-normal fits (orange), and the fits to the observed data of \citetalias{gladders_13_icbs4} (dashed orange) and \citet[][dashed purple]{madau_14}. The offset in normalization between \illustris and the observations is largely due to the contribution of galaxies with $M_* < 10^9 \msun$ that are not included in our sample. The gray shaded area highlights the future, and dotted lines show the extrapolation of the log-normal fits (see Section \ref{sec:discussion:sfrd} for a detailed discussion and caveats regarding this extrapolation).}
\label{fig:sfrd}
\end{figure}

\begin{figure*}
\centering
\includegraphics[trim = 4mm 2mm 6mm 0mm, clip, scale=0.56]{\figdir/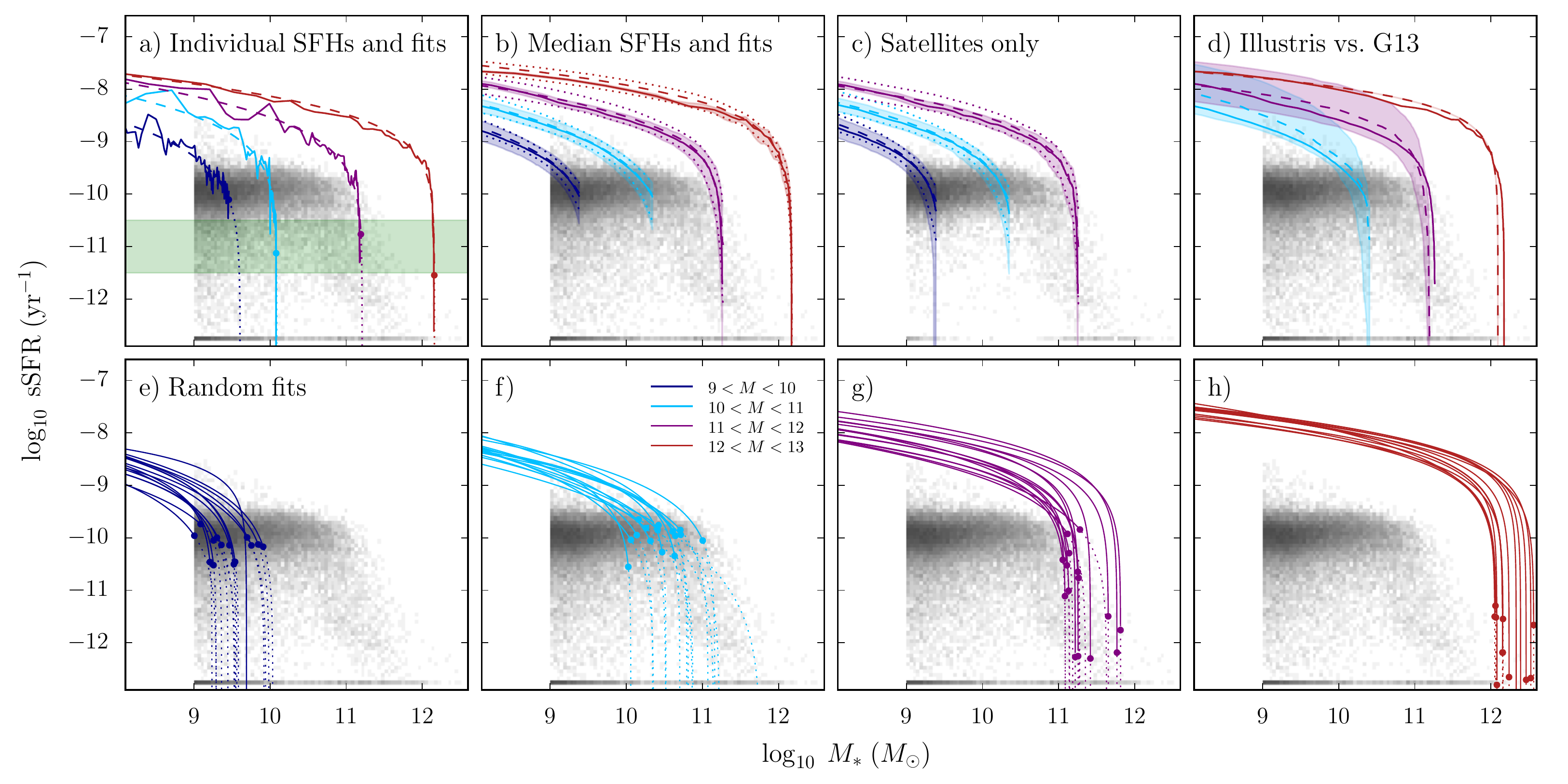}
\caption{Evolution of SFHs and log-normal fits with respect to the star formation main sequence. In each panel, the gray histogram shows the distribution of \illustris sSFR and stellar mass at $z = 0$ (in panel (c), only satellites are included). All galaxies with $\log_{10} \mathrm{sSFR} < -13$ are included in the thin line at the bottom of the histogram. In all panels, the colors refer to the same four mass bins. Panel (a): representative \illustris SFHs (solid lines) from each mass bin and the corresponding log-normal fit (dashed lines). The evolution of the fitted trajectory after $z = 0$ is shown with dotted lines, the points indicate the location of the fit at $z = 0$. The green shaded band represents our definition of the green valley. Panel (b): the median (solid lines) and 68\% scatter (shaded areas) of the trajectories in each mass bin, compared to the median fit (dashed lines) and 68\% scatter of the fits (dotted lines). On average, the fits track the evolution of the SFHs very well. Panel (c): same as panel (b), but only for galaxies that are satellites at $z = 0$. Panel (d): the solid lines show the same medians as in the previous panels, the dashed lines and shaded areas the median and 68\% scatter of the \citetalias{gladders_13_icbs4} sample for the same mass bin. Panels (e) through (h): randomly chosen fits from each mass bin. Virtually all trajectories end in a steep drop to low sSFR.}
\label{fig:ssfr}
\end{figure*}

We have demonstrated that the log-normal functional form describes the majority of simulated and observationally inferred SFHs in the \illustris and \citetalias{pacifici_16} samples. We have shown that the log-normal parameter space of normalization, peak time, and width is useful for comparing the stellar histories of simulated and observed galaxies, and to infer the physical parameters that influence star formation in galaxies. In this section, we test the performance of the log-normal framework using more advanced metrics and discuss the theoretical motivation for log-normal SFHs. 

\subsection{The Global SFRD}
\label{sec:discussion:sfrd}

Figure~\ref{fig:sfrd} compares various estimates of the global SFRD of the universe. The dashed lines show the observational estimates of \citetalias{gladders_13_icbs4} and \citet{madau_14}. The SFRD of our \illustris sample (dark blue) underestimates the observed normalization due to contributions from galaxies with $M_* < 10^9 \msun$, which we do not account for. The \illustris SFRD is well matched by the sum of all log-normal fits (light blue). The SFRD of the \citetalias{gladders_13_icbs4} log-normal fits (orange line in Figure~\ref{fig:sfrd}) matches their log-normal fit to the data almost perfectly. This match is expected because the SFRD was used as a constraint in their fits (any deviations are due to the re-sampling of the \citetalias{gladders_13_icbs4} galaxies, see Section~\ref{sec:methods:g13}).

We extrapolate the fitted SFRDs into the future (dotted lines), but keep two important caveats in mind. First, log-normal SFHs with significant future star formation cannot react to quenching events, meaning their limiting stellar mass is likely an overestimate of what one would find if \illustris were run into the future. Second, the prediction is affected by our prior on peak time: if left unconstrained, the most extreme cases of late-forming, high-width SFHs dominate the prediction. Even with very strict priors, however, the future SFRD of the log-normal fits does not decrease significantly compared to that shown in Figure~\ref{fig:sfrd}, implying that the high future SFRD is not solely due to unconstrained fits. In fact, the log-normal extrapolation is not too surprising given that the slope of the \illustris SFRD is already shallower than the observations at $z \leq 1$. This deviation is not an artifact of the log-normal fit, and has been noted before (see Figure 8 of \citealt{vogelsberger_14_illustris} and Figure 6 of \citealt{torrey_14}, or Figure 2 of \citealt{sparre_15}).

\subsection{Evolution along and across the Main Sequence}
\label{sec:discussion:sfms}

While we have shown that log-normals faithfully capture the overall, cumulative evolution of SFHs, we have yet to investigate metrics that are more sensitive to the instantaneous SFR. For example, the distinction between star-forming and quiescent galaxies is often drawn based on their position in $M_*$--sSFR space, where star-forming galaxies fall onto a relatively tight, evolving relation called the main sequence \citep[e.g.,][]{noeske_07}. In this section, we investigate whether the log-normal fits correctly reproduce the $M_*$--sSFR evolution of \illustris galaxies. 

Figure~\ref{fig:ssfr} summarizes various aspects of the $M_*$--sSFR trajectories. The gray background histogram in each panel shows the $z = 0$ distribution of all \illustris galaxies in our sample, where the line at the bottom contains galaxies with $\log_{10}$ SFR below $-13$. The main sequence in \illustris is somewhat flatter than observations suggest \citep{sparre_15}, preventing us from comparing the evolution of the SFHs relative to some main sequence fitting function \citep[e.g.,][]{speagle_14}. Instead, we consider trajectories in $M_*$--sSFR space itself.

Panel (a) of Figure~\ref{fig:ssfr} shows the trajectories of four individual galaxies, which were randomly chosen from four mass bins indicated by different colors. The log-normal fits result in trajectories very similar to those of the SFHs themselves. Panel (b) addresses the same question in a more quantitative manner, by showing the median trajectories of all SFHs and fits. Not only do the fits reproduce the medians very well, the 68\% scatter in the fits is also very similar to that in the SFHs. We do, however, note that the scatter of the fits in the high-mass bins ($M_* > 11$) exceeds the scatter in the SFHs at early times. This difference can be understood by considering that the stellar mass of the galaxies is growing more rapidly than at late times, meaning that fit errors will manifest themselves in a larger deviation from the median.

In Section~\ref{sec:results:fitquality}, we found that log-normals sometimes have trouble fitting the SFHs of satellites that cease their star formation abruptly. Thus, we separately investigate the median SFHs and fits of satellites in panel (c). Their trajectories do, as expected, reach lower sSFRs at late times, on average, but the fit quality is as good as for the entire sample. Even if the SFR decreases too slowly in a log-normal fit, this decrease will lead to a steep trajectory away from the main sequence because $M_*$ evolves very little. This means that the trajectories are essentially insensitive to the duration of quenching. We thus conclude that the good match of our fits in this parameter space is reassuring, but that quenching trajectories are easy to match regardless of how good the SFH fit is at late times. 

In the same vein, we find that the $M_*$--sSFR trajectories of \illustris and \citetalias{gladders_13_icbs4} galaxies agree better than one might expect from the somewhat different best-fit parameter distributions (panel (d) of Figure~\ref{fig:ssfr}). This agreement makes sense considering that \citet{abramson_16} showed that the \citetalias{gladders_13_icbs4} sSFR evolution is compatible with the observed evolution of the main sequence normalization and slope, and that \illustris also roughly matches the observed evolution \citep{sparre_15}.
 
Finally, the bottom panels of Figure~\ref{fig:ssfr} provide an intuition for where the log-normal trajectories in the different mass bins typically end up (dots indicate the end points at $z = 0$, dotted lines the future development). The sSFR of log-normals decreases monotonically, and Figure~\ref{fig:ssfr} reaffirms that all log-normal trajectories end in a steep drop to low sSFRs eventually. In this picture, the scatter in the main sequence is naturally explained as an age gradient at fixed stellar mass \citep{abramson_16, oemler_16}. 

\subsection{Can Log-normals Capture Rapid Quenching?}
\label{sec:discussion:quench}

In the previous section, we concluded that $M_*$--sSFR trajectories are very well reproduced by log-normal fits, but that these trajectories are insensitive to the timing of the sSFR evolution. In this section, we test this timing explicitly by considering the timescale over which star formation ceases. A galaxy is generally defined to be quiescent (or have quenched) if its sSFR that has fallen below a particular value well below the star formation main sequence, with the so-called green valley as an intermediate space between star-forming and quiescent galaxies. To assess the time evolution of log-normal fits, we measure their green valley crossing time, which we define to be the time an SFH spends between logarithmic sSFRs of $-10.5$ and $-11.5$ (see panel (a) in Figure~\ref{fig:ssfr}). For this test, we consider only SFHs whose sSFRs have fallen below $-11.5$ in the last five time bins to ensure that they have reliably crossed the green valley. Furthermore, we require that the log-normal fit has also fallen below an sSFR of $-11.5$ (though this criterion makes relatively little difference to the selection). According to these criteria, about 3\% of the \illustris galaxies have quenched today. While this fraction is lower than observed \citep{wetzel_13, bluck_16}, the actual number of galaxies does not matter for our comparison. We compute the time an \illustris SFH spent in the green valley by linearly interpolating between time bins. For the fits, we calculate the time spent in the green valley exactly. 

\begin{figure}
\centering
\includegraphics[trim = 2mm 3mm 3mm 0mm, clip, scale=0.5]{\figdir/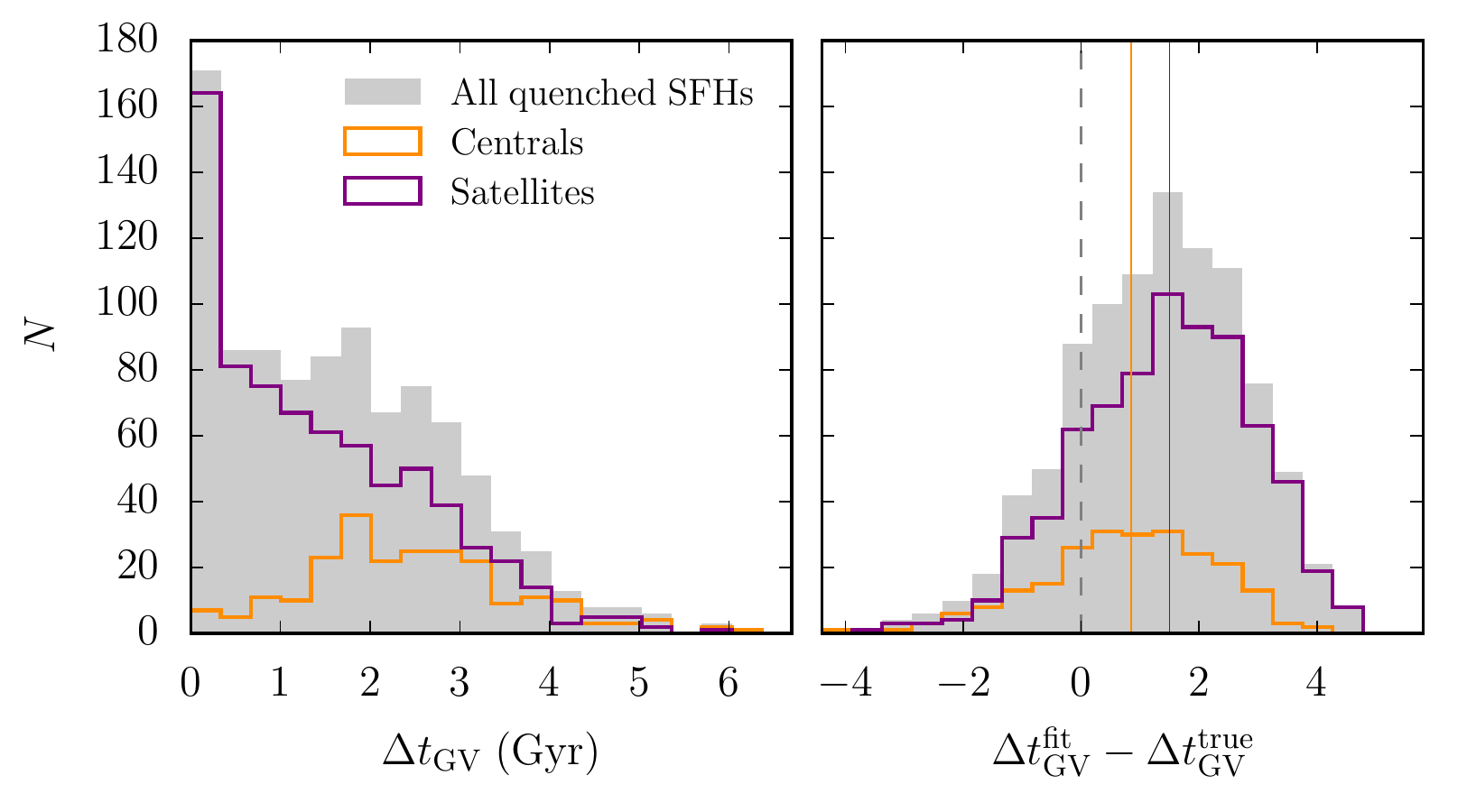}
\caption{Log-normal fits overestimate the time galaxies spend in the green valley. The left panel shows the green valley crossing times of all SFHs that are quenched at $z = 0$ (gray), split into centrals (orange) and satellites (purple). The right panel shows the differences between these crossing times and those measured from the corresponding log-normal fits. On average, the log-normal fits overestimate the time spent in the green valley by $0.85$ Gyr for centrals and $1.5$ Gyr for satellites (vertical lines).}
\label{fig:gv}
\end{figure}

Figure~\ref{fig:gv} shows the results of this exercise. The left panel shows the distribution of the actual green valley crossing times as measured from the SFHs. Satellites quench faster, on average, with a significant population that crosses the green valley in less than half a gigayear. The different timescales for centrals and satellites are expected and hint at different quenching mechanisms \citep[e.g.,][]{hahn_16}. The right panel shows the difference between the fitted and actual crossing times. For centrals, the distribution is relatively symmetric, but systematically offset by $0.85$ Gyr. For satellites, this difference increases to $1.5$ Gyr, largely because the log-normals fail to reproduce the population of rapidly quenching SFHs. On the other hand, \citet{abramson_16} showed that the \citetalias{gladders_13_icbs4} log-normals do reproduce the general increase of the average green valley crossing time with increasing age of the universe.

The overestimated quenching times represent the most serious limitation of log-normal fits that we have discovered. However, how important is this limitation? Some insight can be gained from abundance matching studies such as \citealt{wetzel_13} (see also \citealt{vandenbosch_08} and \citealt{hahn_16}). They find that above $M_* = 10^{10} \msun$, the dominant avenue by which galaxies cease to form stars is intrinsic rather than satellite quenching. This result explains why only a small fraction of SFHs in this mass range were poorly fitted by the log-normal form. Below $10^{10} \msun$, satellite quenching accounts for a larger and larger fraction of quiescent galaxies \citep{wetzel_13}. We did not investigate galaxies less massive than $10^{9} \msun$, but the importance of satellite quenching would presumably continue to increase with decreasing mass. For example, \citet{geha_12} find that dwarf galaxies are more or less only quenched if they are satellites, indicating that the slower, intrinsic cessation of star formation becomes less important at low masses. Moreover, the model of \citet{wetzel_13} infers very short quenching timescales for satellites, raising the possibility that quenching could occur even more rapidly than in \illustris. Unfortunately, their conclusions are hard to compare to our data because they rely critically on the observed fractions of green valley and quenched galaxies, which are not very well matched by \illustris (e.g., \citealt{bluck_16}; see also, however, the recent observational discussion regarding the density of galaxies in the green valley, e.g. \citealt{chang_15} or \citealt{eales_17}).

In summary, log-normal fits reproduce very well how galaxies evolve in $M_*$--sSFR space, but in some cases do not accurately capture how fast this evolution happens. As expected, this failure is most pronounced in rapidly quenched satellites, which represent a small fraction of the population at $M_* > 10^{10} \msun$, but are more frequent at lower masses.

\subsection{The SFH of the Milky Way}
\label{sec:discussion:history}

\begin{figure}
\centering
\includegraphics[trim = 5mm 5mm 5mm 5mm, clip, scale=0.64]{\figdir/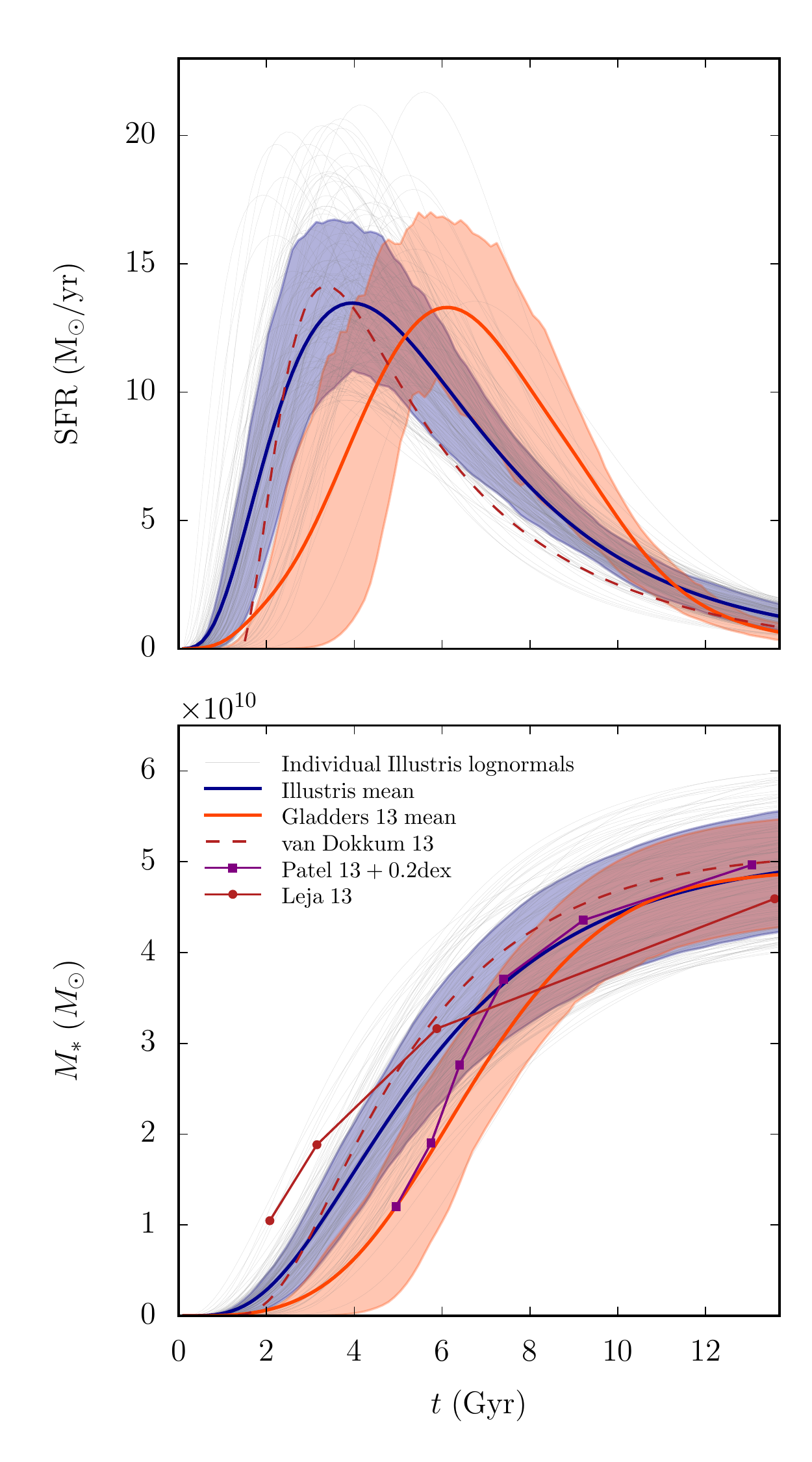}
\caption{The SFR (top panel) and cumulative SFH (bottom panel) of Milky Way analogs, selected to have $4 < M_* < 6 \times 10^{10} \msun$ and $0.5 < {\rm SFR} < 2 \msun / {\rm yr}$. The light gray lines show all log-normals from \illustris that fall into this range at $z = 0$, the blue line shows their mean, and the blue shaded area the 68\% scatter. The same is shown for the \citetalias{gladders_13_icbs4} fits in orange. We compare to predictions assuming a constant number density of galaxies, both from observations \citep[][red dashed line]{vandokkum_13} and from semi-analytical modeling \citep[][red round points]{leja_13}. We also compare to the observed SFH of \citet[][purple square points, scaled up by 0.2 dex]{patel_13_mw} who used the star formation main sequence to determine MW progenitors.}
\label{fig:history}
\end{figure}

One possible application of the log-normal framework is to infer the stellar history of a particular type of galaxy. For example, Figure~\ref{fig:history} shows the SFR and SFH of Milky-Way-like galaxies (hereafter MW analogs) as inferred from the log-normal fits to the \illustris (dark blue) and \citetalias{gladders_13_icbs4} (orange) samples. We have selected MW analogs to have $4 < M_* < 6 \times 10^{10} \msun$ and $0.5 < {\rm SFR} < 2 \msun / {\rm yr}$, resulting in samples of $136$ \illustris and $76$ \citetalias{gladders_13_icbs4} galaxies. The \illustris SFH is, of course, almost identical to what one would obtain by just summing the SFHs of the galaxies in the sub-sample without log-normal fitting. However, the log-normal fits to the \citetalias{gladders_13_icbs4} sample allow us to compare their prediction to the simulation. 

Interestingly, the \citetalias{gladders_13_icbs4} Milky Way analogs are inferred to form later, with $\tpeak = 6.1$ Gyr compared to $4.0$ Gyr for the \illustris sample. We further compare these predictions to the observational result of \citet{vandokkum_13}. They assumed a fixed number density of galaxies in order to infer which high-$z$ galaxies correspond to MW-like galaxies today (red dashed lines in Figure~\ref{fig:history}). While the assumption of constant number density does not hold, in general \citep{behroozi_13_numberdensity, torrey_15_numberdensity, torrey_16_numberdensity, wellons_16_numberdensity}, tests with synthetic SFHs created using the \citet{guo_11} semi-analytic model indicate that the systematic issues of the method, namely merging galaxies and scatter, lead to moderate errors on the inferred SFHs \citep{leja_13}. The \citet{vandokkum_13} result is in good agreement with the \illustris sample, with an even slightly earlier peak time of $3.4$ Gyr. We compare their measurement to the SFH prediction of \citet[][red points]{leja_13}. 

We also show the SFH inferred by \citet{patel_13_mw} who used fits to the observed star formation main sequence and its evolution in order to determine MW progenitors. This selection is systematically different from the assumption of constant number density because the main sequence only takes star-forming galaxies into account. They investigated galaxies with a final mass of about $0.2$ dex lower than that of the MW, an offset we have added to their results in Figure~\ref{fig:history}. Interestingly, \citet{patel_13_mw} suggest a slightly later-forming MW than \citet{vandokkum_13}, in agreement with the \citetalias{gladders_13_icbs4} sample.

\subsection{The Galaxy--Halo Connection Revisited}
\label{sec:discussion:galaxyhalo}

In Section~\ref{sec:res:correlations:halo}, we established that the halo MAH is the most important factor determining the peak time and width of an SFH. As an extension of this result, we investigate whether the peak time--width relation arises simply as a consequence of halo MAHs in a $\Lambda$CDM universe. 

The simplest possible connection we can assume is a steady-state model, where the SFR is proportional to the halo MAR \citep[e.g.,][]{bouche_10, lilly_13, dekel_13, becker_15, tacchella_16_profiles}. For concreteness, we quantify the MAH using the function of \citet{wechsler_02_halo_assembly}. At $z \gsim 1$, we can neglect the influence of dark energy and use $a \propto t^{2/3}$ to convert Equation~\ref{eq:wechsler} into an expression for the MAR \citep[e.g.,][]{dekel_13},
\begin{equation}
\frac{dM}{dt} \propto e^{- \alpha z} (1+z)^{5/2} \,.
\end{equation}
Because this expression has only two free parameters, it predicts a particular relation between the peak time and width of the MAH (where peak time now refers to the time when the MAR was highest). Interestingly, this relation is a power law with a slope of $1$, shallower than the value of $3/2$ measured for the \illustris SFHs. At $z < 1$, we compute the peak time and width numerically, and find slopes even shallower than one. The results of this calculation are shown with gray dashed lines in Figure~\ref{fig:correlations}. The SFR $\propto$ MAR assumption significantly underpredicts the peak time at all formation redshifts, meaning that there is an offset between the formation times of halos and galaxies. However, it qualitatively explains the trend with respect to the mean peak time--width relation: at early formation times, the width of the halo MAH is larger than that of the SFH, but due to the shallow slope of the halo peak time--width relation this difference is reversed at late formation times. 

We note that the shape of the \citet{wechsler_02_halo_assembly} expression for the halo MAR is similar to a log-normal for certain parameters. Thus, we attempt to fit the \illustris SFHs with log-normals. As mentioned in Section~\ref{sec:methods:mah}, a significant fraction of these fits fail. As before, the successful fits exhibit earlier peak times than the SFHs and obey a power-law relation between the peak time and width of the MAH. In the case of log-normal MAH fits, however, the slope is $1.35$, closer to the slope of $3/2$ measured for the SFHs, implying that the correlation between MAHs and SFHs might be even tighter than indicated by the best-fit parameters of the \citet{wechsler_02_halo_assembly} function.

In summary, both the shape of the SFHs and the peak time--width relation are encoded in halo MAHs to some degree, but the simple assumption that the SFR is proportional to the MAR cannot explain the SFHs quantitatively. It seems that baryonic processes act to delay the peak time of the SFHs, and to increase the width at later formation times. For example, a decrease in star formation is expected to follow a decrease in gas supply with some time delay, which might not be constant \citep[e.g.,][see also the simulation results of \citet{feldmann_16_a, feldmann_16_b}]{wetzel_13, schawinski_14}. In general, baryonic processes are expected to decouple galaxy and halo growth to some extent in lower-mass systems \citep[e.g.,][]{brook_12, uebler_14}. Furthermore, the strength of the galaxy--halo connection in \illustris could be artificially enhanced because several physical processes in the Illustris model (such as black hole seeding and wind velocities) are tied to the dark matter properties of galaxies.


\section{Conclusions}
\label{sec:conclusions}

\begin{figure*}
\centering
\includegraphics[trim = 4mm 5mm 7mm 0mm, clip, scale=0.53]{\figdir/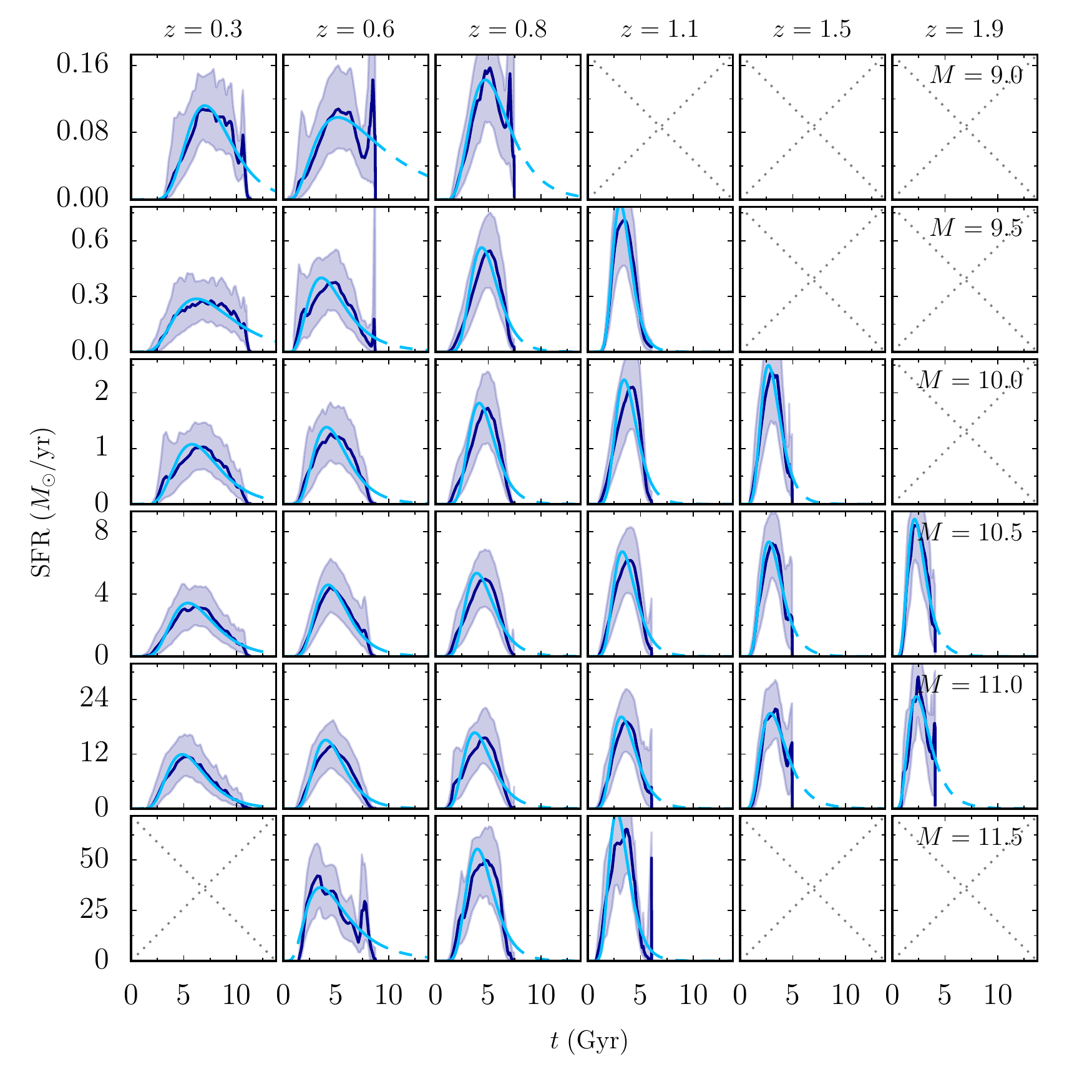}
\includegraphics[trim = 2mm 5mm 7mm 0mm, clip, scale=0.53]{\figdir/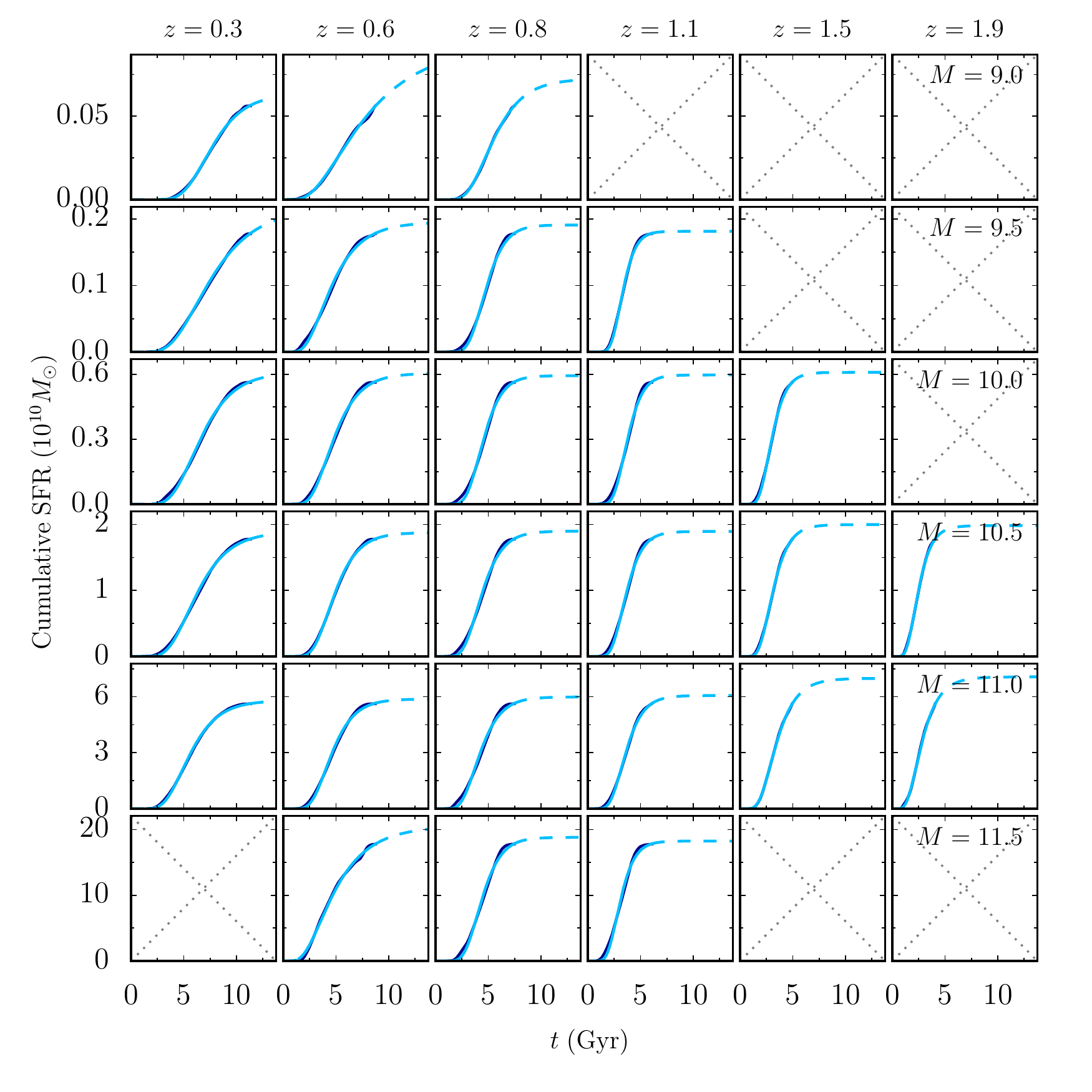}
\caption{Log-normal fits (light blue) to the median SFHs of \citet[][dark blue, shaded areas indicate the 50\% confidence contours]{pacifici_16}. The left block of panels shows the SFR, the right block the cumulative SFR. Within each block, each column corresponds to a redshift and each row to a mass. The labels in the rightmost panels give the mid-point of each mass interval, such that the first interval corresponds to $8.75 \leq \log_{10} M_* \leq 9.25$ and so on (compare to Figure 5 in \citetalias{pacifici_16}). The SFHs only run to the redshift where they were observed, whereas the fit is plotted over the entire age of the universe. }
\label{fig:pacifici_fits}
\end{figure*}

Inspired by the work of \citet{gladders_13_icbs4}, we have investigated the log-normal function as a description of the SFHs of individual galaxies in both simulations and observations, and shown it to be a good fit in the majority of cases. We have discussed the strengths and failures of the log-normal picture, and connected the log-normal parameter space to the physical properties of galaxies and their halos. Our main conclusions are as follows.

\begin{enumerate}
\item Log-normals are an excellent fit to the SFHs of the vast majority of \illustris galaxies. While log-normals ignore short-term fluctuations in the SFR, the cumulative SFH is fit to within 5\% deviation from the total mass formed for 85\% of \illustris galaxies, and to within 10\% for 99\% of galaxies. The most poorly fit SFHs belong to rapidly quenched satellite galaxies. We find similarly good fit qualities for the observationally inferred median SFHs of \citetalias{pacifici_16}. 
\item Comparing different fitting functions, we find that the log-normal performs slightly better than the popular delayed-$\tau$ model, though both models predict steeply rising and slowly declining SFHs. While models with more than three free parameters (such as the double power law) result in even better fits, they are less predictive and more difficult to constrain in observational analyses like \citetalias{gladders_13_icbs4}.
\item The \illustris, \citetalias{gladders_13_icbs4}, and \citetalias{pacifici_16} galaxy samples occupy similar regions in the log-normal parameter space of total stellar mass, peak time, and full width at half maximum, though \illustris predicts more extended SFHs than the observations. \illustris galaxies exhibit a tight correlation between peak time and width, $\tpeak \propto \width^{3/2}$, with only 0.18 dex scatter.
\item Log-normal fits correctly reproduce many global aspects of star formation, such as the SFRD and the evolution of galaxies along and across the star formation main sequence. The timing of this evolution, however, can be poorly matched for rapidly quenched satellite SFHs, leading to significantly overestimated quenching timescales.
\item Considering the physical properties of galaxies, we find that the formation history of the halo has the strongest influence on peak time and width, and that the starvation redshift of \citet{hearin_13} is a particularly good predictor of peak time. Generally, the trends are as expected: at fixed stellar mass, earlier-forming galaxies live in early forming halos, inhabit dense environments, host massive black holes, and have small sizes. Assuming that the SFR is proportional to the dark matter accretion rate can qualitatively explain the peak time--width relation, but does not quantitatively predict the correct peak times or widths. In general, SFHs are a diverse population with large scatter in all parameters, meaning that they are not uniquely determined by any one factor.
\end{enumerate}
We have left a number of possible investigations for future work. For example, we did not evaluate the fidelity of the \citetalias{gladders_13_icbs4} fitting procedure on an object-to-object basis, a test that could be performed using a simulated mock dataset. Conversely, we did not investigate whether the log-normal assumption can help in recovering masses and SFRs from photometry or spectra \citep[e.g.,][]{simha_14}. Finally, any analysis based on a single simulation is subject to the systematic uncertainty in the galaxy formation models used. Our analysis should be repeated using future cosmological simulations to ascertain the extent to which our results are specific to \illustris.


\vspace{0.5cm}

We are deeply grateful to all those who shared their data with us, namely the authors of \citetalias{gladders_13_icbs4} (Mike Gladders, Augustus Oemler, Alan Dressler, Bianca Poggianti and Benedetta Vulcani), Camilla Pacifici for sharing her SFHs, and the Illustris team for making their simulation public (Mark Vogelsberger, Shy Genel, Volker Springel, Debora Sijacki, Dandan Xu, Greg Snyder, Dylan Nelson, Lars Hernquist, and, in particular, Vicente Rodriguez-Gomez for his help with the merger trees). We thank Peter Behroozi, Charlie Conroy, Andrew Hearin, Lars Hernquist, Joel Leja, Camilla Pacifici, Ana-Roxana Pop, Laura Sales, Josh Speagle, and Sandro Tacchella for fruitful discussions and/or comments on a draft of this paper. B.D. gratefully acknowledges the financial support of an Institute for Theory and Computation Fellowship.


\appendix

\section{Mathematical Details of SFH Fitting Functions}
\label{sec:app:funcs}

\begin{figure*}
\centering
\includegraphics[trim = 0mm 2mm 2mm 0mm, clip, scale=0.5]{\figdir/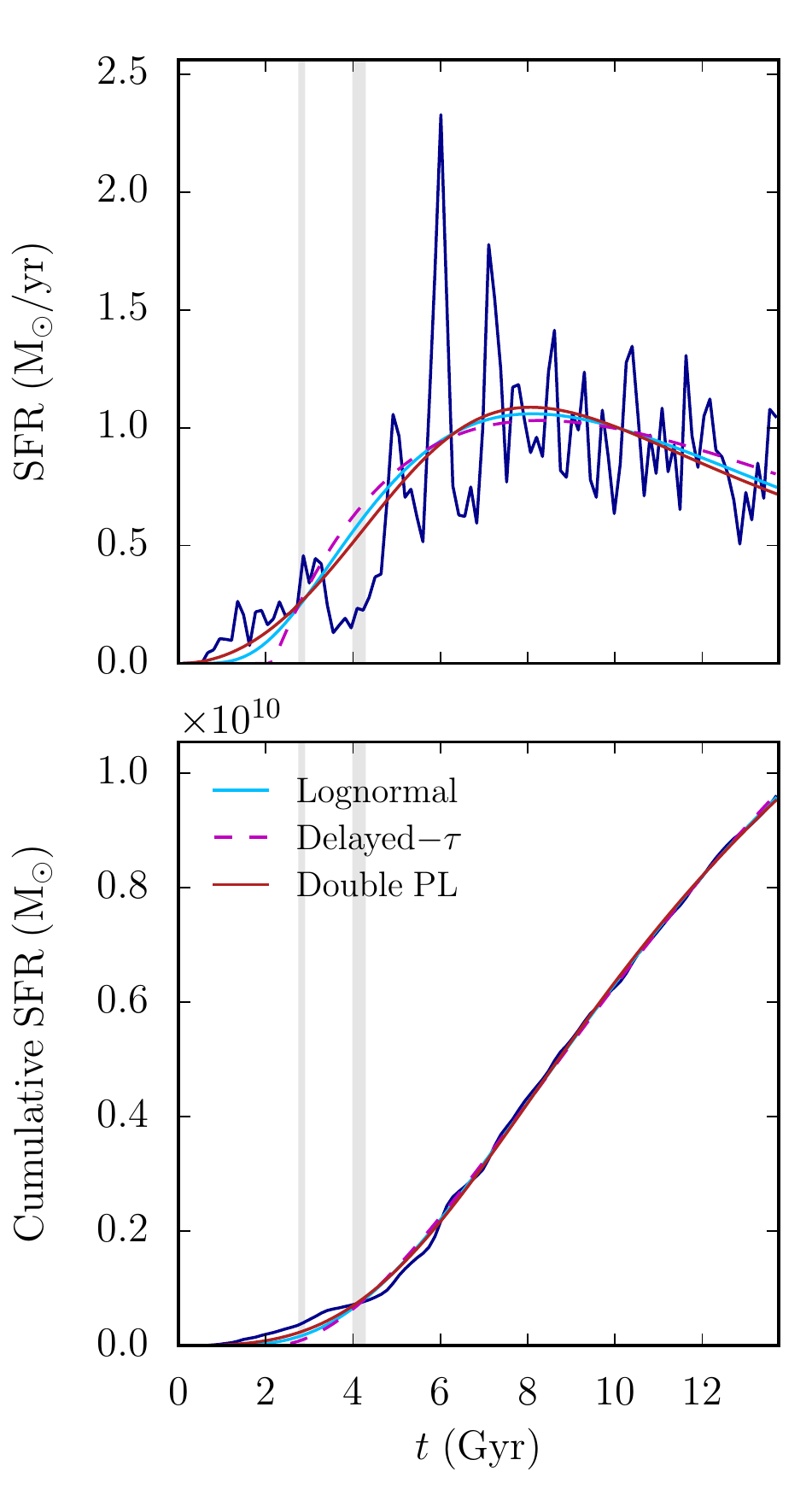}
\includegraphics[trim = 8mm 2mm 2mm 0mm, clip, scale=0.5]{\figdir/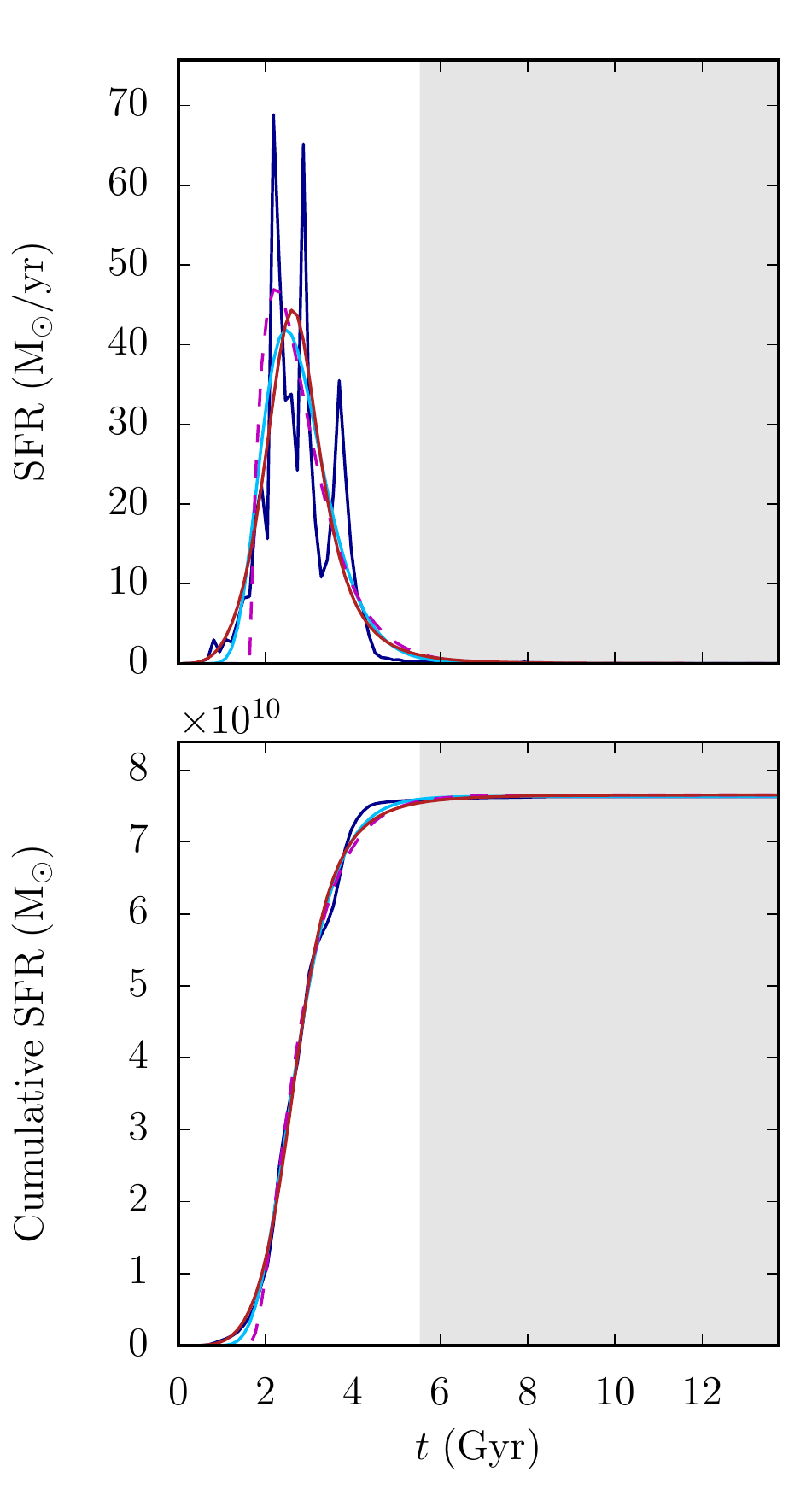}
\includegraphics[trim = 8mm 2mm 2mm 0mm, clip, scale=0.5]{\figdir/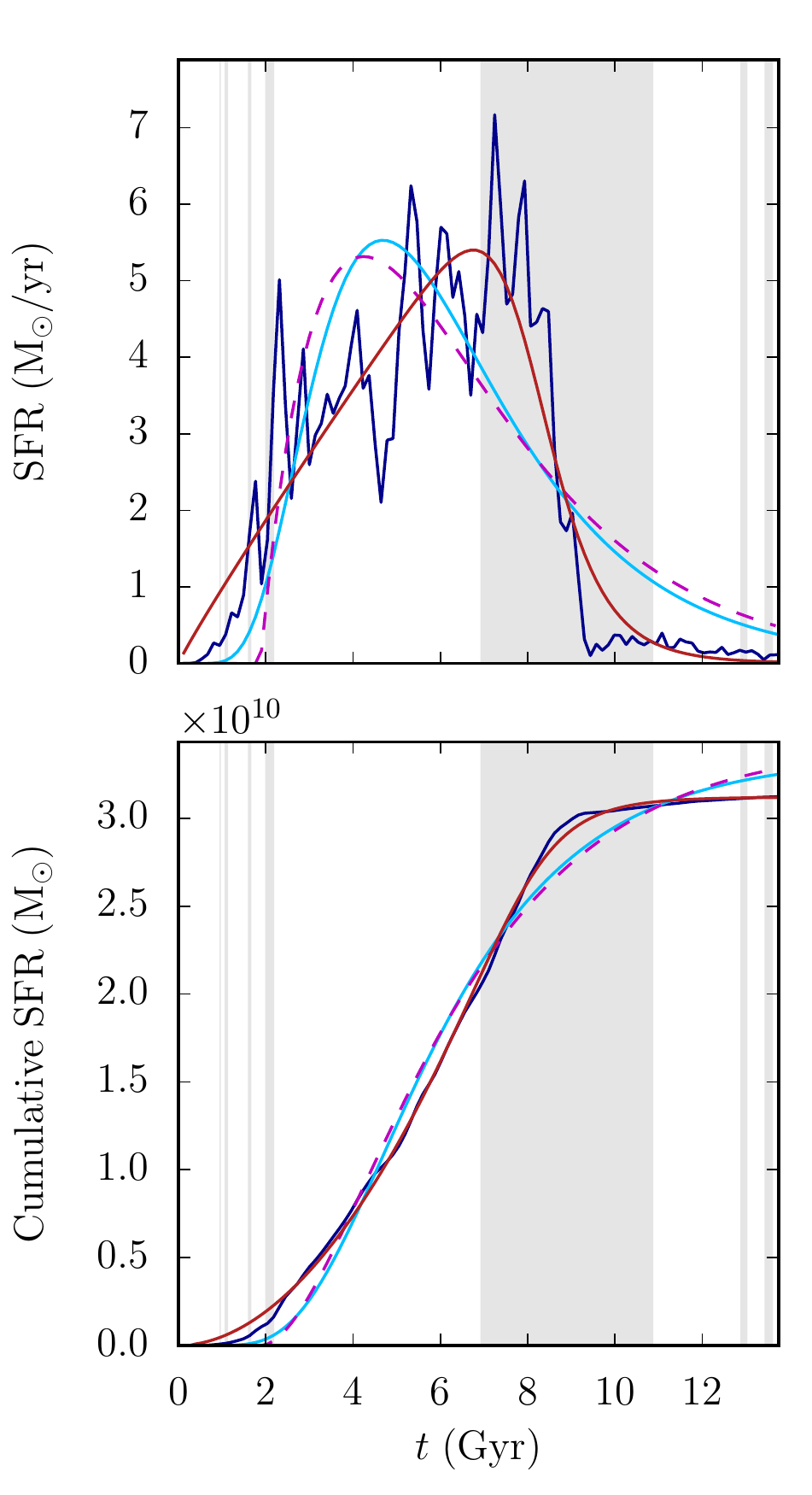}
\includegraphics[trim = 8mm 2mm 2mm 0mm, clip, scale=0.5]{\figdir/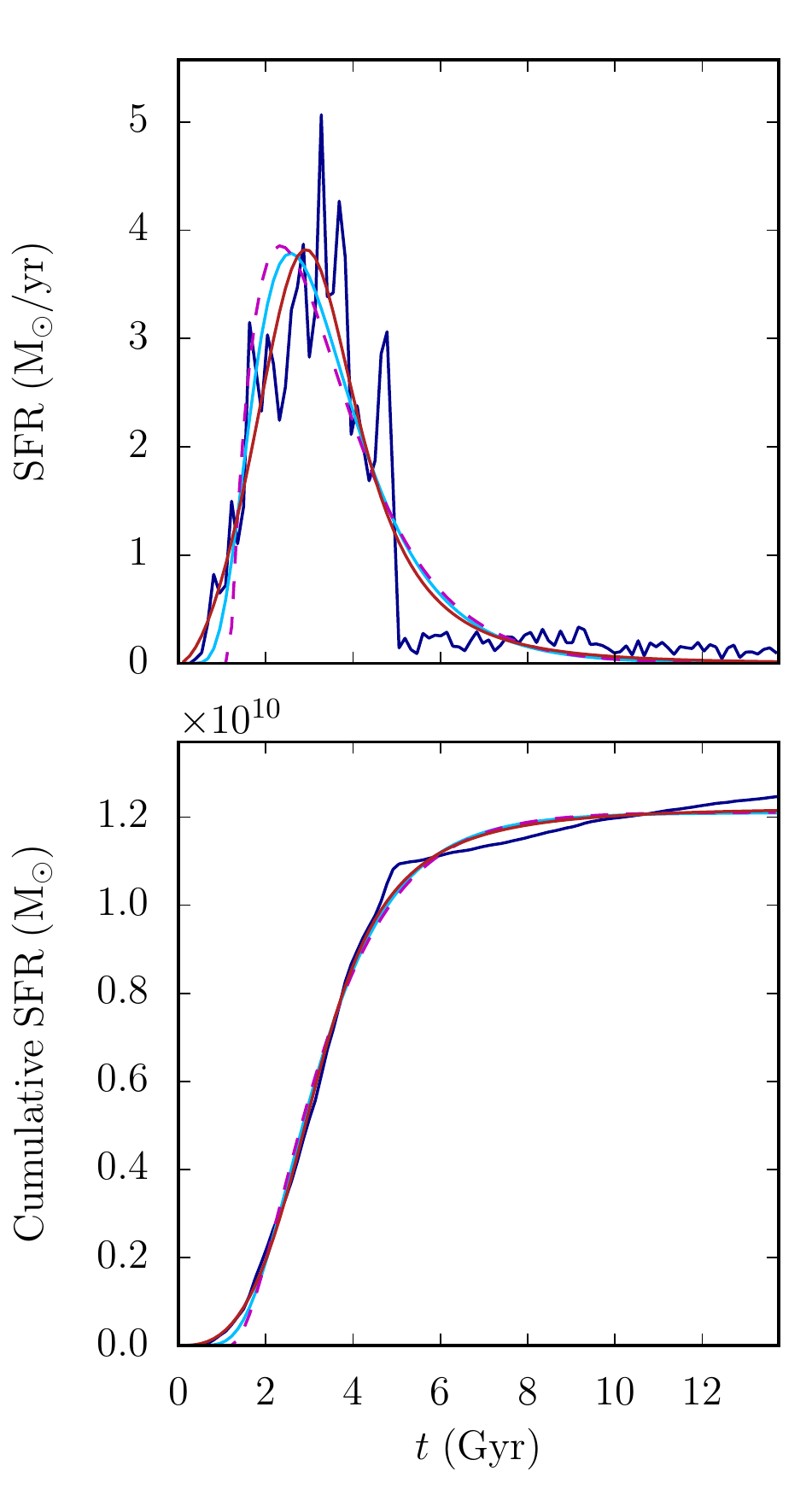}
\caption{Comparison of fits with different SFH models. The dark blue line shows the SFR (top panels) and cumulative SFH (bottom panels), fitted with a log-normal (light blue), delayed-$\tau$ model (dashed pink), and double power law (solid red). The fits shown here are representative examples of a few typical cases. From the left: (1) one of many SFHs for which all three fits more or less agree; (2) an SFH for which the very steep rising part of the delayed-$\tau$ model is a poor fit; (3) a galaxy that suddenly quenched after infall is best fit by the double-power-law model; (4) another case of rapid quenching where all three models fail.}
\label{fig:fitfuncs}
\end{figure*}

\begin{figure}
\centering
\includegraphics[trim = 0mm 5mm 5mm 0mm, clip, scale=0.65]{\figdir/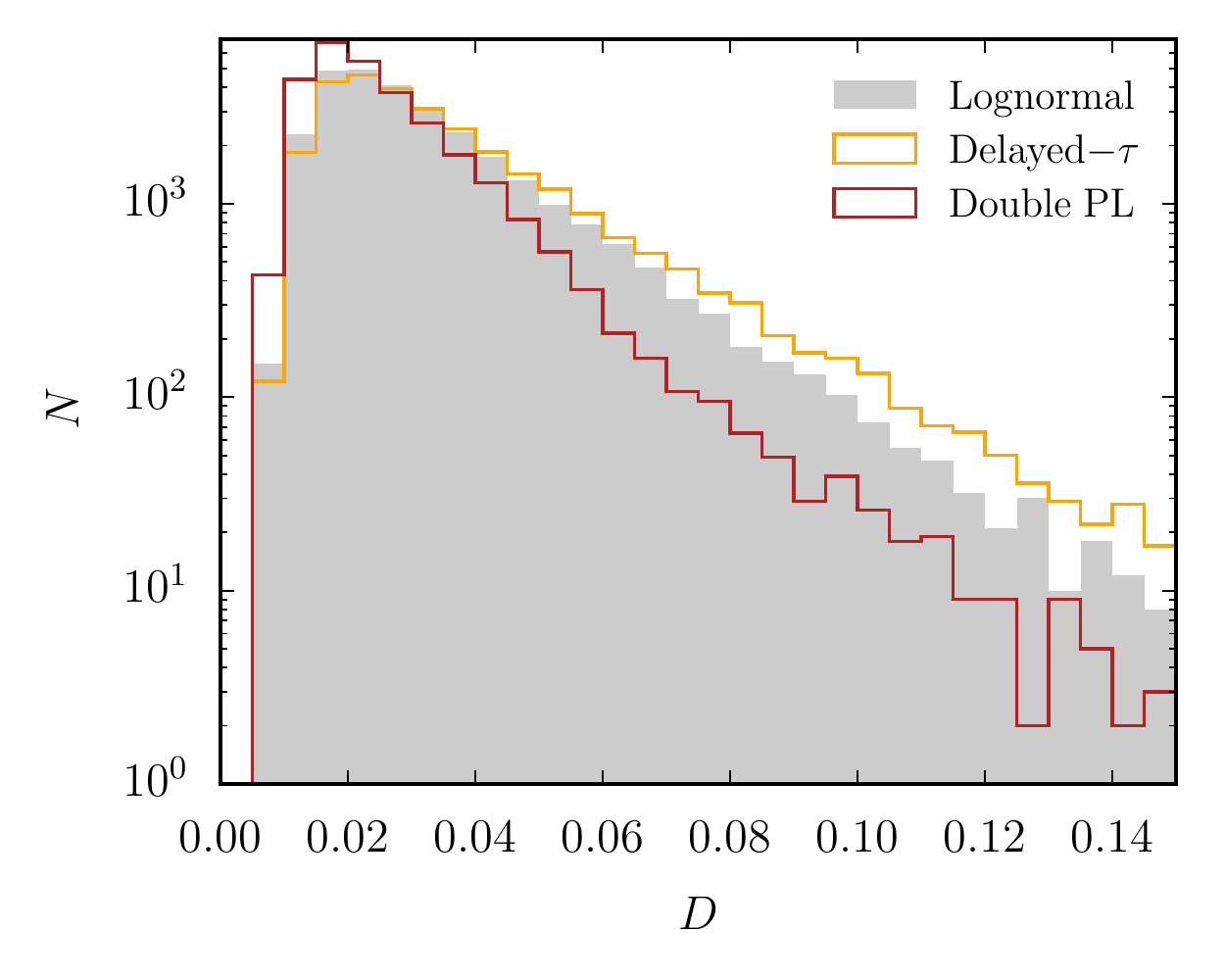}
\caption{Distribution of fit quality as quantified by the maximum deviation $D$ for different fitting models. The distribution is shown on a logarithmic scale in order to emphasize the tail at poor fit qualities. The delayed-$\tau$ model results in somewhat worse fits than the lognormal function. The double power law is an even better fit which is expected since it has one more free parameter.}
\label{fig:fit_quality_fitfuncs}
\end{figure}

In this Appendix, we compute certain properties of a number of commonly used SFH fitting functions, such as their peak time, peak SFR, and cumulative SFR.

\subsection{The Log-normal Function}
\label{sec:app:funcs:lognormal}

The log-normal functional form is given in Equation \ref{eq:lognormal}. Counter-intuitively, $T_0$ is not the logarithm of its peak time, which is given by
\begin{equation}
\label{eq:tpeak}
\tpeak = e^{T_0 - \tau^2} \,,
\end{equation}
leading to a peak SFR of 
\begin{equation}
\label{eq:sfrpeak}
\rm SFR_{\rm peak} = \frac{A}{\sqrt{2 \pi \tau^2}} e^{-T_0 + \tau^2 / 2} \,.
\end{equation}
By integrating the log-normal, we find its cumulative SFH,
\begin{equation}
\rm cSFR(t) \equiv \int_0^t \rm SFR(t') dt' = \frac{A}{2} \left[ 1 - \rm erf \left( -\frac{\ln t - T_0}{\tau \sqrt{2}} \right)\right]
\end{equation}
where erf denotes the error function,
\begin{equation}
{\rm erf(x)} = \frac{1}{\sqrt{\pi}} \int_{-x}^{x} e^{-t^2} dt \,.
\end{equation}
The simplicity of this expression is extremely convenient and allows an intuitive interpretation of $T_0$ as the logarithm of the time when half the stars have formed, while $\tau$ determines how wide the error function extends before and after $T_0$ in logarithmic time units.

In addition to peak time, we derive a more intuitive description of the width of an SFH in time, namely the interval between the times when the SFH first and last reaches a fraction $1/\beta$ of its peak value (Equation \ref{eq:sfrpeak}). The interval between these times in logarithmic space is 
\begin{equation}
\Delta(\ln t) = 2 \sqrt{2 \ln(\beta)} \tau\,.
\end{equation}
Translating back into linear space, 
\begin{align}
\Delta t_{\beta} &= e^{T_0 + \Delta(\ln t) / 2} - e^{T_0 - \Delta(\ln t) / 2} \nonumber \\
&= 2 \tpeak \sinh \left(\sqrt{2\ \ln(\beta)} \tau \right) \,.
\end{align}
In this paper, we use the full width at half maximum in linear time, or $\beta = 2$, to determine the width of the SFH,
\begin{equation}
\width = 2 \tpeak \sinh \left(\sqrt{2 \ln(2)} \tau \right) \,.
\end{equation}

\subsection{The Delayed-$\tau$ Model}
\label{sec:app:funcs:linexp}

One of the oldest suggestions for the form of SFHs is an exponentially declining SFR \citep[e.g.,][]{tinsley_72, bruzual_83}, physically motivated by the idea of a fixed gas supply that is gradually being exhausted. However, it is clear that this function will not fit simulated SFHs well because it corresponds to an infinitely sharp rise of the SFR from zero to a peak value. An extension of this model is the delayed-$\tau$ model \citep[e.g.,][]{gavazzi_02, lee_10, simha_14},
\begin{equation}
{\rm SFR}(t) = \frac{A}{\tau^2} (t-t_{\rm i}) \exp\left( -\frac{t-t_{\rm i}}{\tau} \right) \,,
\end{equation}
which represents a linear increase in the SFR at early times and an exponential decrease at late times. Before the initial time $t_{\rm i}$, no stars are formed. Here, we have normalized the function such that, as for the log-normal, the total stellar mass formed is $10^9 \times A$. The SFR peaks at a time $\tpeak = t_{\rm i} + \tau$ with ${\rm SFR}_{\rm peak} = A / \tau / e$. The cumulative SFR is
\begin{equation}
{\rm cSFR}(t) = A \times 10^9 \left[ \left(-\frac{t-t_{\rm i}}{\tau} - 1\right) \exp\left( -\frac{t-t_{\rm i}}{\tau} \right) + 1 \right] \,.
\end{equation}
The delayed-$\tau$ model can be simplified further by fixing $t_{\rm i}$, for example, to 1 Gyr as suggested by \citet{simha_14}. However, in Section \ref{sec:app:fits:comp}, we will show that even the delayed-$\tau$ model with three free parameters has trouble capturing the rising SFHs in a number of cases, meaning that a fixed $t_{\rm i}$ would lead to unacceptable fits.

\subsection{The Double-power-law Model}
\label{sec:app:funcs:dpl}

The double power law \citep[e.g.,][]{behroozi_13_shmr} models the SFH as
\begin{equation}
{\rm SFR}(t) = A \left[ \left( \frac{t}{\tau}\right)^B + \left( \frac{t}{\tau}\right)^{-C} \right]^{-1} \,.
\end{equation}
Due to its four free parameters, this function is more flexible than the log-normal or delayed-$\tau$ models. Unfortunately, some of the following mathematical expressions are somewhat cumbersome. We find a peak time of 
\begin{equation}
\tpeak = \tau \left( \frac{B}{C} \right)^{-\frac{1}{B + C}}
\end{equation}
and a resulting peak SFR of 
\begin{equation}
{\rm SFR}_{\rm peak} = A \left[ \left( \frac{B}{C} \right)^{-\frac{B}{B + C}} + \left( \frac{B}{C} \right)^{\frac{C}{B + C}} \right]^{-1} \,.
\end{equation}
The most inconvenient expression is that for the cSFR. If we define $x\equiv t/\tau$,
\begin{equation}
{\rm cSFR}(t) = A \times 10^9 \times \tau \times f(x,B,C) \nonumber
\end{equation}
where
\begin{equation}
f(x,B,C) \equiv \frac{x^{1-B} \left[(B-1) B x^{B+C} f_1 + C (C+1) f_2 - C(C+1)\right]}{(B-1)(C+1)(B+C)} \nonumber
\end{equation}
and 
\begin{align}
f_1 \equiv\ & _2F_1\left(1, \frac{C+1}{B+C}, \frac{C+1}{B+C} + 1, -x^{B+C} \right) \nonumber \\
f_2 \equiv\ & _2F_1\left(1, \frac{1-B}{B+C}, \frac{C+1}{B+C}, -x^{B+C} \right)
\end{align}
where $_2F_1$ is the hypergeometric function. While relatively straight-forward to implement numerically, this function is not exactly intuitive. Moreover, we could not derive a simple expression for the total stellar mass formed, making the interpretation of the normalization $A$ more complicated than in the cases of the log-normal or delayed-$\tau$ models.

\section{More Details on the Log-normal Fits}
\label{sec:app:fits}

In this appendix, we discuss certain aspects of our SFH fits in detail, including the fits to the \citetalias{pacifici_16} inferred SFHs, the relative fit qualities of different fitting functions, and an alternative parameter space for the log-normal functional form.

\subsection{The Pacifici et al. Sample}
\label{sec:app:fits:p16}

Figure~\ref{fig:pacifici_fits} shows log-normal fits to the median SFHs inferred by \citetalias{pacifici_16} in six redshift and six mass bins. Some of the bins are omitted because they contained too few objects. Visually, the quality of the fits appears satisfactory: the log-normals tend to rise slightly faster than the inferred median SFHs, but the fits are well within the 50\% confidence curves at all times. The distribution of the fit quality indicator $D$ is similar to the \illustris sample, with $D$ between 1.4\% and 5\%, and a median of 2.8\% (compared to 3.1\% for the \illustris sample).

\subsection{Comparison of Fit Qualities}
\label{sec:app:fits:comp}

\begin{figure}
\centering
\includegraphics[trim = 105mm 8mm 130mm 80mm, clip, scale=0.69]{\figdir/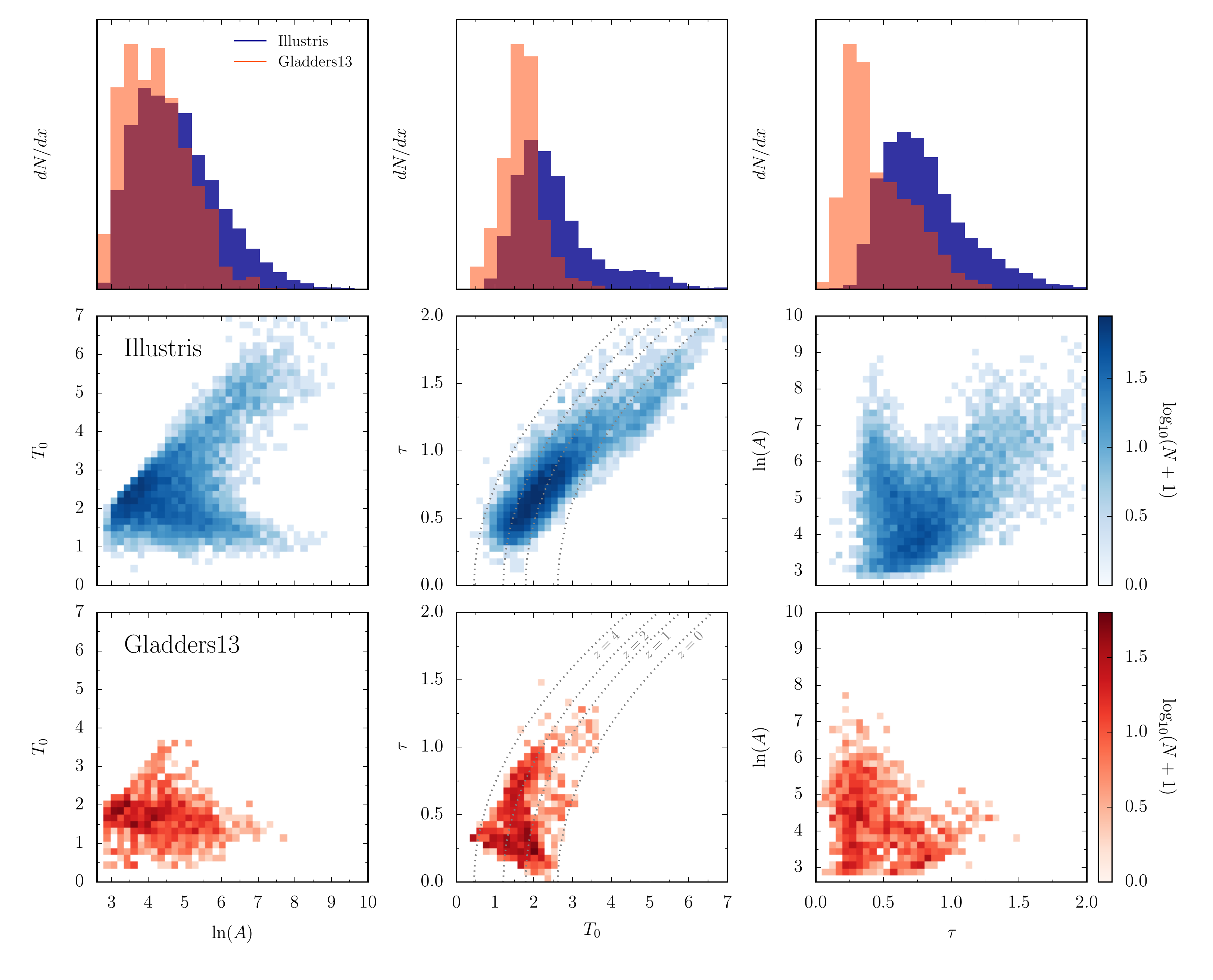}
\caption{Log-normal parameters in $T_0$--$\tau$ space for \citetalias{gladders_13_icbs4} (orange) and the \illustris high-mass sample ($M_* > 10^{10} \msun$, blue). The dotted gray lines delineate constant peak redshifts such that galaxies to the right of the $z = 0$ line will peak in the future. Comparing to Figure \ref{fig:params_alt}, the correlation between $T_0$ and $\tau$ appears slightly weaker than that between $\tpeak$ and $\width$.}
\label{fig:params_std}
\end{figure}

We have fit all galaxies in the \illustris sample with the function forms discussed in Appendix~\ref{sec:app:funcs}. Except for a very small number of failed delayed-$\tau$ fits, best-fit parameters could be obtained for each SFH and fitting model. Figure~\ref{fig:fitfuncs} shows a few representative cases. For the majority of SFHs, all three fitting functions give similar results, though with subtle differences. For example, the delayed-$\tau$ model generally struggles to capture the earliest times because of the sudden onset of star formation in that model (second SFH in Figure~\ref{fig:fitfuncs}). Due to its free declining slope, the double power law significantly outperforms the other models when sudden quenching leads to a sharp drop in the SFH (third SFH in Figure~\ref{fig:fitfuncs}). However, even the double power law cannot capture all types of quenching as shown in the rightmost panels of Figure~\ref{fig:fitfuncs}.

We quantify the goodness of fit for each model in Figure~\ref{fig:fit_quality_fitfuncs}. The fit quality of the log-normals is systematically better than for the delayed-$\tau$ model, but the log-normal is outperformed by the double power law. This difference is expected because the double power law has an extra free parameter: while the slope of the rise and decline in star formation are linked in the log-normal and delayed-$\tau$ models, they are independent in the double power law. This freedom allows for more accurate fits, but also reduces the predictive power of the model. For example, if data is available for only the early, rising part of an SFH, the declining slope of a double-power-law fit is unconstrained, meaning the function makes no prediction for the declining time scale (e.g., for very late-forming galaxies in \illustris). This issue makes the double power law somewhat less attractive for observational investigations such as \citetalias{gladders_13_icbs4} where limited data is available for each galaxy.

\subsection{The Original Log-normal Parameter space}
\label{sec:app:fits:stdparams}

Throughout the paper, we have used a log-normal parameter space of peak time and full width at half maximum. \citetalias{gladders_13_icbs4}, however, use the original parameter space given by Equation~\ref{eq:lognormal}, i.e. the half-mass time $T_0$ and the logarithmic width $\tau$. In order to allow a direct comparison between the \illustris results and the figures in \citetalias{gladders_13_icbs4}, Figure~\ref{fig:params_std} shows the same comparison as in Figure~\ref{fig:params_alt}, but in $T_0$--$\tau$ space. Our normalization $\mlimit$ only differs from \citetalias{gladders_13_icbs4}'s $A$ by a multiplicative factor, meaning the correlations with $A$ are very similar to those shown in Figure~\ref{fig:params_alt}. In $T_0$--$\tau$ space, the peak time--width correlation appears somewhat weaker, supporting the notion that the $\tpeak$--$\width$ parameter space is easier to interpret.


\bibliographystyle{aasjournal}
\bibliography{gf}

\begin{thebibliography}{}
\expandafter\ifx\csname natexlab\endcsname\relax\def\natexlab#1{#1}\fi
\providecommand{\url}[1]{\href{#1}{#1}}

\bibitem[{{Abramson} {et~al.}(2015){Abramson}, {Gladders}, {Dressler},
  {Oemler}, {Poggianti}, \& {Vulcani}}]{abramson_15}
{Abramson}, L.~E., {Gladders}, M.~D., {Dressler}, A., {et~al.} 2015, \apjl,
  801, L12

\bibitem[{{Abramson} {et~al.}(2016){Abramson}, {Gladders}, {Dressler},
  {Oemler}, {Poggianti}, \& {Vulcani}}]{abramson_16}
---. 2016, \apj, 832, 7

\bibitem[{{Becker}(2015)}]{becker_15}
{Becker}, M.~R. 2015, arXiv e-Prints, arXiv:1507.03605

\bibitem[{{Behroozi} {et~al.}(2013{\natexlab{a}}){Behroozi}, {Marchesini},
  {Wechsler}, {Muzzin}, {Papovich}, \& {Stefanon}}]{behroozi_13_numberdensity}
{Behroozi}, P.~S., {Marchesini}, D., {Wechsler}, R.~H., {et~al.}
  2013{\natexlab{a}}, \apjl, 777, L10

\bibitem[{{Behroozi} {et~al.}(2013{\natexlab{b}}){Behroozi}, {Wechsler}, \&
  {Conroy}}]{behroozi_13_shmr}
{Behroozi}, P.~S., {Wechsler}, R.~H., \& {Conroy}, C. 2013{\natexlab{b}}, \apj,
  770, 57

\bibitem[{{Bluck} {et~al.}(2016){Bluck}, {Mendel}, {Ellison}, {Patton},
  {Simard}, {Henriques}, {Torrey}, {Teimoorinia}, {Moreno}, \&
  {Starkenburg}}]{bluck_16}
{Bluck}, A.~F.~L., {Mendel}, J.~T., {Ellison}, S.~L., {et~al.} 2016, \mnras,
  462, 2559

\bibitem[{{Bouch{\'e}} {et~al.}(2010){Bouch{\'e}}, {Dekel}, {Genzel}, {Genel},
  {Cresci}, {F{\"o}rster Schreiber}, {Shapiro}, {Davies}, \&
  {Tacconi}}]{bouche_10}
{Bouch{\'e}}, N., {Dekel}, A., {Genzel}, R., {et~al.} 2010, \apj, 718, 1001

\bibitem[{{Brinchmann} {et~al.}(2004){Brinchmann}, {Charlot}, {White},
  {Tremonti}, {Kauffmann}, {Heckman}, \& {Brinkmann}}]{brinchmann_04}
{Brinchmann}, J., {Charlot}, S., {White}, S.~D.~M., {et~al.} 2004, \mnras, 351,
  1151

\bibitem[{{Brook} {et~al.}(2012){Brook}, {Stinson}, {Gibson}, {Ro{\v s}kar},
  {Wadsley}, \& {Quinn}}]{brook_12}
{Brook}, C.~B., {Stinson}, G., {Gibson}, B.~K., {et~al.} 2012, \mnras, 419, 771

\bibitem[{{Bruzual} \& {Charlot}(2003)}]{bruzual_03}
{Bruzual}, G., \& {Charlot}, S. 2003, \mnras, 344, 1000

\bibitem[{{Bruzual A.}(1983)}]{bruzual_83}
{Bruzual A.}, G. 1983, \apj, 273, 105

\bibitem[{{Bundy} {et~al.}(2006){Bundy}, {Ellis}, {Conselice}, {Taylor},
  {Cooper}, {Willmer}, {Weiner}, {Coil}, {Noeske}, \& {Eisenhardt}}]{bundy_06}
{Bundy}, K., {Ellis}, R.~S., {Conselice}, C.~J., {et~al.} 2006, \apj, 651, 120

\bibitem[{{Calvi} {et~al.}(2011){Calvi}, {Poggianti}, \& {Vulcani}}]{calvi_11}
{Calvi}, R., {Poggianti}, B.~M., \& {Vulcani}, B. 2011, \mnras, 416, 727

\bibitem[{{Carollo} {et~al.}(2013){Carollo}, {Bschorr}, {Renzini}, {Lilly},
  {Capak}, {Cibinel}, {Ilbert}, {Onodera}, {Scoville}, {Cameron}, {Mobasher},
  {Sanders}, \& {Taniguchi}}]{carollo_13}
{Carollo}, C.~M., {Bschorr}, T.~J., {Renzini}, A., {et~al.} 2013, \apj, 773,
  112

\bibitem[{{Chabrier}(2003)}]{chabrier_03}
{Chabrier}, G. 2003, \pasp, 115, 763

\bibitem[{{Chang} {et~al.}(2015){Chang}, {van der Wel}, {da Cunha}, \&
  {Rix}}]{chang_15}
{Chang}, Y.-Y., {van der Wel}, A., {da Cunha}, E., \& {Rix}, H.-W. 2015, \apjs,
  219, 8

\bibitem[{{Cohn} \& {van de Voort}(2015)}]{cohn_15}
{Cohn}, J.~D., \& {van de Voort}, F. 2015, \mnras, 446, 3253

\bibitem[{{Conroy} \& {Wechsler}(2009)}]{conroy_09}
{Conroy}, C., \& {Wechsler}, R.~H. 2009, \apj, 696, 620

\bibitem[{{Conroy} {et~al.}(2006){Conroy}, {Wechsler}, \&
  {Kravtsov}}]{conroy_06}
{Conroy}, C., {Wechsler}, R.~H., \& {Kravtsov}, A.~V. 2006, \apj, 647, 201

\bibitem[{{Cowie} {et~al.}(1996){Cowie}, {Songaila}, {Hu}, \&
  {Cohen}}]{cowie_96}
{Cowie}, L.~L., {Songaila}, A., {Hu}, E.~M., \& {Cohen}, J.~G. 1996, \aj, 112,
  839

\bibitem[{{Croton} {et~al.}(2007){Croton}, {Gao}, \& {White}}]{croton_07}
{Croton}, D.~J., {Gao}, L., \& {White}, S.~D.~M. 2007, \mnras, 374, 1303

\bibitem[{{Dalal} {et~al.}(2008){Dalal}, {White}, {Bond}, \&
  {Shirokov}}]{dalal_08}
{Dalal}, N., {White}, M., {Bond}, J.~R., \& {Shirokov}, A. 2008, \apj, 687, 12

\bibitem[{{Dav{\'e}} {et~al.}(2012){Dav{\'e}}, {Finlator}, \&
  {Oppenheimer}}]{dave_12}
{Dav{\'e}}, R., {Finlator}, K., \& {Oppenheimer}, B.~D. 2012, \mnras, 421, 98

\bibitem[{{Dav{\'e}} {et~al.}(2016){Dav{\'e}}, {Thompson}, \&
  {Hopkins}}]{dave_16}
{Dav{\'e}}, R., {Thompson}, R., \& {Hopkins}, P.~F. 2016, \mnras, 462, 3265

\bibitem[{{Davis} {et~al.}(1985){Davis}, {Efstathiou}, {Frenk}, \&
  {White}}]{davis_85}
{Davis}, M., {Efstathiou}, G., {Frenk}, C.~S., \& {White}, S.~D.~M. 1985, \apj,
  292, 371

\bibitem[{{De Lucia} \& {Blaizot}(2007)}]{delucia_07}
{De Lucia}, G., \& {Blaizot}, J. 2007, \mnras, 375, 2

\bibitem[{{Dekel} {et~al.}(2013){Dekel}, {Zolotov}, {Tweed}, {Cacciato},
  {Ceverino}, \& {Primack}}]{dekel_13}
{Dekel}, A., {Zolotov}, A., {Tweed}, D., {et~al.} 2013, \mnras, 435, 999

\bibitem[{{Dolag} {et~al.}(2009){Dolag}, {Borgani}, {Murante}, \&
  {Springel}}]{dolag_09}
{Dolag}, K., {Borgani}, S., {Murante}, G., \& {Springel}, V. 2009, \mnras, 399,
  497

\bibitem[{{Dressler}(1980)}]{dressler_80}
{Dressler}, A. 1980, \apj, 236, 351

\bibitem[{{Dressler} {et~al.}(2013){Dressler}, {Oemler}, {Poggianti},
  {Gladders}, {Abramson}, \& {Vulcani}}]{dressler_13_icbs2}
{Dressler}, A., {Oemler}, Jr., A., {Poggianti}, B.~M., {et~al.} 2013, \apj,
  770, 62

\bibitem[{{Dressler} {et~al.}(2016){Dressler}, {Kelson}, {Abramson},
  {Gladders}, {Oemler}, {Poggianti}, {Mulchaey}, {Vulcani}, {Shectman},
  {Williams}, \& {McCarthy}}]{dressler_16}
{Dressler}, A., {Kelson}, D.~D., {Abramson}, L.~E., {et~al.} 2016, \apj, 833,
  251

\bibitem[{{Eales} {et~al.}(2017){Eales}, {de Vis}, {W.~L.~Smith}, {Appah},
  {Ciesla}, {Duffield}, \& {Schofield}}]{eales_17}
{Eales}, S., {de Vis}, P., {W.~L.~Smith}, M., {et~al.} 2017, \mnras, 465, 3125

\bibitem[{{Elbaz} {et~al.}(2007){Elbaz}, {Daddi}, {Le Borgne}, {Dickinson},
  {Alexander}, {Chary}, {Starck}, {Brandt}, {Kitzbichler}, {MacDonald},
  {Nonino}, {Popesso}, {Stern}, \& {Vanzella}}]{elbaz_07}
{Elbaz}, D., {Daddi}, E., {Le Borgne}, D., {et~al.} 2007, \aap, 468, 33

\bibitem[{{Fagioli} {et~al.}(2016){Fagioli}, {Carollo}, {Renzini}, {Lilly},
  {Onodera}, \& {Tacchella}}]{fagioli_16}
{Fagioli}, M., {Carollo}, C.~M., {Renzini}, A., {et~al.} 2016, \apj, 831, 173

\bibitem[{{Feldmann} {et~al.}(2016{\natexlab{a}}){Feldmann}, {Hopkins},
  {Quataert}, {Faucher-Gigu{\`e}re}, \& {Kere{\v s}}}]{feldmann_16_a}
{Feldmann}, R., {Hopkins}, P.~F., {Quataert}, E., {Faucher-Gigu{\`e}re}, C.-A.,
  \& {Kere{\v s}}, D. 2016{\natexlab{a}}, \mnras, 458, L14

\bibitem[{{Feldmann} {et~al.}(2016{\natexlab{b}}){Feldmann}, {Quataert},
  {Hopkins}, {Faucher-Gigu{\`e}re}, \& {Kere{\v s}}}]{feldmann_16_b}
{Feldmann}, R., {Quataert}, E., {Hopkins}, P.~F., {Faucher-Gigu{\`e}re}, C.-A.,
  \& {Kere{\v s}}, D. 2016{\natexlab{b}}, \mnras\ submitted, arXiv:1610.02411

\bibitem[{{Gallagher} {et~al.}(1984){Gallagher}, {Hunter}, \&
  {Tutukov}}]{gallagher_84}
{Gallagher}, III, J.~S., {Hunter}, D.~A., \& {Tutukov}, A.~V. 1984, \apj, 284,
  544

\bibitem[{{Gao} {et~al.}(2005){Gao}, {Springel}, \& {White}}]{gao_05_assembly}
{Gao}, L., {Springel}, V., \& {White}, S.~D.~M. 2005, \mnras, 363, L66

\bibitem[{{Gavazzi} {et~al.}(2002){Gavazzi}, {Bonfanti}, {Sanvito}, {Boselli},
  \& {Scodeggio}}]{gavazzi_02}
{Gavazzi}, G., {Bonfanti}, C., {Sanvito}, G., {Boselli}, A., \& {Scodeggio}, M.
  2002, \apj, 576, 135

\bibitem[{{Geha} {et~al.}(2012){Geha}, {Blanton}, {Yan}, \& {Tinker}}]{geha_12}
{Geha}, M., {Blanton}, M.~R., {Yan}, R., \& {Tinker}, J.~L. 2012, \apj, 757, 85

\bibitem[{{Genel} {et~al.}(2014){Genel}, {Vogelsberger}, {Springel}, {Sijacki},
  {Nelson}, {Snyder}, {Rodriguez-Gomez}, {Torrey}, \& {Hernquist}}]{genel_14}
{Genel}, S., {Vogelsberger}, M., {Springel}, V., {et~al.} 2014, \mnras, 445,
  175

\bibitem[{{Gladders} {et~al.}(2013){Gladders}, {Oemler}, {Dressler},
  {Poggianti}, {Vulcani}, \& {Abramson}}]{gladders_13_icbs4}
{Gladders}, M.~D., {Oemler}, A., {Dressler}, A., {et~al.} 2013, \apj, 770, 64

\bibitem[{{Gu} {et~al.}(2016){Gu}, {Conroy}, \& {Behroozi}}]{gu_16}
{Gu}, M., {Conroy}, C., \& {Behroozi}, P. 2016, \apj, 833, 2

\bibitem[{{Guo} {et~al.}(2011){Guo}, {White}, {Boylan-Kolchin}, {De Lucia},
  {Kauffmann}, {Lemson}, {Li}, {Springel}, \& {Weinmann}}]{guo_11}
{Guo}, Q., {White}, S., {Boylan-Kolchin}, M., {et~al.} 2011, \mnras, 413, 101

\bibitem[{{Gutcke} {et~al.}(2017){Gutcke}, {Macci{\`o}}, {Dutton}, \&
  {Stinson}}]{gutcke_17}
{Gutcke}, T.~A., {Macci{\`o}}, A.~V., {Dutton}, A.~A., \& {Stinson}, G.~S.
  2017, \mnras\ submitted, arXiv:1701.01130

\bibitem[{{Hahn} {et~al.}(2016){Hahn}, {Tinker}, \& {Wetzel}}]{hahn_16}
{Hahn}, C., {Tinker}, J.~L., \& {Wetzel}, A.~R. 2016, arXiv e-Prints,
  arXiv:1609.04398

\bibitem[{{Hayward} {et~al.}(2011){Hayward}, {Kere{\v s}}, {Jonsson},
  {Narayanan}, {Cox}, \& {Hernquist}}]{hayward_11}
{Hayward}, C.~C., {Kere{\v s}}, D., {Jonsson}, P., {et~al.} 2011, \apj, 743,
  159

\bibitem[{{Hearin} \& {Watson}(2013)}]{hearin_13}
{Hearin}, A.~P., \& {Watson}, D.~F. 2013, \mnras, 435, 1313

\bibitem[{{Hearin} {et~al.}(2016){Hearin}, {Zentner}, {van den Bosch},
  {Campbell}, \& {Tollerud}}]{hearin_16_decoratedhod}
{Hearin}, A.~P., {Zentner}, A.~R., {van den Bosch}, F.~C., {Campbell}, D., \&
  {Tollerud}, E. 2016, \mnras, 460, 2552

\bibitem[{{Heavens} {et~al.}(2004){Heavens}, {Panter}, {Jimenez}, \&
  {Dunlop}}]{heavens_04}
{Heavens}, A., {Panter}, B., {Jimenez}, R., \& {Dunlop}, J. 2004, \nat, 428,
  625

\bibitem[{{Hinshaw} {et~al.}(2013){Hinshaw}, {Larson}, {Komatsu}, {Spergel},
  {Bennett}, {Dunkley}, {Nolta}, {Halpern}, {Hill}, {Odegard}, {Page}, {Smith},
  {Weiland}, {Gold}, {Jarosik}, {Kogut}, {Limon}, {Meyer}, {Tucker}, {Wollack},
  \& {Wright}}]{hinshaw_13}
{Hinshaw}, G., {Larson}, D., {Komatsu}, E., {et~al.} 2013, \apjs, 208, 19

\bibitem[{{Hopkins} \& {Beacom}(2006)}]{hopkins_06}
{Hopkins}, A.~M., \& {Beacom}, J.~F. 2006, \apj, 651, 142

\bibitem[{{Karim} {et~al.}(2011){Karim}, {Schinnerer},
  {Mart{\'{\i}}nez-Sansigre}, {Sargent}, {van der Wel}, {Rix}, {Ilbert},
  {Smol{\v c}i{\'c}}, {Carilli}, {Pannella}, {Koekemoer}, {Bell}, \&
  {Salvato}}]{karim_11}
{Karim}, A., {Schinnerer}, E., {Mart{\'{\i}}nez-Sansigre}, A., {et~al.} 2011,
  \apj, 730, 61

\bibitem[{{Kauffmann} {et~al.}(1993){Kauffmann}, {White}, \&
  {Guiderdoni}}]{kauffmann_93}
{Kauffmann}, G., {White}, S.~D.~M., \& {Guiderdoni}, B. 1993, \mnras, 264, 201

\bibitem[{{Kauffmann} {et~al.}(2003){Kauffmann}, {Heckman}, {White}, {Charlot},
  {Tremonti}, {Brinchmann}, {Bruzual}, {Peng}, {Seibert}, {Bernardi},
  {Blanton}, {Brinkmann}, {Castander}, {Cs{\'a}bai}, {Fukugita}, {Ivezic},
  {Munn}, {Nichol}, {Padmanabhan}, {Thakar}, {Weinberg}, \&
  {York}}]{kauffmann_03}
{Kauffmann}, G., {Heckman}, T.~M., {White}, S.~D.~M., {et~al.} 2003, \mnras,
  341, 33

\bibitem[{{Kelson}(2014)}]{kelson_14}
{Kelson}, D.~D. 2014, arXiv e-Prints, arXiv:1406.5191

\bibitem[{{Kelson} {et~al.}(2016){Kelson}, {Benson}, \& {Abramson}}]{kelson_16}
{Kelson}, D.~D., {Benson}, A.~J., \& {Abramson}, L.~E. 2016, arXiv e-Prints,
  arXiv:1610.06566

\bibitem[{{Kennicutt} \& {Evans}(2012)}]{kennicutt_12}
{Kennicutt}, R.~C., \& {Evans}, N.~J. 2012, \araa, 50, 531

\bibitem[{{Kennicutt}(1998)}]{kennicutt_98_review}
{Kennicutt}, Jr., R.~C. 1998, \araa, 36, 189

\bibitem[{{Kravtsov} {et~al.}(2004){Kravtsov}, {Berlind}, {Wechsler}, {Klypin},
  {Gottl{\"o}ber}, {Allgood}, \& {Primack}}]{kravtsov_04}
{Kravtsov}, A.~V., {Berlind}, A.~A., {Wechsler}, R.~H., {et~al.} 2004, \apj,
  609, 35

\bibitem[{{Kriek} {et~al.}(2007){Kriek}, {van Dokkum}, {Franx}, {Illingworth},
  {Coppi}, {F{\"o}rster Schreiber}, {Gawiser}, {Labb{\'e}}, {Lira},
  {Marchesini}, {Quadri}, {Rudnick}, {Taylor}, {Urry}, \& {van der
  Werf}}]{kriek_07}
{Kriek}, M., {van Dokkum}, P.~G., {Franx}, M., {et~al.} 2007, \apj, 669, 776

\bibitem[{{Lee} {et~al.}(2010){Lee}, {Ferguson}, {Somerville}, {Wiklind}, \&
  {Giavalisco}}]{lee_10}
{Lee}, S.-K., {Ferguson}, H.~C., {Somerville}, R.~S., {Wiklind}, T., \&
  {Giavalisco}, M. 2010, \apj, 725, 1644

\bibitem[{{Leitner}(2012)}]{leitner_12}
{Leitner}, S.~N. 2012, \apj, 745, 149

\bibitem[{{Leja} {et~al.}(2017){Leja}, {Johnson}, {Conroy}, {van Dokkum}, \&
  {Byler}}]{leja_16}
{Leja}, J., {Johnson}, B.~D., {Conroy}, C., {van Dokkum}, P.~G., \& {Byler}, N.
  2017, \apj, 837, 170

\bibitem[{{Leja} {et~al.}(2013){Leja}, {van Dokkum}, \& {Franx}}]{leja_13}
{Leja}, J., {van Dokkum}, P., \& {Franx}, M. 2013, \apj, 766, 33

\bibitem[{{Leja} {et~al.}(2015){Leja}, {van Dokkum}, {Franx}, \&
  {Whitaker}}]{leja_15}
{Leja}, J., {van Dokkum}, P.~G., {Franx}, M., \& {Whitaker}, K.~E. 2015, \apj,
  798, 115

\bibitem[{{Lewis} {et~al.}(2015){Lewis}, {Dolphin}, {Dalcanton}, {Weisz},
  {Williams}, {Bell}, {Seth}, {Simones}, {Skillman}, {Choi}, {Fouesneau},
  {Guhathakurta}, {Johnson}, {Kalirai}, {Leroy}, {Monachesi}, {Rix}, \&
  {Schruba}}]{lewis_15}
{Lewis}, A.~R., {Dolphin}, A.~E., {Dalcanton}, J.~J., {et~al.} 2015, \apj, 805,
  183

\bibitem[{{Lilly} {et~al.}(2013){Lilly}, {Carollo}, {Pipino}, {Renzini}, \&
  {Peng}}]{lilly_13}
{Lilly}, S.~J., {Carollo}, C.~M., {Pipino}, A., {Renzini}, A., \& {Peng}, Y.
  2013, \apj, 772, 119

\bibitem[{{Lilly} {et~al.}(1996){Lilly}, {Le Fevre}, {Hammer}, \&
  {Crampton}}]{lilly_96}
{Lilly}, S.~J., {Le Fevre}, O., {Hammer}, F., \& {Crampton}, D. 1996, \apjl,
  460, L1

\bibitem[{{Lin} {et~al.}(2016){Lin}, {Mandelbaum}, {Huang}, {Huang}, {Dalal},
  {Diemer}, {Jian}, \& {Kravtsov}}]{lin_16}
{Lin}, Y.-T., {Mandelbaum}, R., {Huang}, Y.-H., {et~al.} 2016, \apj, 819, 119

\bibitem[{{Madau} \& {Dickinson}(2014)}]{madau_14}
{Madau}, P., \& {Dickinson}, M. 2014, \araa, 52, 415

\bibitem[{{Madau} {et~al.}(1998){Madau}, {Pozzetti}, \& {Dickinson}}]{madau_98}
{Madau}, P., {Pozzetti}, L., \& {Dickinson}, M. 1998, \apj, 498, 106

\bibitem[{{McBride} {et~al.}(2009){McBride}, {Fakhouri}, \& {Ma}}]{mcbride_09}
{McBride}, J., {Fakhouri}, O., \& {Ma}, C.-P. 2009, \mnras, 398, 1858

\bibitem[{{McDermid} {et~al.}(2015){McDermid}, {Alatalo}, {Blitz}, {Bournaud},
  {Bureau}, {Cappellari}, {Crocker}, {Davies}, {Davis}, {de Zeeuw}, {Duc},
  {Emsellem}, {Khochfar}, {Krajnovi{\'c}}, {Kuntschner}, {Morganti}, {Naab},
  {Oosterloo}, {Sarzi}, {Scott}, {Serra}, {Weijmans}, \& {Young}}]{mcdermid_15}
{McDermid}, R.~M., {Alatalo}, K., {Blitz}, L., {et~al.} 2015, \mnras, 448, 3484

\bibitem[{{Mistani} {et~al.}(2016){Mistani}, {Sales}, {Pillepich},
  {Sanchez-Janssen}, {Vogelsberger}, {Nelson}, {Rodriguez-Gomez}, {Torrey}, \&
  {Hernquist}}]{mistani_16}
{Mistani}, P.~A., {Sales}, L.~V., {Pillepich}, A., {et~al.} 2016, \mnras, 455,
  2323

\bibitem[{{Mitra} {et~al.}(2017){Mitra}, {Dav{\'e}}, {Simha}, \&
  {Finlator}}]{mitra_17}
{Mitra}, S., {Dav{\'e}}, R., {Simha}, V., \& {Finlator}, K. 2017, \mnras, 464,
  2766

\bibitem[{{Miyatake} {et~al.}(2016){Miyatake}, {More}, {Takada}, {Spergel},
  {Mandelbaum}, {Rykoff}, \& {Rozo}}]{miyatake_16}
{Miyatake}, H., {More}, S., {Takada}, M., {et~al.} 2016, Physical Review
  Letters, 116, 041301

\bibitem[{{More} {et~al.}(2009){More}, {van den Bosch}, {Cacciato}, {Mo},
  {Yang}, \& {Li}}]{more_09_ii}
{More}, S., {van den Bosch}, F.~C., {Cacciato}, M., {et~al.} 2009, \mnras, 392,
  801

\bibitem[{{Moster} {et~al.}(2013){Moster}, {Naab}, \& {White}}]{moster_13}
{Moster}, B.~P., {Naab}, T., \& {White}, S.~D.~M. 2013, \mnras, 428, 3121

\bibitem[{{Neistein} {et~al.}(2006){Neistein}, {van den Bosch}, \&
  {Dekel}}]{neistein_06}
{Neistein}, E., {van den Bosch}, F.~C., \& {Dekel}, A. 2006, \mnras, 372, 933

\bibitem[{{Noeske} {et~al.}(2007){Noeske}, {Weiner}, {Faber}, {Papovich},
  {Koo}, {Somerville}, {Bundy}, {Conselice}, {Newman}, {Schiminovich}, {Le
  Floc'h}, {Coil}, {Rieke}, {Lotz}, {Primack}, {Barmby}, {Cooper}, {Davis},
  {Ellis}, {Fazio}, {Guhathakurta}, {Huang}, {Kassin}, {Martin}, {Phillips},
  {Rich}, {Small}, {Willmer}, \& {Wilson}}]{noeske_07}
{Noeske}, K.~G., {Weiner}, B.~J., {Faber}, S.~M., {et~al.} 2007, \apjl, 660,
  L43

\bibitem[{{Oemler}(1974)}]{oemler_74}
{Oemler}, Jr., A. 1974, \apj, 194, 1

\bibitem[{{Oemler} {et~al.}(2016){Oemler}, {Abramson}, {Gladders}, {Dressler},
  {Poggianti}, \& {Vulcani}}]{oemler_16}
{Oemler}, Jr, A., {Abramson}, L.~E., {Gladders}, M.~D., {et~al.} 2016, arXiv
  e-Prints, arXiv:1611.05932

\bibitem[{{Oemler} {et~al.}(2013{\natexlab{a}}){Oemler}, {Dressler},
  {Gladders}, {Fritz}, {Poggianti}, {Vulcani}, \& {Abramson}}]{oemler_13_icbs3}
{Oemler}, Jr., A., {Dressler}, A., {Gladders}, M.~G., {et~al.}
  2013{\natexlab{a}}, \apj, 770, 63

\bibitem[{{Oemler} {et~al.}(2013{\natexlab{b}}){Oemler}, {Dressler},
  {Gladders}, {Rigby}, {Bai}, {Kelson}, {Villanueva}, {Fritz}, {Rieke},
  {Poggianti}, \& {Vulcani}}]{oemler_13_icbs1}
---. 2013{\natexlab{b}}, \apj, 770, 61

\bibitem[{{Oesch} {et~al.}(2010){Oesch}, {Bouwens}, {Carollo}, {Illingworth},
  {Trenti}, {Stiavelli}, {Magee}, {Labb{\'e}}, \& {Franx}}]{oesch_10}
{Oesch}, P.~A., {Bouwens}, R.~J., {Carollo}, C.~M., {et~al.} 2010, \apjl, 709,
  L21

\bibitem[{{Pacifici} {et~al.}(2012){Pacifici}, {Charlot}, {Blaizot}, \&
  {Brinchmann}}]{pacifici_12}
{Pacifici}, C., {Charlot}, S., {Blaizot}, J., \& {Brinchmann}, J. 2012, \mnras,
  421, 2002

\bibitem[{{Pacifici} {et~al.}(2013){Pacifici}, {Kassin}, {Weiner}, {Charlot},
  \& {Gardner}}]{pacifici_13}
{Pacifici}, C., {Kassin}, S.~A., {Weiner}, B., {Charlot}, S., \& {Gardner},
  J.~P. 2013, \apjl, 762, L15

\bibitem[{{Pacifici} {et~al.}(2016{\natexlab{a}}){Pacifici}, {Oh}, {Oh}, {Lee},
  \& {Yi}}]{pacifici_16_timing}
{Pacifici}, C., {Oh}, S., {Oh}, K., {Lee}, J., \& {Yi}, S.~K.
  2016{\natexlab{a}}, \apj, 824, 45

\bibitem[{{Pacifici} {et~al.}(2016{\natexlab{b}}){Pacifici}, {Kassin},
  {Weiner}, {Holden}, {Gardner}, {Faber}, {Ferguson}, {Koo}, {Primack}, {Bell},
  {Dekel}, {Gawiser}, {Giavalisco}, {Rafelski}, {Simons}, {Barro}, {Croton},
  {Dav{\'e}}, {Fontana}, {Grogin}, {Koekemoer}, {Lee}, {Salmon}, {Somerville},
  \& {Behroozi}}]{pacifici_16}
{Pacifici}, C., {Kassin}, S.~A., {Weiner}, B.~J., {et~al.} 2016{\natexlab{b}},
  \apj, 832, 79

\bibitem[{{Patel} {et~al.}(2013){Patel}, {Fumagalli}, {Franx}, {van Dokkum},
  {van der Wel}, {Leja}, {Labb{\'e}}, {Brammer}, {Skelton}, {Momcheva},
  {Whitaker}, {Lundgren}, {Muzzin}, {Quadri}, {Nelson}, {Wake}, \&
  {Rix}}]{patel_13_mw}
{Patel}, S.~G., {Fumagalli}, M., {Franx}, M., {et~al.} 2013, \apj, 778, 115

\bibitem[{{Peacock} \& {Smith}(2000)}]{peacock_00}
{Peacock}, J.~A., \& {Smith}, R.~E. 2000, \mnras, 318, 1144

\bibitem[{{Peng} {et~al.}(2015){Peng}, {Maiolino}, \& {Cochrane}}]{peng_15}
{Peng}, Y., {Maiolino}, R., \& {Cochrane}, R. 2015, \nat, 521, 192

\bibitem[{{Peng} {et~al.}(2012){Peng}, {Lilly}, {Renzini}, \&
  {Carollo}}]{peng_12}
{Peng}, Y.-j., {Lilly}, S.~J., {Renzini}, A., \& {Carollo}, M. 2012, \apj, 757,
  4

\bibitem[{{Peng} {et~al.}(2010){Peng}, {Lilly}, {Kova{\v c}}, {Bolzonella},
  {Pozzetti}, {Renzini}, {Zamorani}, {Ilbert}, {Knobel}, {Iovino}, {Maier},
  {Cucciati}, {Tasca}, {Carollo}, {Silverman}, {Kampczyk}, {de Ravel},
  {Sanders}, {Scoville}, {Contini}, {Mainieri}, {Scodeggio}, {Kneib}, {Le
  F{\`e}vre}, {Bardelli}, {Bongiorno}, {Caputi}, {Coppa}, {de la Torre},
  {Franzetti}, {Garilli}, {Lamareille}, {Le Borgne}, {Le Brun}, {Mignoli},
  {Perez Montero}, {Pello}, {Ricciardelli}, {Tanaka}, {Tresse}, {Vergani},
  {Welikala}, {Zucca}, {Oesch}, {Abbas}, {Barnes}, {Bordoloi}, {Bottini},
  {Cappi}, {Cassata}, {Cimatti}, {Fumana}, {Hasinger}, {Koekemoer},
  {Leauthaud}, {Maccagni}, {Marinoni}, {McCracken}, {Memeo}, {Meneux}, {Nair},
  {Porciani}, {Presotto}, \& {Scaramella}}]{peng_10}
{Peng}, Y.-j., {Lilly}, S.~J., {Kova{\v c}}, K., {et~al.} 2010, \apj, 721, 193

\bibitem[{{Poggianti} {et~al.}(2013){Poggianti}, {Calvi}, {Bindoni},
  {D'Onofrio}, {Moretti}, {Valentinuzzi}, {Fasano}, {Fritz}, {De Lucia},
  {Vulcani}, {Bettoni}, {Gullieuszik}, \& {Omizzolo}}]{poggianti_13}
{Poggianti}, B.~M., {Calvi}, R., {Bindoni}, D., {et~al.} 2013, \apj, 762, 77

\bibitem[{{Postman} \& {Geller}(1984)}]{postman_84}
{Postman}, M., \& {Geller}, M.~J. 1984, \apj, 281, 95

\bibitem[{{Reddick} {et~al.}(2013){Reddick}, {Wechsler}, {Tinker}, \&
  {Behroozi}}]{reddick_13}
{Reddick}, R.~M., {Wechsler}, R.~H., {Tinker}, J.~L., \& {Behroozi}, P.~S.
  2013, \apj, 771, 30

\bibitem[{{Rodriguez-Gomez} {et~al.}(2015){Rodriguez-Gomez}, {Genel},
  {Vogelsberger}, {Sijacki}, {Pillepich}, {Sales}, {Torrey}, {Snyder},
  {Nelson}, {Springel}, {Ma}, \& {Hernquist}}]{rodriguezgomez_15}
{Rodriguez-Gomez}, V., {Genel}, S., {Vogelsberger}, M., {et~al.} 2015, \mnras,
  449, 49

\bibitem[{{Rodriguez-Gomez} {et~al.}(2016){Rodriguez-Gomez}, {Pillepich},
  {Sales}, {Genel}, {Vogelsberger}, {Zhu}, {Wellons}, {Nelson}, {Torrey},
  {Springel}, {Ma}, \& {Hernquist}}]{rodriguezgomez_16}
{Rodriguez-Gomez}, V., {Pillepich}, A., {Sales}, L.~V., {et~al.} 2016, \mnras,
  458, 2371

\bibitem[{{Rodriguez-Gomez} {et~al.}(2017){Rodriguez-Gomez}, {Sales}, {Genel},
  {Pillepich}, {Zjupa}, {Nelson}, {Griffen}, {Torrey}, {Snyder},
  {Vogelsberger}, {Springel}, {Ma}, \& {Hernquist}}]{rodriguezgomez_17_spin}
{Rodriguez-Gomez}, V., {Sales}, L.~V., {Genel}, S., {et~al.} 2017, \mnras, 467,
  3083

\bibitem[{{Sandage}(1986)}]{sandage_86}
{Sandage}, A. 1986, \aap, 161, 89

\bibitem[{{Schawinski} {et~al.}(2014){Schawinski}, {Urry}, {Simmons},
  {Fortson}, {Kaviraj}, {Keel}, {Lintott}, {Masters}, {Nichol}, {Sarzi},
  {Skibba}, {Treister}, {Willett}, {Wong}, \& {Yi}}]{schawinski_14}
{Schawinski}, K., {Urry}, C.~M., {Simmons}, B.~D., {et~al.} 2014, \mnras, 440,
  889

\bibitem[{{Schaye} {et~al.}(2015){Schaye}, {Crain}, {Bower}, {Furlong},
  {Schaller}, {Theuns}, {Dalla Vecchia}, {Frenk}, {McCarthy}, {Helly},
  {Jenkins}, {Rosas-Guevara}, {White}, {Baes}, {Booth}, {Camps}, {Navarro},
  {Qu}, {Rahmati}, {Sawala}, {Thomas}, \& {Trayford}}]{schaye_15}
{Schaye}, J., {Crain}, R.~A., {Bower}, R.~G., {et~al.} 2015, \mnras, 446, 521

\bibitem[{{Seljak}(2000)}]{seljak_00_wl}
{Seljak}, U. 2000, \mnras, 318, 203

\bibitem[{{Sijacki} {et~al.}(2015){Sijacki}, {Vogelsberger}, {Genel},
  {Springel}, {Torrey}, {Snyder}, {Nelson}, \& {Hernquist}}]{sijacki_15}
{Sijacki}, D., {Vogelsberger}, M., {Genel}, S., {et~al.} 2015, \mnras, 452, 575

\bibitem[{{Simha} {et~al.}(2014){Simha}, {Weinberg}, {Conroy}, {Dave},
  {Fardal}, {Katz}, \& {Oppenheimer}}]{simha_14}
{Simha}, V., {Weinberg}, D.~H., {Conroy}, C., {et~al.} 2014, arXiv e-Prints,
  arXiv:1404.0402

\bibitem[{{Skillman} {et~al.}(2014){Skillman}, {Hidalgo}, {Weisz}, {Monelli},
  {Gallart}, {Aparicio}, {Bernard}, {Boylan-Kolchin}, {Cassisi}, {Cole},
  {Dolphin}, {Ferguson}, {Mayer}, {Navarro}, {Stetson}, \&
  {Tolstoy}}]{skillman_14}
{Skillman}, E.~D., {Hidalgo}, S.~L., {Weisz}, D.~R., {et~al.} 2014, \apj, 786,
  44

\bibitem[{{Somerville} {et~al.}(2001){Somerville}, {Primack}, \&
  {Faber}}]{somerville_01}
{Somerville}, R.~S., {Primack}, J.~R., \& {Faber}, S.~M. 2001, \mnras, 320, 504

\bibitem[{{Sparre} {et~al.}(2017){Sparre}, {Hayward}, {Feldmann},
  {Faucher-Gigu{\`e}re}, {Muratov}, {Kere{\v s}}, \& {Hopkins}}]{sparre_17}
{Sparre}, M., {Hayward}, C.~C., {Feldmann}, R., {et~al.} 2017, \mnras, 466, 88

\bibitem[{{Sparre} {et~al.}(2015){Sparre}, {Hayward}, {Springel},
  {Vogelsberger}, {Genel}, {Torrey}, {Nelson}, {Sijacki}, \&
  {Hernquist}}]{sparre_15}
{Sparre}, M., {Hayward}, C.~C., {Springel}, V., {et~al.} 2015, \mnras, 447,
  3548

\bibitem[{{Speagle} {et~al.}(2014){Speagle}, {Steinhardt}, {Capak}, \&
  {Silverman}}]{speagle_14}
{Speagle}, J.~S., {Steinhardt}, C.~L., {Capak}, P.~L., \& {Silverman}, J.~D.
  2014, \apjs, 214, 15

\bibitem[{{Springel}(2010)}]{springel_10}
{Springel}, V. 2010, \mnras, 401, 791

\bibitem[{{Springel} \& {Hernquist}(2003)}]{springel_03}
{Springel}, V., \& {Hernquist}, L. 2003, \mnras, 339, 289

\bibitem[{{Springel} {et~al.}(2001){Springel}, {White}, {Tormen}, \&
  {Kauffmann}}]{springel_01_subfind}
{Springel}, V., {White}, S.~D.~M., {Tormen}, G., \& {Kauffmann}, G. 2001,
  \mnras, 328, 726

\bibitem[{{Springel} {et~al.}(2005){Springel}, {White}, {Jenkins}, {Frenk},
  {Yoshida}, {Gao}, {Navarro}, {Thacker}, {Croton}, {Helly}, {Peacock}, {Cole},
  {Thomas}, {Couchman}, {Evrard}, {Colberg}, \&
  {Pearce}}]{springel_05_millennium}
{Springel}, V., {White}, S.~D.~M., {Jenkins}, A., {et~al.} 2005, \nat, 435, 629

\bibitem[{{Tacchella} {et~al.}(2016{\natexlab{a}}){Tacchella}, {Dekel},
  {Carollo}, {Ceverino}, {DeGraf}, {Lapiner}, {Mandelker}, \&
  {Primack}}]{tacchella_16_profiles}
{Tacchella}, S., {Dekel}, A., {Carollo}, C.~M., {et~al.} 2016{\natexlab{a}},
  \mnras, 458, 242

\bibitem[{{Tacchella} {et~al.}(2016{\natexlab{b}}){Tacchella}, {Dekel},
  {Carollo}, {Ceverino}, {DeGraf}, {Lapiner}, {Mandelker}, \& {Primack
  Joel}}]{tacchella_16_ms}
---. 2016{\natexlab{b}}, \mnras, 457, 2790

\bibitem[{{Tacchella} {et~al.}(2013){Tacchella}, {Trenti}, \&
  {Carollo}}]{tacchella_13}
{Tacchella}, S., {Trenti}, M., \& {Carollo}, C.~M. 2013, \apjl, 768, L37

\bibitem[{{Tasitsiomi} {et~al.}(2004){Tasitsiomi}, {Kravtsov}, {Gottl{\"o}ber},
  \& {Klypin}}]{tasitsiomi_04_clusterprof}
{Tasitsiomi}, A., {Kravtsov}, A.~V., {Gottl{\"o}ber}, S., \& {Klypin}, A.~A.
  2004, \apj, 607, 125

\bibitem[{{Thomas} {et~al.}(2005){Thomas}, {Maraston}, {Bender}, \& {Mendes de
  Oliveira}}]{thomas_05}
{Thomas}, D., {Maraston}, C., {Bender}, R., \& {Mendes de Oliveira}, C. 2005,
  \apj, 621, 673

\bibitem[{{Tinker} {et~al.}(2016){Tinker}, {Wetzel}, {Conroy}, \&
  {Mao}}]{tinker_16}
{Tinker}, J., {Wetzel}, A., {Conroy}, C., \& {Mao}, Y.-Y. 2016, arXiv e-Prints,
  arXiv:1609.03388

\bibitem[{{Tinsley}(1968)}]{tinsley_68}
{Tinsley}, B.~M. 1968, \apj, 151, 547

\bibitem[{{Tinsley}(1972)}]{tinsley_72}
---. 1972, \aap, 20, 383

\bibitem[{{Tojeiro} {et~al.}(2007){Tojeiro}, {Heavens}, {Jimenez}, \&
  {Panter}}]{tojeiro_07}
{Tojeiro}, R., {Heavens}, A.~F., {Jimenez}, R., \& {Panter}, B. 2007, \mnras,
  381, 1252

\bibitem[{{Tojeiro} {et~al.}(2009){Tojeiro}, {Wilkins}, {Heavens}, {Panter}, \&
  {Jimenez}}]{tojeiro_09}
{Tojeiro}, R., {Wilkins}, S., {Heavens}, A.~F., {Panter}, B., \& {Jimenez}, R.
  2009, \apjs, 185, 1

\bibitem[{{Torrey} {et~al.}(2012){Torrey}, {Cox}, {Kewley}, \&
  {Hernquist}}]{torrey_12}
{Torrey}, P., {Cox}, T.~J., {Kewley}, L., \& {Hernquist}, L. 2012, \apj, 746,
  108

\bibitem[{{Torrey} {et~al.}(2014){Torrey}, {Vogelsberger}, {Genel}, {Sijacki},
  {Springel}, \& {Hernquist}}]{torrey_14}
{Torrey}, P., {Vogelsberger}, M., {Genel}, S., {et~al.} 2014, \mnras, 438, 1985

\bibitem[{{Torrey} {et~al.}(2017){Torrey}, {Wellons}, {Ma}, {Hopkins}, \&
  {Vogelsberger}}]{torrey_16_numberdensity}
{Torrey}, P., {Wellons}, S., {Ma}, C.-P., {Hopkins}, P.~F., \& {Vogelsberger},
  M. 2017, \mnras, 467, 4872

\bibitem[{{Torrey} {et~al.}(2015){Torrey}, {Wellons}, {Machado}, {Griffen},
  {Nelson}, {Rodriguez-Gomez}, {McKinnon}, {Pillepich}, {Ma}, {Vogelsberger},
  {Springel}, \& {Hernquist}}]{torrey_15_numberdensity}
{Torrey}, P., {Wellons}, S., {Machado}, F., {et~al.} 2015, \mnras, 454, 2770

\bibitem[{{Treu} {et~al.}(2005){Treu}, {Ellis}, {Liao}, \& {van
  Dokkum}}]{treu_05}
{Treu}, T., {Ellis}, R.~S., {Liao}, T.~X., \& {van Dokkum}, P.~G. 2005, \apjl,
  622, L5

\bibitem[{{{\"U}bler} {et~al.}(2014){{\"U}bler}, {Naab}, {Oser}, {Aumer},
  {Sales}, \& {White}}]{uebler_14}
{{\"U}bler}, H., {Naab}, T., {Oser}, L., {et~al.} 2014, \mnras, 443, 2092

\bibitem[{{Valentinuzzi} {et~al.}(2010){Valentinuzzi}, {Fritz}, {Poggianti},
  {Cava}, {Bettoni}, {Fasano}, {D'Onofrio}, {Couch}, {Dressler}, {Moles},
  {Moretti}, {Omizzolo}, {Kj{\ae}rgaard}, {Vanzella}, \&
  {Varela}}]{valentinuzzi_10}
{Valentinuzzi}, T., {Fritz}, J., {Poggianti}, B.~M., {et~al.} 2010, \apj, 712,
  226

\bibitem[{{van den Bosch} {et~al.}(2008){van den Bosch}, {Aquino}, {Yang},
  {Mo}, {Pasquali}, {McIntosh}, {Weinmann}, \& {Kang}}]{vandenbosch_08}
{van den Bosch}, F.~C., {Aquino}, D., {Yang}, X., {et~al.} 2008, \mnras, 387,
  79

\bibitem[{{van der Wel} {et~al.}(2014){van der Wel}, {Franx}, {van Dokkum},
  {Skelton}, {Momcheva}, {Whitaker}, {Brammer}, {Bell}, {Rix}, {Wuyts},
  {Ferguson}, {Holden}, {Barro}, {Koekemoer}, {Chang}, {McGrath},
  {H{\"a}ussler}, {Dekel}, {Behroozi}, {Fumagalli}, {Leja}, {Lundgren},
  {Maseda}, {Nelson}, {Wake}, {Patel}, {Labb{\'e}}, {Faber}, {Grogin}, \&
  {Kocevski}}]{vanderwel_14}
{van der Wel}, A., {Franx}, M., {van Dokkum}, P.~G., {et~al.} 2014, \apj, 788,
  28

\bibitem[{{van Dokkum} {et~al.}(2013){van Dokkum}, {Leja}, {Nelson}, {Patel},
  {Skelton}, {Momcheva}, {Brammer}, {Whitaker}, {Lundgren}, {Fumagalli},
  {Conroy}, {F{\"o}rster Schreiber}, {Franx}, {Kriek}, {Labb{\'e}},
  {Marchesini}, {Rix}, {van der Wel}, \& {Wuyts}}]{vandokkum_13}
{van Dokkum}, P.~G., {Leja}, J., {Nelson}, E.~J., {et~al.} 2013, \apjl, 771,
  L35

\bibitem[{{Vogelsberger} {et~al.}(2013){Vogelsberger}, {Genel}, {Sijacki},
  {Torrey}, {Springel}, \& {Hernquist}}]{vogelsberger_13}
{Vogelsberger}, M., {Genel}, S., {Sijacki}, D., {et~al.} 2013, \mnras, 436,
  3031

\bibitem[{{Vogelsberger} {et~al.}(2014{\natexlab{a}}){Vogelsberger}, {Genel},
  {Springel}, {Torrey}, {Sijacki}, {Xu}, {Snyder}, {Bird}, {Nelson}, \&
  {Hernquist}}]{vogelsberger_14_nature}
{Vogelsberger}, M., {Genel}, S., {Springel}, V., {et~al.} 2014{\natexlab{a}},
  \nat, 509, 177

\bibitem[{{Vogelsberger} {et~al.}(2014{\natexlab{b}}){Vogelsberger}, {Genel},
  {Springel}, {Torrey}, {Sijacki}, {Xu}, {Snyder}, {Nelson}, \&
  {Hernquist}}]{vogelsberger_14_illustris}
---. 2014{\natexlab{b}}, \mnras, 444, 1518

\bibitem[{{Wechsler} {et~al.}(2002){Wechsler}, {Bullock}, {Primack},
  {Kravtsov}, \& {Dekel}}]{wechsler_02_halo_assembly}
{Wechsler}, R.~H., {Bullock}, J.~S., {Primack}, J.~R., {Kravtsov}, A.~V., \&
  {Dekel}, A. 2002, \apj, 568, 52

\bibitem[{{Wechsler} {et~al.}(2006){Wechsler}, {Zentner}, {Bullock},
  {Kravtsov}, \& {Allgood}}]{wechsler_06}
{Wechsler}, R.~H., {Zentner}, A.~R., {Bullock}, J.~S., {Kravtsov}, A.~V., \&
  {Allgood}, B. 2006, \apj, 652, 71

\bibitem[{{Weisz} {et~al.}(2014){Weisz}, {Dolphin}, {Skillman}, {Holtzman},
  {Gilbert}, {Dalcanton}, \& {Williams}}]{weisz_14}
{Weisz}, D.~R., {Dolphin}, A.~E., {Skillman}, E.~D., {et~al.} 2014, \apj, 789,
  147

\bibitem[{{Weisz} {et~al.}(2011){Weisz}, {Dalcanton}, {Williams}, {Gilbert},
  {Skillman}, {Seth}, {Dolphin}, {McQuinn}, {Gogarten}, {Holtzman}, {Rosema},
  {Cole}, {Karachentsev}, \& {Zaritsky}}]{weisz_11}
{Weisz}, D.~R., {Dalcanton}, J.~J., {Williams}, B.~F., {et~al.} 2011, \apj,
  739, 5

\bibitem[{{Wellons} \& {Torrey}(2017)}]{wellons_16_numberdensity}
{Wellons}, S., \& {Torrey}, P. 2017, \mnras, 467, 3887

\bibitem[{{Wetzel} {et~al.}(2013){Wetzel}, {Tinker}, {Conroy}, \& {van den
  Bosch}}]{wetzel_13}
{Wetzel}, A.~R., {Tinker}, J.~L., {Conroy}, C., \& {van den Bosch}, F.~C. 2013,
  \mnras, 432, 336

\bibitem[{{Whitaker} {et~al.}(2012){Whitaker}, {van Dokkum}, {Brammer}, \&
  {Franx}}]{whitaker_12}
{Whitaker}, K.~E., {van Dokkum}, P.~G., {Brammer}, G., \& {Franx}, M. 2012,
  \apjl, 754, L29

\bibitem[{{Williams} {et~al.}(2011){Williams}, {Dalcanton}, {Johnson}, {Weisz},
  {Seth}, {Dolphin}, {Gilbert}, {Skillman}, {Rosema}, {Gogarten}, {Holtzman},
  \& {de Jong}}]{williams_11}
{Williams}, B.~F., {Dalcanton}, J.~J., {Johnson}, L.~C., {et~al.} 2011, \apjl,
  734, L22

\bibitem[{{Wuyts} {et~al.}(2011){Wuyts}, {F{\"o}rster Schreiber}, {van der
  Wel}, {Magnelli}, {Guo}, {Genzel}, {Lutz}, {Aussel}, {Barro}, {Berta},
  {Cava}, {Graci{\'a}-Carpio}, {Hathi}, {Huang}, {Kocevski}, {Koekemoer},
  {Lee}, {Le Floc'h}, {McGrath}, {Nordon}, {Popesso}, {Pozzi}, {Riguccini},
  {Rodighiero}, {Saintonge}, \& {Tacconi}}]{wuyts_11}
{Wuyts}, S., {F{\"o}rster Schreiber}, N.~M., {van der Wel}, A., {et~al.} 2011,
  \apj, 742, 96

\bibitem[{{Zentner} {et~al.}(2014){Zentner}, {Hearin}, \& {van den
  Bosch}}]{zentner_14}
{Zentner}, A.~R., {Hearin}, A.~P., \& {van den Bosch}, F.~C. 2014, \mnras, 443,
  3044

\end{thebibliography}

\end{document}